\documentclass[11pt]{article}
\usepackage[hidelinks]{hyperref}
\usepackage{tikz}
\usepackage[first=0,last=9]{lcg}
\usetikzlibrary{calc}
\usepackage{multicol}
\usepackage{ctable}
\usepackage{mdframed}
\usepackage{color, colortbl}
\usepackage{epsfig,subcaption}
\usepackage{caption}
\usepackage{amssymb, amsmath}
\usepackage{scrextend}
\usepackage{color}
\usepackage{bm}
\usepackage{amsthm}
\usepackage{authblk}
\usepackage{cite}
\usepackage{pgfplots}
\usepackage{palatino}
\usepackage{arydshln}

\usetikzlibrary{shapes,decorations,arrows,calc,arrows.meta,fit,positioning}

\usepackage{graphicx}
\usepackage{color}

\usepackage{amssymb}
\usepackage{amsmath,amsfonts,amssymb}
\usepackage{verbatim}
\usepackage{stfloats}
\usepackage[bookmarks=false]{}

\newtheorem{theorem}{Theorem}

\newtheorem{lemma}{Lemma}

\newcommand\Tstrut{\rule{0pt}{2ex}}         %
\newcommand\Bstrut{\rule[-1ex]{0pt}{0pt}}   %

\DeclareCaptionLabelFormat{adja-page}{\hrulefill\\#1 #2 \emph{(previous page)}}

\usetikzlibrary{decorations.markings}

\tikzstyle{interrupt}=[
    postaction={
        decorate,
        decoration={markings,
                    mark= at position 0.25
                          with
                          {
                          \begin{scope}[-]
                            \fill[white] (0,0) rectangle (0,0);
                            \draw (-0.15,0.25) -- (0.15,-0.25);
                             \draw (0.15,0.25) -- (-0.15,-0.25);
                             \end{scope}
                          }
                    }
                }
]

\def \Ja {100}
\def \Jb {1000}

\pgfplotsset{compat = 1.3}

\begin{document}

\title{Cross Subspace Alignment Codes \\ for Coded Distributed Batch Computation}
\author{Zhuqing Jia and Syed A. Jafar}
\affil{Center for Pervasive Communications and Computing (CPCC), UC Irvine\\
Email: \{zhuqingj, syed\}@uci.edu}
\date{}
\maketitle

\begin{abstract}
The goal of coded distributed computation is to efficiently distribute a computation task, such as matrix multiplication, $N$-linear computation, or multivariate polynomial evaluation, across $S$ servers through a coding scheme, such that the response from any $R$ servers ($R$ is called the recovery threshold) is sufficient for the user to recover the desired computed value. Current state-of-art approaches are based on either exclusively matrix-partitioning (Entangled Polynomial (EP) Codes for matrix multiplication), or exclusively batch processing  (Lagrange Coded Computing (LCC) for $N$-linear computations or multivariate polynomial evaluations). We present three related classes of codes, based on the idea of Cross-Subspace Alignment (CSA) which was introduced originally in the context of secure and private information retrieval. CSA codes are characterized by a Cauchy-Vandermonde matrix structure that facilitates interference alignment along Vandermonde terms, while the desired computations remain resolvable along the Cauchy terms. These codes are shown to unify, generalize and improve upon the state-of-art codes for distributed computing. First we introduce CSA codes for matrix multiplication, which yield LCC codes as a special case, and are shown to outperform LCC codes in general in download-limited settings. While matrix-partitioning approaches (EP codes) for distributed matrix multiplication have the advantage of flexible server computation latency,  batch processing approaches (CSA, LCC)  have significant advantages in communication costs as well as encoding and decoding complexity per matrix multiplication. In order to combine the benefits of these approaches,  we introduce Generalized CSA (GCSA) codes for matrix multiplication that bridge the extremes of matrix-partitioning and batch processing approaches and demonstrate synergistic gains due to cross subspace alignment. Finally, we introduce $N$-CSA codes for $N$-linear distributed batch computations and multivariate batch polynomial evaluations. $N$-CSA codes include LCC codes as a special case, and are in general capable of outperforming LCC codes in download-constrained settings by upto a factor of $N$. Generalizations of $N$-CSA codes to include $X$-secure data and $B$-byzantine servers are also provided.

\end{abstract}

\pagebreak

\section{Introduction}
\begin{figure}[!h]
\centering
    \begin{tikzpicture}
    \node[rectangle, help lines, fill=black!5, text=black, draw=black, minimum size=0.7cm, inner sep=0.2cm, rounded corners=0.5em] (S1) at (2cm, 1cm) { ${\mathbf{A}=(\mathbf{A}_{1},\dots,\mathbf{A}_{L})}$};

\node[rectangle, draw=black, fill=black!5, text=black,minimum size=0.7cm,  inner sep=0.2cm, rounded corners=0.5em] (Si) at (14cm, 1cm) { ${\mathbf{B}=(\mathbf{B}_{1},\dots,\mathbf{B}_{L})}$};
\node [draw, rectangle,fill=black!5, text=black, inner sep =0.2cm] (D1) at (2cm, -1.5cm) {\footnotesize Server $1$};
\node [rectangle, inner sep =0.2cm] (Ddots1) at (4cm, -1.5cm) {$\cdots$};
\node [draw, rectangle, fill=black!5, text=black, inner sep =0.2cm] (Dm) at (6cm, -1.5cm) {\footnotesize Server $i$};
\node [rectangle, inner sep =0.2cm] (Ddots2) at (8cm, -1.5cm) {$\cdots$};
\node [draw, rectangle, fill=black!5, text=black, inner sep =0.2cm] (Dn) at (10cm, -1.5cm) {\footnotesize Server $j$};
\node [rectangle, inner sep =0.2cm] (Ddots2) at (12cm, -1.5cm) {$\cdots$};
\node [draw, rectangle, fill=black!5, text=black, inner sep =0.2cm] (DN) at (14cm, -1.5cm) {\footnotesize Server $S$};

\draw [violet, thick,  ->] (S1)--(DN) node[draw, rectangle, fill=white, pos=0.3]{\scriptsize $\widetilde{A}^S$};
\draw [violet, thick,  ->] (S1)--(D1) node[draw, rectangle, fill=white, pos=0.3]{\scriptsize $\widetilde{A}^1$};
\draw [violet, thick,  ->] (S1)--(Dn) node[draw, rectangle, fill=white, pos=0.3]{\scriptsize $\widetilde{A}^j$};
\draw [violet, thick,  ->] (S1)--(Dm) node[draw, rectangle, fill=white, pos=0.3]{\scriptsize $\widetilde{A}^i$};

\draw [blue, thick, ->] (Si)--(D1) node[draw, rectangle, fill=white, pos=0.3]{\scriptsize $\widetilde{B}^1$};
\draw [blue, thick, ->] (Si)--(Dm) node[draw, rectangle, fill=white, pos=0.3]{\scriptsize $\widetilde{B}^i$};
\draw [blue, thick, ->] (Si)--(Dn) node[draw, rectangle, fill=white, pos=0.3]{\scriptsize $\widetilde{B}^j$};
\draw [blue, thick, ->] (Si)--(DN) node[draw, rectangle, fill=white, pos=0.3]{\scriptsize $\widetilde{B}^S$};

\node[circle, draw=black, fill=black!5, text=black, minimum size=0.9cm, inner sep=0] (Uj) at (8cm, -4.5cm) { \small User};

\draw [red, thick, ->] (D1)--(Uj) node[draw, rectangle, fill=white, pos=0.25]{\scriptsize $Y_1$};
\draw [red, thick, ->] (Dn)--(Uj) node[draw, rectangle, fill=white, pos=0.25]{\scriptsize $Y_j$};
\draw [red, thick, interrupt, ->] (Dm)--(Uj);
\draw [red, thick, ->] (DN)--(Uj) node[draw, rectangle, fill=white, pos=0.25]{\scriptsize $Y_S$};
\node[below=0.5cm of Uj, minimum size=0.3cm, inner sep=0.1cm] (P) {\scriptsize  $\mathbf{A}\mathbf{B}=(\mathbf{A}_{1}\mathbf{B}_{1},\dots,\mathbf{A}_{L}\mathbf{B}_{L})$};
\draw[->](Uj)--(P);
\node [rectangle, rounded corners=0.5em, inner sep =0.2cm, fill=yellow, fill opacity=0.75, text opacity=1] at (8cm, -3.25cm) {A total of $R$ answers downloaded};
\end{tikzpicture}
\caption{\it The CDBMM problem. Source (master) nodes generate matrices $\mathbf{A}=(\mathbf{A}_{1},\mathbf{A}_{2},\cdots,\mathbf{A}_{L})$ and $\mathbf{B}=(\mathbf{B}_{1},\mathbf{B}_{2},\cdots,\mathbf{B}_{L})$, and upload them to $S$ distributed servers in coded form $\widetilde{A}^{[s]}$, $\widetilde{B}^{[s]}$, respectively. For all $l\in[L]$, ${\bf A}_l$ and ${\bf B}_l$ are $\lambda\times\mu$ and $\mu\times \kappa$ matrices, respectively, over a field $\mathbb{F}$. The $s^{th}$ server computes the answer $Y_s$, which is a function of all information available to it, i.e., $\widetilde{A}^{s}$ and $\widetilde{B}^{s}$. For effective straggler (e.g., Server $i$ in the figure) mitigation, upon downloading answers from any $R$ servers, where $R<S$, the user must be able to recover the product $\mathbf{A}\mathbf{B}=(\mathbf{A}_{1}\mathbf{B}_{1},\mathbf{A}_{2}\mathbf{B}_{2},\dots,\mathbf{A}_{L}\mathbf{B}_{L})$. }
    \label{fig:DBMM}
\end{figure}
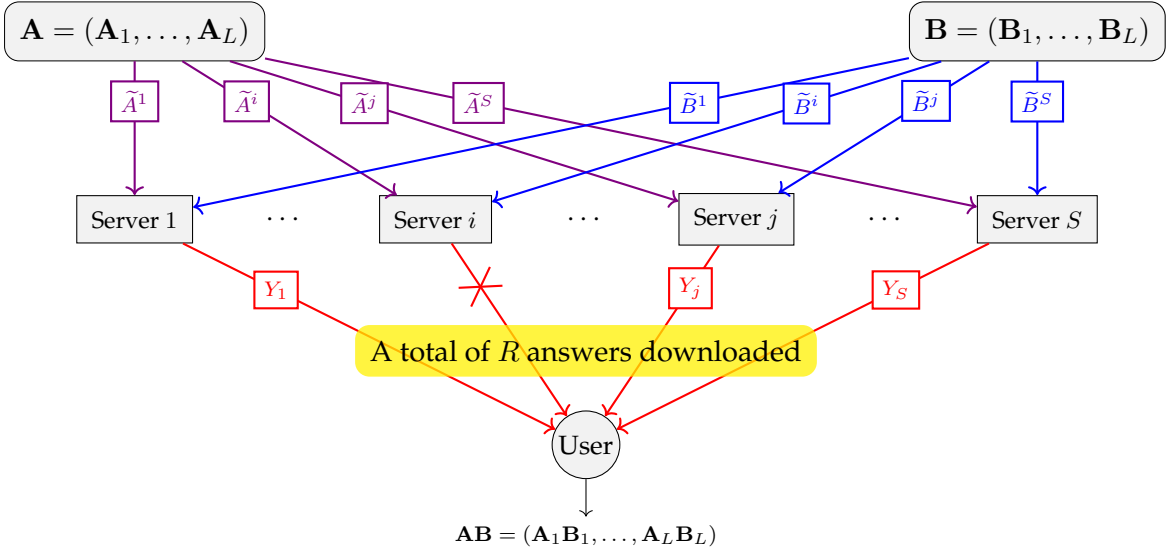

 In the era of big data and cloud computing along with massive parallelization, there is particular interest in algorithms for  coded distributed computation that are resilient to stragglers  \cite{Yu_Maddah-Ali_Avestimehr_Polynomial,Dutta_Fahim_Haddadpour,GPolyDot, Yu_Maddah-Ali_Avestimehr, Yu_Lagrange, Reisizadeh_Prakash_Pedarsani, Lee_Suh_Ramchandran, Lee_Lam_Pedarsani, Dutta_Cadambe_Short, Dutta_Cadambe_Codedconv, Yu_Maddah-Ali_CodedDFT, Jahani-Nezhad_Maddah-Ali, Baharav_Lee_Ocal, Suh_Lee_Msparse, Wang_Liu_CLT, Mallick_Chaudhari_Joshi, Wang_Liu_Sparse, Severinson_iAmat_Rosnes, Haddadpour_Cadambe_Finite,Sheth_Dutta_Chaudhari, Jeong_Ye_Grover,Kim_Sohn_Moon_Group,Park_Lee_Sohn,Li_Maddah-Ali_Fog}. 
The goal in coded distributed computation is to distribute the computation task according to a coding scheme across $S$  servers (also known as workers or processors), such that the response from any $R$ servers is sufficient for the user to recover (decode) the result of the computation. The parameter $R$ is called the recovery threshold. Coded distributed computing offers the advantage of reduced latency from massive parallelization, because the tasks assigned to each server are smaller, and the redundancy added by coding helps avoid bottlenecks due to stragglers. The main metrics of interest for coded distributed computation include:  the encoding and decoding complexity, latency\footnote{Latency is the time it takes a server to complete a specific computation job. Unlike server computation complexity, it is not normalized by the size of the job, so it depends on the size of the job assigned to the server. Latency constraints are explored in the discussion following Theorem \ref{thm:gcsa} in Section \ref{sec:gcsa}.} and complexity of server computation, the recovery threshold, and the upload and download costs (communication costs). With high end communication speeds approaching Gbps and  computing speeds (processor clock speeds) commonly of the order of GHz, communication and computation costs may be comparable for many applications, allowing meaningful tradeoffs between the two. On the other hand, since communication bottlenecks are quite common, communication costs remain a key concern in distributed computing. Note that even with higher communication costs distributed computing may be necessary if, e.g., the computation task is too large to be efficiently carried out locally, or if the sources that generate the inputs for computation are not the same as the destination where the output of computation is desired, i.e., communication is unavoidable.  Figure \ref{fig:DBMM} shows such a setting for coded distributed batch matrix multiplication (CDBMM).
Another notable aspect of such settings is that the cost dynamics for uploads and downloads may be different, e.g., if the input data is relatively static and multiple users request  computations on different parts of the same dataset, then the download cost may be much more of a concern than upload cost. This will be significant when we compare different coding schemes in this work.

Distributed coded computing can be applied to a myriad of computational tasks. Of particular interest to this work are matrix multiplications, $N$-linear computations (e.g., computing the determinants of $N\times N$ matrices, or the product of $N$ matrices), and evaluations of multivariate polynomials. These are some of the most fundamental building blocks of  computation. Moreover, these problems are closely related. Indeed matrix multiplications are bilinear operations, so they are special cases of multilinear computations, and multilinear computations may be seen as special cases of multivariate polynomial evaluations. Several elegant coding schemes, or codes, have been proposed for solving these problems. Codes for distributed matrix multiplication evolved through MDS codes \cite{Lee_Lam_Pedarsani}, Polynomial codes \cite{Yu_Maddah-Ali_Avestimehr_Polynomial}, MatDot  and PolyDot codes \cite{Dutta_Fahim_Haddadpour} to the current state of art reflected in Generalized PolyDot codes \cite{GPolyDot} and Entangled Polynomial (EP) codes \cite{Yu_Maddah-Ali_Avestimehr}. For multilinear computations and evaluations of multivariate polynomials, the state of art is represented by Lagrange Coded Computing (LCC), introduced in \cite{Yu_Lagrange}. 

It is interesting to note that the solutions to these problems fall into two distinct categories --- those based on partitioning of a single computation task \cite{Yu_Maddah-Ali_Avestimehr_Polynomial, Dutta_Fahim_Haddadpour, GPolyDot, Yu_Maddah-Ali_Avestimehr}, and those based on batch processing of multiple computation tasks \cite{Yu_Lagrange}. For example, consider the CDBMM problem shown in Figure \ref{fig:DBMM} where the goal is to efficiently multiply  $L$ instances of $\lambda\times\kappa$ matrices, ${\bf A}=({\bf A}_1, {\bf A}_2,\cdots, {\bf A}_L)$,  with $L$ instances of $\kappa\times\mu$ matrices ${\bf B}=({\bf B}_1,{\bf B}_2,\cdots, {\bf B}_L)$,  to compute the batch of $L$ matrix products, ${\bf AB}=({\bf A}_1{\bf B}_1, {\bf A}_2{\bf B}_2,\cdots,{\bf A}_L{\bf B}_L)$. Matrix-partitioning approaches compute each of the $L$ products ${\bf A}_l{\bf B}_l$ one at a time by partitioning individual matrices ${\bf A}_l$ and ${\bf B}_l$ and coding across these partitions. Batch processing approaches do not partition individual matrices, instead they code across the batch of ${\bf A}$ matrices and across  the batch of ${\bf B}$ matrices. The state-of-art for matrix-partitioning approaches is represented by Entangled Polynomial Codes (EP codes) \cite{Yu_Maddah-Ali_Avestimehr}, while Lagrange Coded Computing (LCC) \cite{Yu_Lagrange} represents the state of art for batch processing.  Since the problems are related, it is natural to ask, how do the matrix-partitioning solutions  compare with the batch-processing solutions? Furthermore, can these solutions be improved, unified, generalized? These are the questions that we address in this work.

The essential ingredient in this work that allows us to compare, improve, unify and generalize the solutions to matrix multiplication, multilinear computation and multivariate polynomial evaluation, turns out to be the idea of cross-subspace alignment. Cross-subspace alignment (CSA) was originally introduced in the context of $X$-Secure $T$-Private Information Retrieval (XSTPIR) \cite{Jia_Sun_Jafar_XSTPIR}. Coding schemes that exploit CSA have been used to  improve upon  and generalize the best known schemes for PIR with $X$-secure data, $T$-private queries and various forms of storage, e.g., fully replicated \cite{Jia_Sun_Jafar}, graph based replicated storage with limited replication factor \cite{Jia_Jafar_GXSTPIR}, or MDS coded  storage \cite{Jia_Jafar_MDSXSTPIR}. CSA schemes have also recently been shown to be useful to minimize download communication cost for   secure and/or private matrix multiplication \cite{Kakar_Ebadifar_Sezgin_CSA, Kakar_Ebadifar_Sezgin, Jia_Jafar_SDMM, Jia_Jafar_MDSXSTPIR}. Building upon these efforts, in this work we introduce a new and generalized class of coded distributed computation codes, called CSA codes, that are inspired by the idea of cross-subspace alignment. The contributions of this work are summarized as follows. 

\begin{enumerate}
\item {\bf CSA Codes.} In Theorem \ref{thm:csacodes} of this paper that appears in Section \ref{sec:csacodes}, we introduce CSA codes for coded distributed batch matrix multiplication.  These codes are used to multiply a batch of matrices ${\bf A}_1, {\bf A}_2, \cdots, {\bf A}_L$ with ${\bf B}_1, {\bf B}_2, \cdots, {\bf B}_L$ to recover the $L$ desired matrix products ${\bf A}_1{\bf B}_1, {\bf A}_2{\bf B}_2, \cdots, {\bf A}_L{\bf B}_L$. There is no partitioning of individual matrices. Instead,  coding is done across the matrices within a batch. CSA codes partition a batch of $L$ matrices into $\ell$ sub-batches of $K_c$ matrices each ($L=\ell K_c$).  Due to cross-subspace alignment, the interference is limited to $K_c-1$ dimensions regardless of the number of sub-batches $\ell$, so that the recovery threshold $R=L+K_c-1$. The download per server does not depend on $\ell$, although the upload and server computation complexity do scale with $\ell$. Surprisingly, setting $\ell=1$ recovers the Lagrange Coded Computing (LCC) solution to coded distributed batch matrix multiplication as a special case of CSA codes. Besides the additional flexibility, the main advantage of choosing $\ell>1$ in CSA codes is to reduce the download cost relative to LCC codes (see Fig. \ref{fig:gcsa} in Section \ref{sec:gcsamain}). This advantage is especially significant   in settings where the download cost is the primary bottleneck.
\item {\bf EP vs CSA Codes.} We compare matrix partitioning approaches (say EP codes that generalize MatDot and Polynomial codes) with batch processing approaches (CSA codes that generalize Lagrange Coded Computing) for distributed matrix multiplication (see Fig. \ref{fig:compareud} in Section \ref{sec:csaobs}). Remarkably, we find  that batch processing presents a significant advantage in communication cost per matrix multiplication (i.e., normalized by the batch size $L$). As a function of the recovery threshold $R$, and for any fixed recovery ratio $R/S$, CSA codes  have the same server computation complexity per matrix multiplication as EP codes, but CSA codes simultaneously achieve normalized (upload cost, download cost)=$(\mathcal{O}(1),\mathcal{O}(1))$, overcoming a key barrier of existing matrix-partitioning codes where upload cost of $\mathcal{O}(1)$ can only be achieved with download cost of $\mathcal{O}(R)$ and download cost of $\mathcal{O}(1)$ can only be achieved with upload cost of $\mathcal{O}(\sqrt{R})$. A corresponding improvement in the tradeoff between encoding and decoding complexity is also observed. %

\item {\bf GCSA Codes.} Since there is no partitioning of  individual matrices in the aforementioned CSA codes, this means that each server must carry out a computational load equivalent to at least one full matrix multiplication before it can respond with an answer. This presents a latency barrier for batch processing schemes that cannot be overcome regardless of the number of servers and the batch size. For  applications with stricter latency requirements such a solution may be infeasible, making it necessary to reduce the computational load per server by further parallelization, i.e., partitioning of individual matrices. To this end, in Theorem \ref{thm:gcsa} that appears in Section \ref{sec:gcsamain} of this paper, we present Generalized CSA codes (GCSA codes in short) that combine the matrix partitioning approach of, say EP codes, with the batch processing of CSA codes. 
GCSA codes bridge the two extremes by efficiently combining both matrix-partitioning and batch processing, and offer flexibility in how much of each approach is used. Both EP codes and LCC codes can be recovered as special cases of GCSA codes, but GCSA codes are capable of outperforming both EP and LCC codes in general (see Fig. \ref{fig:job} and Fig. \ref{fig:gcsa} in Section \ref{sec:gcsamain}). When no matrix partitioning is used, GCSA codes reduce to CSA codes, and if no batch processing is used then GCSA codes reduce to EP codes. With GCSA codes, the degree of matrix partitioning controls the server latency by limiting the computational load per server, while the batch partitioning on top yields the advantage of batch processing in communication costs.  The combination is far from trivial.  For example, consider a matrix partitioning approach that splits the task among $10$ servers such that any $R_1=7$ need to respond, and a similar batch processing approach that also splits the task among $10$ servers such that any $R_2=7$ need to respond. Then if we simply take the $10$ matrix-partitioned tasks and use batch processing on top  to distribute each task among $10$ servers, for a total of $100$ servers, then the recovery threshold of the naive combination is $6\times 10+ 4\times 6+1=85$. However, GCSA codes achieve a significantly lower recovery threshold ($R\leq R_1R_2=49$).

\item {\bf $N$-CSA Codes.} As noted, CSA codes are a generalization of LCC codes for distributed batch matrix multiplication. However, the applications of LCC codes extend beyond matrix multiplication, to $N$-linear batch computation and multivariate polynomial batch evaluations, raising the question whether corresponding generalizations of LCC codes to CSA type codes exist for these applications as well. We answer this question in the affirmative, by introducing $N$-CSA codes for the problem of coded distributed $N$-linear batch computation as well as multivariate polynomial evaluations, that are strictly generalizations of LCC codes for both of these applications. This generalization for batch size $L=\ell K_c$ is done as follows. For all $n\in[N]$, the batch of $L$ realizations of the $n^{th}$ variable is split into $\ell$ sub-batches, each containing $K_c$ realizations. The $K_c$ realizations within each sub-batch are coded into an MDS $(S,K_c)$ code according to a Cauchy structure, and distributed to the $S$ servers. Each server evaluates the $N$-linear map function with the coded variables of each sub-batch, and returns a weighted sum of evaluations of these $\ell$ sub-batches.  By cross-subspace alignment, undesired evaluations only occupy $(N-1)(K_c-1)$ dimensions, so that the recovery threshold is $R=L+(N-1)(K_c-1)$. Finally, because $N$-linear maps are fundamental construction blocks of multivariate polynomials of total degree $N$, it is straightforward to apply $N$-CSA codes  for multivariate polynomial batch evaluation. Specifically, we can regard any multivariate polynomial of total degree $N$ as a linear combination of various restricted evaluations of $N$-linear maps. Each server prepares answers for various $N$-linear maps that  constitute the given multivariate polynomial, then returns the user with the linear combination of these answers according to the given polynomial. Once again, the $N$-CSA code based scheme for multivariate polynomial batch evaluation thus obtained, generalizes LCC codes, which can be recovered by setting $\ell=1$. The main advantage of choosing $\ell>1$ with CSA codes remains the download cost.  $N$-CSA codes achieve normalized download cost $D=\frac{R}{L}=1+\left(\frac{N-1}{\ell}\right)\left(\frac{K_c-1}{K_c}\right)$. The special case of $\ell=1$ which gives us LCC codes corresponds to download cost of $\mathcal{O}(N)$, but by using the full scope of values of $\ell$ the download cost can be reduced by up to a factor of $N$, albeit with increasing recovery threshold. Reducing download cost generally also reduces decoding complexity, which can be important when downlink and/or computational resource at the user side is limited.

\end{enumerate}

Next we provide an overview of the state of art approaches for coded distributed computing, summarize the key ideas behind cross-subspace alignment, and tabulate the comparisons between the codes proposed in this work and the prior state of art.

\section{EP Codes, LCC Codes, CSA Codes}
\subsection{Matrix Partitioning: EP Codes} 
EP codes  \cite{Yu_Maddah-Ali_Avestimehr} for coded distributed matrix multiplication problem are based on matrix partitioning. The constituent matrices $\mathbf{A}$ and $\mathbf{B}$ are partitioned into $m\times p$ blocks and $p\times n$ blocks, respectively, as shown below, so that the desired matrix product involves a total of $mn$ linear combinations of products of block matrices. 
\begin{align}
{\bf A}&=\left(
\begin{matrix}
{\bf A}^{1,1}&{\bf A}^{1,2}&\cdots&{\bf A}^{1,p}\\
{\bf A}^{2,1}&{\bf A}^{2,2}&\cdots&{\bf A}^{2,p}\\
\vdots&\vdots&\vdots&\vdots\\
{\bf A}^{m,1}&{\bf A}^{m,2}&\cdots&{\bf A}^{m,p}\\
\end{matrix}\right)
\hspace{1cm}
{\bf B}=\left(
\begin{matrix}
{\bf B}^{1,1}&{\bf B}^{1,2}&\cdots&{\bf B}^{1,n}\\
{\bf B}^{2,1}&{\bf B}^{2,2}&\cdots&{\bf B}^{2,n}\\
\vdots&\vdots&\vdots&\vdots\\
{\bf B}^{p,1}&{\bf B}^{p,2}&\cdots&{\bf B}^{p,n}\\
\end{matrix}\right)\\
{\bf AB}&=\left(
\begin{matrix}
\sum_{j=1}^p{\bf A}^{1,j}{\bf B}^{j,1}&\sum_{j=1}^p{\bf A}^{1,j}{\bf B}^{j,2}&\cdots&\sum_{j=1}^p{\bf A}^{1,j}{\bf B}^{j,n}\\
\sum_{j=1}^p{\bf A}^{2,j}{\bf B}^{j,1}&\sum_{j=1}^p{\bf A}^{2,j}{\bf B}^{j,2}&\cdots&\sum_{j=1}^p{\bf A}^{2,j}{\bf B}^{j,n}\\
\vdots&\vdots&\vdots&\vdots\\
\sum_{j=1}^p{\bf A}^{m,j}{\bf B}^{j,1}&\sum_{j=1}^p{\bf A}^{m,j}{\bf B}^{j,2}&\cdots&\sum_{j=1}^p{\bf A}^{m,j}{\bf B}^{j,n}\\
\end{matrix}
\right)
\end{align}
Coded matrices  are constructed as follows,
\begin{align}
    \widetilde{A}(\alpha)&=\sum_{m'\in[m]}\sum_{p'\in[p]}\mathbf{A}^{m',p'}\alpha^{p'-1+p(m'-1)},\\
    \widetilde{B}(\alpha)&=\sum_{p'\in[p]}\sum_{n'\in[n]}\mathbf{B}^{p',n'}\alpha^{p-p'+pm(n'-1)},
\end{align}
and the $s^{th}$ server is sent  the values $\widetilde{A}(\alpha_s)$ and $\widetilde{B}(\alpha_s)$. Here $\alpha_1, \alpha_2,\cdots,\alpha_S$ are distinct elements from the operating field $\mathbb{F}$. Each server produces the answer $\widetilde{A}(\alpha_s)\widetilde{B}(\alpha_s)$, which can be expressed as
\begin{align}
    \widetilde{A}(\alpha)\widetilde{B}(\alpha)=\sum_{i=1}^{R}\mathbf{C}^{(i)}\alpha^{i-1},
\end{align}
where $R=pmn+p-1$ is the recovery threshold, and $\mathbf{C}^{(1)},\mathbf{C}^{(2)},\cdots,\mathbf{C}^{(R)}$ are various linear combinations of products of matrix blocks. Note that for all $i\in[R]$, $\mathbf{C}^{(i)}$ are distributed over $1,\alpha,\cdots,\alpha^{R-1}$, thus from the answers of any $R$ servers, $\mathbf{C}^{(1)},\mathbf{C}^{(2)},\cdots,\mathbf{C}^{(R)}$ are recoverable by inverting a Vandermonde matrix. Furthermore, it is proved in \cite{Yu_Maddah-Ali_Avestimehr} that by the construction of $\widetilde{A}(\alpha)$ and $\widetilde{B}(\alpha)$, the $\mathbf{C}^{(1)},\mathbf{C}^{(2)},\cdots,\mathbf{C}^{(R)}$ terms include the $mn$ desired terms, while the remaining undesired terms (interference) align into the remaining $R-mn$ dimensions. 

For example, suppose $p=m=n=2$, so that the coded matrices are constructed as follows.
\begin{align}
    \widetilde{A}(\alpha)&=\mathbf{A}^{1,1}+\alpha\mathbf{A}^{1,2}+\alpha^2\mathbf{A}^{2,1}+\alpha^3\mathbf{A}^{2,2},\\
    \widetilde{B}(\alpha)&=\alpha\mathbf{B}^{1,1}+\alpha^5\mathbf{B}^{1,2}+\mathbf{B}^{2,1}+\alpha^4\mathbf{B}^{2,2}.
\end{align}
And the answer can be expressed as follows.
\begin{align}
    \widetilde{A}(\alpha)\widetilde{B}(\alpha)&=\underbrace{\mathbf{A}^{1,1}\mathbf{B}^{2,1}}_{\mathbf{C}^{(1)}}+\alpha\underbrace{(\mathbf{A}^{1,1}\mathbf{B}^{1,1}+\mathbf{A}^{1,2}\mathbf{B}^{2,1})}_{\mathbf{C}^{(2)}}+\alpha^2\underbrace{(\mathbf{A}^{2,1}\mathbf{B}^{2,1}+\mathbf{A}^{1,2}\mathbf{B}^{1,1})}_{\mathbf{C}^{(3)}}\notag\\
    &\quad\quad+\alpha^3\underbrace{(\mathbf{A}^{2,1}\mathbf{B}^{1,1}+\mathbf{A}^{2,2}\mathbf{B}^{2,1})}_{\mathbf{C}^{(4)}}+\alpha^4\underbrace{(\mathbf{A}^{1,1}\mathbf{B}^{2,2}+\mathbf{A}^{2,2}\mathbf{B}^{1,1})}_{\mathbf{C}^{(5)}}+\alpha^5\underbrace{(\mathbf{A}^{1,1}\mathbf{B}^{1,2}+\mathbf{A}^{1,2}\mathbf{B}^{2,2})}_{\mathbf{C}^{(6)}}\notag\\
    &\quad\quad+\alpha^6\underbrace{(\mathbf{A}^{1,2}\mathbf{B}^{1,2}+\mathbf{A}^{2,1}\mathbf{B}^{2,2})}_{\mathbf{C}^{(7)}}+\alpha^7\underbrace{(\mathbf{A}^{2,1}\mathbf{B}^{1,2}+\mathbf{A}^{2,2}\mathbf{B}^{2,2})}_{\mathbf{C}^{(8)}}+\alpha^8\underbrace{(\mathbf{A}^{2,2}\mathbf{B}^{1,2})}_{\mathbf{C}^{(9)}}.
\end{align}
Note that the desired product $\mathbf{A}\mathbf{B}$ corresponds to the $mn=4$ terms $\mathbf{C}^{(2)},\mathbf{C}^{(4)},\mathbf{C}^{(6)},\mathbf{C}^{(8)}$, which appear along $\alpha, \alpha^3, \alpha^5, \alpha^7$. The remaining $R-mn=(pmn+p-1)-mn=5$ terms, i.e., $\mathbf{C}^{(1)},\mathbf{C}^{(3)},\mathbf{C}^{(5)},\mathbf{C}^{(7)},\mathbf{C}^{(9)}$ are undesired terms (interference). 

In particular, note that the term $\mathbf{C}^{(9)}$, which is interference, has a higher order ($\alpha^8$) than all desired terms. In general, EP codes produce $p-1$ such terms, namely $\mathbf{C}^{(pmn+1)},\cdots,\mathbf{C}^{(R)}$, that have a higher order than all desired terms. It turns out this is useful in the construction of GCSA codes to achieve better Interference Alignment (because these higher order terms produced by EP codes naturally align with the interference terms that result from batch processing).

EP codes may be seen as bridging the extremes of Polynomial codes and MatDot codes. Polynomial codes \cite{Yu_Maddah-Ali_Avestimehr_Polynomial} can be recovered from EP codes by setting $p=1$, and MatDot codes \cite{Dutta_Fahim_Haddadpour} can be obtained from EP codes by setting $m=n=1$. EP codes also represent an improvement of PolyDot codes \cite{Dutta_Fahim_Haddadpour} within a factor of $2$ in terms of recovery threshold, due to  better interference alignment. Finally, EP codes have similar performance as Generalized PolyDot codes \cite{GPolyDot}. Thus, EP codes represent the state of art of prior work in terms of matrix partitioning approaches to coded distributed matrix multiplication.
\subsection{Batch Processing: LCC Codes} 
Lagrange Coded Computing (LCC) codes \cite{Yu_Lagrange} represent the state of art of prior work in terms of batch processing approaches  for coded distributed batch multivariate polynomial evaluation, which includes as special cases distributed batch matrix multiplication as well as distributed batch $N$-linear computation. LCC codes are so named because they exploit the Lagrange interpolation polynomial to encode input data. For example, consider the  multivariate polynomial $\Phi(\cdot)$ of total degree $N$, and suppose we are interested in batch evaluations of the polynomial, $\Phi(\mathbf{x}_1),\Phi(\mathbf{x}_2),\cdots,\Phi(\mathbf{x}_L)$ over the given batch of data points $\mathbf{x}_1,\mathbf{x}_2,\cdots,\mathbf{x}_L$. Note that for matrix multiplication, ${\bf x}_l=({\bf A}_l,{\bf B}_l)$ and $\Phi({\bf x}_l)={\bf A}_l{\bf B}_l$, which is a bilinear operation ($N=2$). LCC codes encode the dataset according to the Lagrange interpolation polynomial,
\begin{align}
    \widetilde{X}(\alpha)=\sum_{l\in[L]}\mathbf{x}_l\prod_{l'\in[L]\setminus\{l\}}\frac{\alpha-\beta_{l'}}{\beta_l-\beta_{l'}},
\end{align}
and the $s^{th}$ server is sent the evaluation $\widetilde{X}(\alpha_s)$. Here $\alpha_1,\alpha_2,\cdots,\alpha_S, \beta_1,\beta_2,\cdots,\beta_L$ are $(S+L)$ distinct elements from the operation field $\mathbb{F}$. The $s^{th}$ server returns the user with the answer $\Phi(\widetilde{X}(\alpha_s))$. Note that the degree of the polynomial $\Phi(\widetilde{X}(\alpha))$ is less than or equal to $N(L-1)=NL-N$. Therefore, from the answers of any $R=NL-N+1$ servers, the user is able to reconstruct the polynomial $\Phi(\widetilde{X}(\alpha))$ by polynomial interpolation. Upon obtaining the polynomial $\Phi(\widetilde{X}(\alpha))$, the user evaluates it at $\beta_l$ for every $l\in[L]$ to obtain $\Phi(\widetilde{X}(\beta_l)) = \Phi(\mathbf{x}_l)$.

\subsection{Cross Subspace Alignment: CSA Codes}
The distinguishing feature of CSA codes is a Cauchy-Vandermonde structure that facilitates a form of interference alignment (labeled cross-subspace-alignment in \cite{Jia_Sun_Jafar_XSTPIR}), such that the desired symbols occupy dimensions corresponding to the Cauchy part,  and everything else (interference) aligns within the higher order terms that constitute the Vandermonde part.  As a simple example of the CSA codes introduced in this work, consider the problem of coded distributed batch matrix multiplication, and suppose we wish to compute the batch of $L=4$ matrix products ${\bf A}_1{\bf B}_1$, ${\bf A}_2{\bf B}_2$, ${\bf A}_3{\bf B}_3$, ${\bf A}_4{\bf B}_4$. For this, the ${\bf A}$ and ${\bf B}$ matrices are encoded into the form 
\begin{align}
\widetilde{A}(\alpha)&=\Delta(\alpha)\left(\frac{1}{\underline{1}-\alpha}{\bf A}_1+\frac{1}{\underline{2}-\alpha}{\bf A}_2+\frac{1}{\underline{3}-\alpha}{\bf A}_3+\frac{1}{\underline{4}-\alpha}{\bf A}_4\right),\\
\widetilde{B}(\alpha)&=\frac{1}{\underline{1}-\alpha}{\bf B}_1+\frac{1}{\underline{2}-\alpha}{\bf B}_2+\frac{1}{\underline{3}-\alpha}{\bf B}_3+\frac{1}{\underline{4}-\alpha}{\bf B}_4,
\end{align}
and the $s^{th}$ server is sent the evaluations $\widetilde{A}(\alpha_s),\widetilde{B}(\alpha_s)$. Here the values $\alpha_1, \alpha_2, \cdots,\alpha_S, \underline{1}, \underline{2},\cdots,\underline{4}$ represent any $S+4$ distinct elements of the operational field $\mathbb{F}$, and $\Delta(\alpha)=(\underline{1}-\alpha)(\underline{2}-\alpha)(\underline{3}-\alpha)(\underline{4}-\alpha)$. Each server multiplies its $\widetilde{A}(\alpha_s)$ with $\widetilde{B}(\alpha_s)$ producing an answer which (after some algebraic manipulation) can be expressed as
\begin{align}
\widetilde{A}(\alpha)\widetilde{B}(\alpha)&=c_1\left(\frac{1}{\underline{1}-\alpha}\right){\bf A}_1{\bf B}_1+c_{2}\left(\frac{1}{\underline{2}-\alpha}\right){\bf A}_2{\bf B}_2+c_{3}\left(\frac{1}{\underline{3}-\alpha}\right){\bf A}_3{\bf B}_3+c_{4}\left(\frac{1}{\underline{4}-\alpha}\right){\bf A}_4{\bf B}_4\nonumber\\
&\hspace{1cm}+ {\bf I}_1+\alpha {\bf I}_2+\alpha^2{\bf I}_3,
\end{align}
where $c_1, c_2, c_3, c_4$ are non-zero constants. The desired matrix products ${\bf A}_i{\bf B}_i$ appear along $\left(\frac{1}{\underline{i}-\alpha}\right)$ (the Cauchy terms), and everything else (interference) can be distributed over the higher order terms $1, \alpha, \alpha^2$ (the Vandermonde terms) and consolidated into  ${\bf I}_1, {\bf I}_2, {\bf I}_3$. The full-rank property of the Cauchy-Vandermonde matrix ensures that the desired symbols are separable from interference provided we have at least $R=7$ responding servers to resolve the $7$ total dimensions ($4$ desired and $3$ interference dimensions). Surprisingly, upon close inspection this special case of CSA codes turns out to be equivalent to the Lagrange Coded computing scheme for distributed matrix multiplication.  However, CSA codes further generalize and improve upon the Lagrange Coded Computing approach as explained next.

Suppose we double the batch size from $L=4$ to $L=8$, i.e., we wish to compute the matrix products ${\bf A}_1{\bf B}_1$, ${\bf A}_2{\bf B}_2$, $\cdots$, ${\bf A}_8{\bf B}_8$. A straightforward extension is to simply use the previous scheme twice, which would double all costs. This could be accomplished equivalently with CSA codes or with Lagrange Coded Computing. However, because CSA codes generalize Lagrange Coded Computing, they offer much more flexibility. For example, we can partition the batch of $L=8$  ${\bf A},{\bf B}$ matrices into $\ell=2$ sub-batches of $K_c=4$ matrices each, and then proceed as before, so that we have,
\begin{align}
\widetilde{A}_1(\alpha)&=\Delta_1(\alpha)\left(\frac{1}{\underline{1}-\alpha}{\bf A}_1+\frac{1}{\underline{2}-\alpha}{\bf A}_2+\frac{1}{\underline{3}-\alpha}{\bf A}_3+\frac{1}{\underline{4}-\alpha}{\bf A}_4\right),\\
\widetilde{A}_2(\alpha)&=\Delta_2(\alpha)\left(\frac{1}{\underline{5}-\alpha}{\bf A}_5+\frac{1}{\underline{6}-\alpha}{\bf A}_6+\frac{1}{\underline{7}-\alpha}{\bf A}_7+\frac{1}{\underline{8}-\alpha}{\bf A}_8\right),\\
\widetilde{B}_1(\alpha)&=\frac{1}{\underline{1}-\alpha}{\bf B}_1+\frac{1}{\underline{2}-\alpha}{\bf B}_2+\frac{1}{\underline{3}-\alpha}{\bf B}_3+\frac{1}{\underline{4}-\alpha}{\bf B}_4,\\
\widetilde{B}_2(\alpha)&=\frac{1}{\underline{5}-\alpha}{\bf B}_5+\frac{1}{\underline{6}-\alpha}{\bf B}_6+\frac{1}{\underline{7}-\alpha}{\bf B}_7+\frac{1}{\underline{8}-\alpha}{\bf B}_8,
\end{align}
where $\Delta_1(\alpha)=(\underline{1}-\alpha)(\underline{2}-\alpha)(\underline{3}-\alpha)(\underline{4}-\alpha)$ and $\Delta_2(\alpha)=(\underline{5}-\alpha)(\underline{6}-\alpha)(\underline{7}-\alpha)(\underline{8}-\alpha)$. Evidently the upload cost is doubled. However, we will see that the download cost remains unchanged. This is because each server computes and (if responsive) returns (for its corresponding realization of $\alpha$)
\begin{align}
&\widetilde{A}_1(\alpha)\widetilde{B}_1(\alpha)+\widetilde{A}_2(\alpha)\widetilde{B}_2(\alpha)\\
&=c_1\left(\frac{1}{\underline{1}-\alpha}\right){\bf A}_1{\bf B}_1+c_{2}\left(\frac{1}{\underline{2}-\alpha}\right){\bf A}_2{\bf B}_2+\cdots+c_{8}\left(\frac{1}{\underline{8}-\alpha}\right){\bf A}_8{\bf B}_8+ {\bf I}'_1+\alpha {\bf I}'_2+\alpha^2{\bf I}'_3.
\end{align}
Since we have $8$ desired dimensions and $3$ interference dimensions, responses from any $R=11$ servers suffice to separate desired matrix products from interference. Remarkably, while the number of desired matrix products has doubled, the number of interference dimensions have not increased at all. This is why $4$ additional responding servers allow us to recover $4$ additional desired matrix products. This is an advantage unique to cross-subspace alignment, that cannot be achieved with other coding approaches, such as Lagrange Coded computing. CSA codes for distributed matrix multiplication based on batch processing are introduced in this work in Theorem \ref{thm:csacodes}, a  generalization to include matrix partitioning is presented in Theorem \ref{thm:gcsa}, and another generalization for $N$-linear batch computations and multivariate batch polynomial evaluations is presented in Theorem \ref{thm:nlcsa}.

For ease of reference, Table \ref{tab:summary1} and Table \ref{tab:summary2} compare  EP codes, LCC codes, CSA codes, GCSA codes and $N$-CSA codes with respect to their recovery thresholds, communication costs for uploads and downloads, encoding and decoding complexity, and server computation complexity.

This paper is organized as follows. Section \ref{sec:probstat} presents the problem statements and definitions for coded distributed batch matrix multiplication (CDBMM), coded distributed $N$-linear batch computation and coded distributed multivariate batch polynomial evaluations. CSA codes for CDBMM are introduced in Section \ref{sec:csacodes}. Section \ref{sec:gcsa} presents GCSA codes. $N$-CSA codes are presented in Section \ref{sec:ncsa}. Appendix \ref{sec:symxs} presents further generalizations to allow $X$-secure data and $B$-byzantine servers. Section \ref{sec:conc} concludes the paper.

\newcolumntype{g}{>{\columncolor{black!10!white}}c}

\begin{table*}[!htbp]
    \scriptsize
    \centering
   \begin{tabular}{|c|g|g|g|g|g|g|}
  \rowcolor{white} &&&\\
   \specialrule{.2em}{1em}{0.7em}\rowcolor{white}
         & \multicolumn{1}{c|}{Recovery Threshold }& Upload Cost & Download Cost \\\rowcolor{white}
         & $(R)$ & $(U_A, U_B)$ & $(D)$ \\
         \hline\rowcolor{white}
       EP  & $pmn+p-1$ & $S/(pm), S/(pn)$ & ${(pmn+p-1)}/{(mn)}$ \Tstrut\Bstrut\\\cline{2-4}
       codes & $R$ & $\mathcal{O}(m)$, $\mathcal{O}(m)$ & $\mathcal{O}(R/m^2)$ \Tstrut\Bstrut\\\hline\rowcolor{white}
       LCC  & $2K_c'-1$ & $S/K_c', S/K_c'$ & ${(2K_c'-1)}/{K_c'}$ \Tstrut\Bstrut\\\cline{2-4}
       codes & $R$ & $\mathcal{O}(1)$, $\mathcal{O}(1)$ & $\mathcal{O}(1)$ \Tstrut\Bstrut\\\hline\rowcolor{white}
       {\bf CSA}  & $(\ell+1)K_c-1$ & ${S}/{K_c}, S/K_c$ & ${((\ell+1)K_c-1)}/{(\ell K_c)}$ \Tstrut\Bstrut\\\cline{2-4}
      codes & \cellcolor{blue!10}$R$ & \cellcolor{blue!10}$\mathcal{O}(1)$, $\mathcal{O}(1)$ & \cellcolor{blue!10}$\mathcal{O}(1)$  \Tstrut\Bstrut\\ 
      \hline\rowcolor{white}
       {\bf GCSA}  & $pmn((\ell+1)K_c''-1)+p-1$ & $S/(K_c''pm), S/(K_c''pn)$ & $\frac{pmn((\ell+1)K_c''-1)+p-1}{mn\ell K_c''}$ \Tstrut\Bstrut\\\cline{2-4}
       codes & \cellcolor{blue!10}$R$ & \cellcolor{blue!10}$\mathcal{O}(m)$, $\mathcal{O}(m)$ &\cellcolor{blue!10} $\mathcal{O}(p)$ \Tstrut\Bstrut\\
       \specialrule{.2em}{.1em}{.1em} \rowcolor{white}
      & Server Computation & Encoding & Decoding\\\rowcolor{white}
      & Complexity $(\mathcal{C}_s)$ & Complexity $(\mathcal{C}_{eA},\mathcal{C}_{eB})$ & Complexity $(\mathcal{C}_d)$\\\hline\rowcolor{white}
       EP  &  $\mathcal{O}\left({\lambda\mu\kappa}/{(pmn)}\right)$& $\widetilde{\mathcal{O}}\left(\dfrac{\lambda\kappa S\log^2 S}{pm}\right)$, $\widetilde{\mathcal{O}}\left(\dfrac{\kappa\mu S\log^2 S}{pn}\right)$ & $\widetilde{\mathcal{O}}(\lambda\mu p\log^2R)$ \Tstrut\Bstrut\\\cline{2-4}
       codes & $\mathcal{O}\left({\lambda^3}/{R}\right)$ &  $\widetilde{\mathcal{O}}\left({\lambda^2 m\log^2 S}\right)$, $\widetilde{\mathcal{O}}\left({\lambda^2 m\log^2 S}\right)$ & $\widetilde{\mathcal{O}}\left(\dfrac{\lambda^2 R\log^2R}{m^2}\right)$\Tstrut\Bstrut\\\hline\rowcolor{white}
       LCC  &  $\mathcal{O}\left({\lambda\mu\kappa}/{K_c'}\right)$ & $\widetilde{\mathcal{O}}\left(\dfrac{\lambda\kappa S\log^2 S}{K_c'}\right)$, $\widetilde{\mathcal{O}}\left(\dfrac{\kappa\mu S\log^2 S}{K_c'}\right)$ & $\widetilde{\mathcal{O}}(\lambda\mu \log^2R)$ \Tstrut\Bstrut\\\cline{2-4}
       codes & $\mathcal{O}\left({\lambda^3}/{R}\right)$ &  $\widetilde{\mathcal{O}}\left({\lambda^2 \log^2 S}\right)$, $\widetilde{\mathcal{O}}\left({\lambda^2 \log^2 S}\right)$ & $\widetilde{\mathcal{O}}(\lambda^2\log^2R)$\Tstrut\Bstrut\\\hline\rowcolor{white}
       {\bf CSA}  & $\mathcal{O}\left({\lambda\mu\kappa}/{K_c}\right)$ & $\widetilde{\mathcal{O}}\left(\dfrac{\lambda\kappa S\log^2S}{K_c}\right)$, $\widetilde{\mathcal{O}}\left(\dfrac{\kappa\mu S\log^2S}{K_c}\right)$ & $\widetilde{\mathcal{O}}(\lambda\mu\log^2R) $ \Tstrut\Bstrut\\\cline{2-4}
      codes & \cellcolor{blue!10}$\mathcal{O}\left({\lambda^3}/{R}\right)$ & \cellcolor{blue!10}$\widetilde{\mathcal{O}}\left(\lambda^2\log^2S\right)$, $\widetilde{\mathcal{O}}\left(\lambda^2\log^2S\right)$ &\cellcolor{blue!10} $\widetilde{\mathcal{O}}(\lambda^2\log^2R)$\Tstrut\Bstrut\\\hline\rowcolor{white}
      {\bf GCSA}  & $\mathcal{O}\left({\lambda\mu\kappa}/{(K_c''pmn)}\right)$ & $\widetilde{\mathcal{O}}\left(\dfrac{\lambda\kappa S\log^2S}{K_c''pm}\right)$, $\widetilde{\mathcal{O}}\left(\dfrac{\kappa\mu S\log^2S}{K_c''pn}\right)$ & $\widetilde{\mathcal{O}}(\lambda\mu p\log^2R) $ \Tstrut\Bstrut\\\cline{2-4}
      codes & \cellcolor{blue!10}$\mathcal{O}\left({\lambda^3}/{R}\right)$ & \cellcolor{blue!10}$\widetilde{\mathcal{O}}\left(\lambda^2m\log^2S\right)$, $\widetilde{\mathcal{O}}\left(\lambda^2m\log^2S\right)$ &\cellcolor{blue!10} $\widetilde{\mathcal{O}}(\lambda^2p\log^2R)$\Tstrut\Bstrut\\\hline%
    \end{tabular}

    \clearpage

    \caption{\it \small Performance summary of EP \cite{Yu_Maddah-Ali_Avestimehr}, LCC \cite{Yu_Lagrange}, CSA and GCSA codes for CDBMM. Note that choosing $\ell=K_c=1$ reduces GCSA codes to EP codes, while setting $m=n=p=1$ reduces GCSA codes to CSA codes (further restricting $\ell=1$ recovers LCC codes). Shaded rows represent balanced settings with $m=n, \lambda=\mu=\kappa$, fixed positive integers $\ell,\ell''$, and fixed ratio $R/S$. The batch size is $L=\ell K_c$ for CSA codes, $L'=K_c'$ for LCC codes, and $L''=\ell K_c''$ for GCSA codes. }
    \label{tab:summary1}

    \scriptsize
    \centering
    \begin{tabular}{|c|g|g|g|g|g|g|}
    \rowcolor{white} &&&\\
   \specialrule{.2em}{0.1em}{0.7em}\rowcolor{white}
         & Recovery Threshold & Upload Cost & Download Cost \\\rowcolor{white}
         & $(R)$ & $(U_{X^{(n)}}, n\in[N])$ & $(D)$ \\
         \hline\rowcolor{white}
       LCC  & $NK_c'-N+1$ & $S/K_c', S/K_c'$ & ${(NK_c'-N+1)}/{K_c'}$ \Tstrut\Bstrut\\\cline{2-4}
       codes & $R$ & $\mathcal{O}(N)$ & $\mathcal{O}(N)$ \Tstrut\Bstrut\\\hline\rowcolor{white}
       {\bf $N$-CSA}  & $(N+\ell-1)K_c-N+1$ & ${S}/{K_c}, S/K_c$ & ${((N+\ell-1)K_c-N+1)}/{(\ell K_c)}$ \Tstrut\Bstrut\\\cline{2-4}
      codes & \cellcolor{blue!10}$R$ & \cellcolor{blue!10}$\mathcal{O}(N+\ell-1)$ & \cellcolor{blue!10}$\mathcal{O}(1+\frac{N-1}{\ell})$  \Tstrut\Bstrut\\ 
      \specialrule{.2em}{.1em}{.1em}\rowcolor{white}
      & Computational & Encoding & Decoding\\\rowcolor{white}
      & Complexity $(\mathcal{C}_s)$ & Complexity $(\mathcal{C}_{eX^{(n)}},n\in[N])$ & Complexity $(\mathcal{C}_d)$\\\hline\rowcolor{white}
       LCC  &  $\mathcal{O}\left({\omega}/{K_c'}\right)$ & $\widetilde{\mathcal{O}}\left(\dfrac{\dim(V_n) S\log^2 S}{K_c'}\right)$ & $\widetilde{\mathcal{O}}(\dim(W)N \log^2R)$ \Tstrut\Bstrut\\\cline{2-4}
       codes & $\mathcal{O}\left({N\omega}/{R}\right)$ &  $\widetilde{\mathcal{O}}\left({N\dim(V_n) \log^2 S}\right)$ & $\widetilde{\mathcal{O}}(\dim(W)N\log^2R)$\Tstrut\Bstrut\\\hline\rowcolor{white}
       {\bf $N$-CSA}  & $\mathcal{O}\left({\omega}/{K_c}\right)$ & $\widetilde{\mathcal{O}}\left(\dfrac{\dim(V_n) S\log^2S}{K_c}\right)$ & $\widetilde{\mathcal{O}}\left(\left(1+\frac{N-1}{\ell}\right)\dim(W)\log^2R\right) $ \Tstrut\Bstrut\\\cline{2-4}
      codes & \cellcolor{blue!10}$\mathcal{O}\left({(N+\ell-1)\omega}/{R}\right)$ & \cellcolor{blue!10}$\widetilde{\mathcal{O}}\left((N+\ell-1)\dim(V_n)\log^2S\right)$ &\cellcolor{blue!10} $\widetilde{\mathcal{O}}\left(\left(1+\frac{N-1}{\ell}\right)\dim(W)\log^2R\right) $\Tstrut\Bstrut\\\hline%
    \end{tabular}

    \caption{\it \small Performance summary of LCC codes \cite{Yu_Lagrange} and $N$-CSA codes for $N$-linear distributed batch computation.  Setting $\ell=1$ reduces $N$-CSA codes to LCC codes as a special case. Shaded rows represent settings with fixed ratio $R/S$. $\omega$ is the number of arithmetic operations required to compute the $N$-linear map $\Omega(\cdot)$. $\dim(V_n)$ is the dimension of the $n^{th}$ variable of $\Omega(\cdot)$, $\dim{W}$ is the dimension of the output of $\Omega(\cdot)$. The batch size is $L=\ell K_c$ for $N$-CSA codes, and $L'=K_c'$ for LCC codes.}
    \label{tab:summary2}
    
\end{table*}

{\it Notation: }For a positive integer $N$, $[N]$ stands for the set $\{1,2,\dots,N\}$. The notation $X_{[N]}$ denotes the set $\{X_1,X_2,\dots,X_N\}$. For  $\mathcal{I}=\{i_1,i_2,\dots,i_N\}$, $X_{\mathcal{I}}$ denotes the set $\{X_{i_1},X_{i_2},\dots,X_{i_N}\}$. The notation $\otimes$ is used to denote the Kronecker product of two matrices, i.e., for two matrices $\mathbf{A}$ and $\mathbf{B}$, where $(\mathbf{A})_{r,s}=a_{rs}$ and $(\mathbf{B})_{v,w}=b_{vw}$, $(\mathbf{A}\otimes \mathbf{B})_{p(r-1)+v, q(s-1)+w} = a_{rs} b_{vw}$. $\mathbf{I}_N$ denotes the $N\times N$ identity matrix. $\mathbf{T}(X_1,X_2,\cdots,X_N)$ denotes the $N\times N$ lower triangular Toeplitz matrix, i.e.,
\begin{align}
    \mathbf{T}(X_1,X_2,\cdots,X_N)=\begin{bmatrix}X_1&&&& &\\X_2&X_1&&&& \\X_3&X_2&\ddots & & & \\\vdots &\ddots &\ddots &\ddots &&\\\vdots &&\ddots &X_2&X_1&\\X_N&\cdots &\cdots &X_3&X_2&X_1\end{bmatrix}.
\end{align}
The notation $\widetilde{\mathcal{O}}(a\log^2 b)$ suppresses polylog terms. It may be replaced with $\mathcal{O}(a\log^2 b)$ if the field supports the Fast Fourier Transform (FFT), and with  $\mathcal{O}(a\log^2b\log\log(b))$ if it does not.

\section{Problem Statement}\label{sec:probstat}

\subsection{Coded Distributed Batch Matrix Multiplication (CDBMM)}
As shown in Figure \ref{fig:DBMM}, consider two source (master) nodes, each of which generates a sequence of $L$ matrices, denoted as $\mathbf{A}=(\mathbf{A}_{1}, \mathbf{A}_{2},\dots, \mathbf{A}_{L})$ and $\mathbf{B}=(\mathbf{B}_{1}, \mathbf{B}_{2},\dots, \mathbf{B}_{L})$, such that for all $l\in[L]$,  we have $\mathbf{A}_{l}\in \mathbb{F}^{\lambda\times\kappa}$ and $\mathbf{B}_{l}\in\mathbb{F}^{\kappa\times\mu}$, i.e., ${\bf A}_l$ and ${\bf B}_l$ are $\lambda\times\kappa$ and $\kappa\times\mu$ matrices, respectively, over a finite\footnote{With the exception of the generalizations to $X$-security presented in Appendix \ref{sec:symxs}, our coding schemes are applicable over infinite fields ($\mathbb{R}, \mathbb{C}$) as well. However, our problem statement assumes that $\mathbb{F}$ is a finite field, because of the difficulty of defining communication costs or computation complexity over infinite fields.} field $\mathbb{F}$. The sink node (user) is interested in the sequence of product matrices, $\mathbf{A}\mathbf{B}=(\mathbf{A}_{1}\mathbf{B}_{1},\mathbf{A}_{2}\mathbf{B}_{2},\dots,\mathbf{A}_{L}\mathbf{B}_{L})$. To help with this computation, there are $S$ servers (worker nodes). Each of the  sources encodes its matrices according to the functions $\textit{\textbf{f}}=(f_1,f_2,\dots,f_S)$ and $\textit{\textbf{g}}=(g_1,g_2,\dots,g_S)$, where $f_s$ and $g_s$ correspond to the $s^{th}$ server. Specifically, let us denote the encoded matrices for the $s^{th}$ server as $\widetilde{A}^s$ and $\widetilde{B}^s$, so we have
\begin{align}
    \widetilde{A}^s&=f_s(\mathbf{A}),\\
    \widetilde{B}^s&=g_s(\mathbf{B}).
\end{align}
The encoded matrices, $ \widetilde{A}^s,  \widetilde{B}^s$,  are uploaded to the $s^{th}$ server. Let us denote the number of elements from $\mathbb{F}$ in $\widetilde{A}^s$ and $\widetilde{B}^s$ as $|\widetilde{A}^s|$ and $|\widetilde{B}^s|$, respectively.

Upon receiving the encoded matrices,  each Server $s$, $s\in[S]$, prepares (computes) a response $Y_s$, that is a function of $\widetilde{A}^s$ and $\widetilde{B}^s$, i.e.,
\begin{equation}
    Y_s=h_s(\widetilde{A}^s, \widetilde{B}^s),
\end{equation}
where $h_s, s\in[S]$ are the functions used to produce the answer, and we denote them collectively as $\textit{\textbf{h}}=(h_1,h_2,\dots,h_S)$. Some servers may fail to respond, such servers are called stragglers. The user downloads the responses from the remaining servers, from which, using a class of decoding functions (denoted $\textit{\textbf{d}}$), he attempts to recover the desired product ${\bf AB}$.  Define
\begin{equation}
    \textit{\textbf{d}}=\{d_{\mathcal{R}}: \mathcal{R}\subset[S]\},
\end{equation}
where $d_{\mathcal{R}}$ is the decoding function used when the set of responsive servers is $\mathcal{R}$.
We say that $(\textit{\textbf{f}},\textit{\textbf{g}},\textit{\textbf{h}},\textit{\textbf{d}})$ form a CDBMM code. A CDBMM code is said to be $r$-recoverable if the user is able to recover the desired products from the answers obtained from any $r$ servers. In particular, a CDBMM code $(\textit{\textbf{f}},\textit{\textbf{g}},\textit{\textbf{h}},\textit{\textbf{d}})$ is $r$-recoverable if for any $\mathcal{R}\subset[S]$, $|\mathcal{R}|=r$, and for any realization of $\mathbf{A}$, $\mathbf{B}$, we have
\begin{equation}
    \mathbf{AB}=d_{\mathcal{R}}(Y_{\mathcal{R}}).
\end{equation}
Define the recovery threshold $R$ of a CDBMM code $(\textit{\textbf{f}},\textit{\textbf{g}},\textit{\textbf{h}},\textit{\textbf{d}})$ to be the minimum integer $r$ such that the CDBMM code is $r$-recoverable. 

The communication cost of CDBMM is comprised of upload and download costs. The (normalized)\footnote{We normalize the upload cost and download cost with the number of elements contained in the constituent matrices ${\bf A}, {\bf B}$, and the desired product ${\bf AB}$, respectively.} upload costs $U_A$ and $U_B$ are defined as follows.
\begin{align}
    U_A&=\frac{\sum_{s\in[S]}|\widetilde{A}^s|}{L\lambda\kappa},\\
    U_B&=\frac{\sum_{s\in[S]}|\widetilde{B}^s|}{L\kappa\mu}.
\end{align}
Similarly, the (normalized) download cost is defined as follows.
\begin{equation}
    D=\max_{\mathcal{R}, \mathcal{R}\subset[S], |\mathcal{R}|=R}\frac{\sum_{s\in\mathcal{R}}|Y_s|}{L\lambda\mu},
\end{equation}
where $|Y_s|$ is the number of elements from $\mathbb{F}$ in $Y_s$.

Next let us consider the complexity of encoding, decoding and server computation. Define the (normalized) computational complexity at each server, $\mathcal{C}_s$, to be the order of the number of arithmetic operations required to compute the function $h_s$ at each server, normalized by $L$. Similarly, define the (normalized) encoding computational complexity $\mathcal{C}_{eA}$ for $\widetilde{A}^{[S]}$ and $\mathcal{C}_{eB}$ for $\widetilde{B}^{[S]}$ as the order of the number of arithmetic operations required to compute the functions $\textit{\textbf{f}}$ and $\textit{\textbf{g}}$, respectively, each normalized by $L$. Finally, define the (normalized) decoding computational complexity $\mathcal{C}_d$ to be the order of the number of arithmetic operations required to compute $d_{\mathcal{R}}(Y_{\mathcal{R}})$, maximized over $\mathcal{R}, \mathcal{R}\subset[S], |\mathcal{R}|=R$, and normalized by $L$. Note that normalizations\footnote{Absolute latency constraints without such normalizations are also quite important in practice. See the discussion following Theorem \ref{thm:gcsa} in Section \ref{sec:gcsa} leading to Fig. \ref{fig:job}.} by $L$ are needed to have fair comparisons between batch processing approaches and individual matrix-partitioning solutions \emph{per matrix multiplication}.

\subsection{Distributed $N$-linear Batch Computation}
Consider an $N$-linear map, which is a function of $N$ variables that is linear separately in each variable. Formally, a map $\Omega:V_1\times V_2\times\cdots\times V_N\rightarrow W$ is called $N$-linear if for all $n\in[N]$, 
\begin{align}
    &\Omega(x^{(1)},\cdots,x^{(n-1)},c_1x^{(n)}+c_2x'^{(n)},x^{(n+1)},\cdots,x^{(N)})\notag\\
    &=c_1\Omega(x^{(1)},\cdots,x^{(n-1)},x^{(n)},x^{(n+1)},\cdots,x^{(N)})+c_2\Omega(x^{(1)},\cdots,x^{(n-1)},x'^{(n)},x^{(n+1)},\cdots,x^{(N)}),
\end{align}
where $V_{[N]}$ and $W$ are vector spaces over the base field $\mathbb{F}$, for all $i\in[N], x^{(i)}\in V_i$, $x'^{(n)}\in V_n$ and $c_1,c_2\in\mathbb{F}$. Consider $N$ sources (master nodes), $n\in[N]$, such that the $n^{th}$ source generates a sequence of $L$ variables $\mathbf{x}^{(n)}=(x_1^{(n)},x_2^{(n)},\cdots,x_L^{(n)})$,   $x_l^{(n)}\in V_n, \forall l\in[L]$. Let us define
\begin{align}
    \mathbf{x}_l=(x^{(1)}_l,x^{(2)}_l,\cdots,x^{(N)}_l),
\end{align}
for all $l\in[L]$. The sink node (user) is interested in the evaluations of the $N$-linear map $\Omega$ over $\mathbf{x}_{[L]}$, i.e., $\Omega(x^{(1)}_l,x^{(2)}_l,\cdots,x^{(N)}_l)=\Omega(\mathbf{x}_l)$, $l\in[L]$. To help with this computation, there are $S$ servers (worker nodes). For all $n\in[N]$, the $n^{th}$ source encodes its variables according to the functions $\textit{\textbf{f}}^{~(n)}=(f_1^{(n)},f_2^{(n)},\cdots,f_S^{(n)})$, where $f_s^{(n)}$ corresponds to the $s^{th}$ server. Let us denote $(\textit{\textbf{f}}^{~(1)},\textit{\textbf{f}}^{~(2)},\cdots, \textit{\textbf{f}}^{~(N)})$ collectively as $\textit{\textbf{f}}$. Like the problem of CDBMM, for all $n\in[N], s\in[S]$, the coded share of the $n^{th}$ source for the $s^{th}$ server is denoted as $\widetilde{X^{(n)}}^s$, and we have
\begin{align}
    \widetilde{X^{(n)}}^s=f_s^{(n)}(\mathbf{x}^{(n)}).
\end{align}
$\left(\widetilde{X^{(n)}}^s\right)_{n\in[N]}$ are uploaded to the $s^{th}$ server. Let us denote the number of elements in $\widetilde{X^{(n)}}^s$ as $\left|\widetilde{X^{(n)}}^s\right|$, $s\in[S], n\in[N]$.

Upon receiving the coded shares, each server $s$, $s\in[S]$ prepares (computes) a response $Y_s$, that is a function of $\widetilde{X^{(n)}}^s, n\in[N]$. 
\begin{align}
    Y_s=h_s(\widetilde{X^{(1)}}^s,\widetilde{X^{(2)}}^s,\cdots,\widetilde{X^{(N)}}^s),
\end{align}
where $h_s,s\in[S]$ are the functions used to produce the answer, and we denote them collectively as $\textbf{\textit{h}}=(h_1,h_2,\cdots,h_S)$. The user downloads the responses from the servers in the set $\mathcal{R}$, and exploits a class of decoding functions (denoted $\textit{\textbf{d}}$) to recover the desired evaluations $\Omega(\mathbf{x}_l)$, $l\in[L]$.  Define
\begin{equation}
    \textit{\textbf{d}}=\{d_{\mathcal{R}}: \mathcal{R}\subset[S]\},
\end{equation}
where $d_{\mathcal{R}}$ is the decoding function used when the set of responsive servers is $\mathcal{R}$.
We say that $(\textit{\textbf{f}},\textit{\textbf{h}},\textit{\textbf{d}})$ form a distributed $N$-linear batch computation code. A distributed $N$-linear batch computation code is said to be $r$-recoverable if the user is able to recover the desired evaluations from the answers obtained from any $r$ servers. In particular, a distributed $N$-linear batch computation code $(\textit{\textbf{f}},\textit{\textbf{h}},\textit{\textbf{d}})$ is $r$-recoverable if for every $\mathcal{R}\subset[S]$, $|\mathcal{R}|=r$, and for every realization of $\mathbf{x}_{[L]}$, we have
\begin{equation}
    (\Omega(\mathbf{x}_l))_{l\in[L]}=d_{\mathcal{R}}(Y_{\mathcal{R}}).
\end{equation}
Define the recovery threshold $R$ of a distributed $N$-linear batch computation code $(\textit{\textbf{f}},\textit{\textbf{h}},\textit{\textbf{d}})$ to be the minimum integer $r$ such that the distributed $N$-linear batch computation code is $r$-recoverable. 

The communication cost of distributed $N$-linear batch computation is comprised of upload and download costs. For all $n\in[N]$, the (normalized) upload cost for $\widetilde{X^{(n)}}^{[S]}$, denoted as $U_{X^{(n)}}$, is defined as follows
\begin{equation}
    U_{X^{(n)}}=\frac{\sum_{s\in[S]}\left|\widetilde{X^{(n)}}^{s}\right|}{L\dim(V_n)}.
\end{equation}
Similarly, the (normalized) download cost is defined as follows.
\begin{equation}
    D=\max_{\mathcal{R},\mathcal{R}\subset[S],|\mathcal{R}=R|}\frac{\sum_{s\in\mathcal{R}}|Y_s|}{L\dim(W)},
\end{equation}
where $|Y_s|$ is the number of elements from $\mathbb{F}$ in $Y_s$.

Define the (normalized) computational complexity at each server, $\mathcal{C}_s$, to be the order of the number of arithmetic operations required to compute the function $h_s$ at each server, normalized by $L$. For all $n\in[N]$, we also define the (normalized) encoding computational complexity $\mathcal{C}_{eX^{(n)}}$ for $\widetilde{X^{(n)}}^{[S]}$ as the order of the number of arithmetic operations required to compute the functions $\textit{\textbf{f}}^{~\!(n)}$, normalized by $L$. Similarly, define the (normalized) decoding computational complexity $\mathcal{C}_d$ to be the order of the number of arithmetic operations required to compute $d_{\mathcal{R}}(Y_{\mathcal{R}})$, maximized over $\mathcal{R}, \mathcal{R}\subset[S], |\mathcal{R}|=R$, and normalized by $L$.

\subsection{Distributed Multivariate Polynomial Batch Evaluation}
Consider a multivariate polynomial $\Phi:V_1\times V_2\times\cdots\times V_M\rightarrow W$ with $M$ variables of total degree $N$, where $V_{[M]}$ and $W$ are vector spaces over the base field $\mathbb{F}$. Consider $M$ sources (master nodes). For all $m\in[M]$, the $m^{th}$ source generates a sequence of $L$ variables $\mathbf{x}^{(m)}=(x_1^{(m)},x_2^{(m)},\cdots,x_L^{(m)})$, such that for all $l\in[L]$, $x_l^{(m)}\in V_m$. Similarly, we define
\begin{align}
    \mathbf{x}_l=(x^{(1)}_l,x^{(2)}_l,\cdots,x^{(M)}_l),
\end{align}
for all $l\in[L]$. The sink node (user) wishes to compute the evaluations of the multivariate polynomial $\Phi$ over $\mathbf{x}_{[L]}$, i.e., $\Phi(x^{(1)}_l,x^{(2)}_l,\cdots,x^{(M)}_l)=\Phi(\mathbf{x}_l)$, $l\in[L]$, with the help of $S$ servers (worker nodes). To this end, for all $m\in[M]$, the $m^{th}$ source encodes its variables according to the functions $\textit{\textbf{f}}^{~\!(m)}=(f_1^{(m)},f_2^{(m)},\cdots,f_S^{(m)})$, where $f_s^{(m)}$ corresponds to the $s^{th}$ server. And  $(\textit{\textbf{f}}^{~\!(1)},\textit{\textbf{f}}^{~\!(2)},\cdots, \textit{\textbf{f}}^{~\!(M)})$ are collectively denoted as $\textit{\textbf{f}}$. For all $n\in[N], s\in[S]$, the coded share of the $m^{th}$ source for the $s^{th}$ server is denoted as $\widetilde{X^{(m)}}^s$, and we have
\begin{align}
    \widetilde{X^{(m)}}^s=f_s^{(m)}(\mathbf{x}^{(m)}).
\end{align}
$\left(\widetilde{X^{(m)}}^s\right)_{m\in[M]}$ are uploaded to the $s^{th}$ server. Let us denote the number of elements in $\widetilde{X^{(m)}}^s$ as $\left|\widetilde{X^{(m)}}^s\right|$, $s\in[S], m\in[M]$.

Upon receiving coded variables, each server $s$, $s\in[S]$ prepares (computes) a response $Y_s$, that is a function of $\widetilde{X^{(m)}}^s, m\in[N]$. 
\begin{align}
    Y_s=h_s(\widetilde{X^{(1)}}^s,\widetilde{X^{(2)}}^s,\cdots,\widetilde{X^{(M)}}^s),
\end{align}
where $h_s,s\in[S]$ are the functions used to produce the answer, and we denote them collectively as $\textbf{\textit{h}}=(h_1,h_2,\cdots,h_S)$. The user downloads the responses from the servers in the set $\mathcal{R}$, and uses a class of decoding functions (denoted $\textit{\textbf{d}}$) to recover the desired evaluations $\Phi(\mathbf{x}_l)$, $l\in[L]$.  Define
\begin{equation}
    \textit{\textbf{d}}=\{d_{\mathcal{R}}: \mathcal{R}\subset[S]\},
\end{equation}
where $d_{\mathcal{R}}$ is the decoding function used when the set of responsive servers is $\mathcal{R}$. We say that $(\textit{\textbf{f}},\textit{\textbf{h}},\textit{\textbf{d}})$ form a distributed multivariate polynomial batch evaluation code. A distributed multivariate polynomial batch evaluation code is said to be $r$-recoverable if the user is able to recover the desired evaluations from the answers obtained from any $r$ servers, i.e., for any $\mathcal{R}\subset[S]$, $|\mathcal{R}|=r$, and for any realization of $\mathbf{x}_{[L]}$, we have
\begin{equation}
    (\Phi(\mathbf{x}_l))_{l\in[L]}=d_{\mathcal{R}}(Y_{\mathcal{R}}).
\end{equation}
Define the recovery threshold $R$ of a distributed multivariate polynomial batch evaluation code $(\textit{\textbf{f}},\textit{\textbf{h}},\textit{\textbf{d}})$ to be the minimum integer $r$ such that the distributed multivariate polynomial batch evaluation code is $r$-recoverable. 

For all $m\in[M]$, the (normalized) upload cost for $\widetilde{X^{(m)}}^{[S]}$, denoted as $U_{X^{(m)}}$, is defined as follows
\begin{equation}
    U_{X^{(m)}}=\frac{\sum_{s\in[S]}\left|\widetilde{X^{(m)}}^{s}\right|}{L\dim(V_m)}.
\end{equation}
Similarly, the (normalized) download cost is defined as follows.
\begin{equation}
    D=\max_{\mathcal{R},\mathcal{R}\subset[S],|\mathcal{R}=R|}\frac{\sum_{s\in\mathcal{R}}|Y_s|}{L\dim(W)},
\end{equation}
where $|Y_s|$ is the number of elements from $\mathbb{F}$ in $Y_s$.

Define the (normalized) computational complexity at each server, $\mathcal{C}_s$, to be the order of the number of arithmetic operations required to compute the function $h_s$ at each server, normalized by $L$. For all $m\in[M]$, we also define the (normalized) encoding computational complexity $\mathcal{C}_{eX^{(m)}}$ for $\widetilde{X^{(m)}}^{[S]}$ as the order of the number of arithmetic operations required to compute the functions $\textit{\textbf{f}}^{(m)}$, normalized by $L$. Similarly, define the (normalized) decoding computational complexity $\mathcal{C}_d$ to be the order of the number of arithmetic operations required to compute $d_{\mathcal{R}}(Y_{\mathcal{R}})$, maximized over $\mathcal{R}, \mathcal{R}\subset[S], |\mathcal{R}|=R$, and normalized by $L$.

\section{CSA Codes for CDBMM}\label{sec:csacodes}
\subsection{CSA Codes: Main Result}
The main result  of this section introduces CSA Codes, and is stated in the following theorem.
\begin{theorem}\label{thm:csacodes}
For CDBMM over a field $\mathbb{F}$ with $S$ servers, and positive integers $\ell$, $K_c$ such that $L=\ell K_c\leq |\mathbb{F}|-S$, the CSA codes introduced in this work achieve
\begin{align}
   \text{Recovery Threshold:}&& R&=(\ell+1)K_c-1,\\
     \text{Upload Cost for $\widetilde{A}^{[S]},\widetilde{B}^{[S]}$:}&&(U_A, U_B)&=\left(\frac{S}{K_c},\frac{S}{K_c}\right),\\
    \text{Download Cost:} &&D&=\frac{(\ell+1)K_c-1}{\ell K_c},\label{eq:dcost}\\
    \text{Server Computation Complexity:}&& \mathcal{C}_s&=\mathcal{O}(\lambda\kappa\mu/K_c),\\
    \text{Encoding Complexity for $\widetilde{A}^{[S]},\widetilde{B}^{[S]}$:}&&(\mathcal{C}_{eA},\mathcal{C}_{eB})&=\left(\widetilde{\mathcal{O}}\left(\frac{\lambda\kappa S\log^2S}{K_c}\right),  \widetilde{\mathcal{O}}\left(\frac{\kappa\mu S\log^2S}{K_c}\right)\right),\\
     \text{Decoding Complexity:}&&\mathcal{C}_d&=\widetilde{\mathcal{O}}\left(\lambda\mu\log^2R\right).
\end{align}
\end{theorem}
The proof of Theorem \ref{thm:csacodes} appears in Section \ref{sec:proof}. A high level summary of the main ideas is provided here. CSA codes split the $L=\ell K_c$ instances of ${\bf A}_l$ matrices into $\ell$ groups, each containing $K_c$ matrices. The $K_c$ matrices within each group are coded into an MDS $(S, K_c)$ code by a Cauchy encoding matrix to create $S$ linear combinations of these $K_c$ matrices.  Multiplication with a  Cauchy encoding matrix corresponds to the well studied Trummer's problem \cite{Golub_Trummer} for which fast algorithms have been found in \cite{Gerasoulis_Grigoriadis_Trummer,Gerasoulis_Trummer,Pan_Tabanjeh_Trummer}  that limit the encoding complexity to $\mathcal{C}_{eA}=\widetilde{\mathcal{O}}(\frac{\lambda\kappa S\log^2S}{K_c})$. The $s^{th}$ coded linear combination from each of the $\ell$ groups is sent to the $s^{th}$ server. The ${\bf B}_l$ matrices are similarly encoded and uploaded to the $S$ servers. Note that because $K_c$ matrices are linearly combined into one linear combination for each server, and there are $S$ servers, the upload cost of CSA codes is $S/K_c$. Each server multiplies the corresponding instances of coded ${\bf A}, {\bf B}$ matrices and returns the sum of these $\ell$ products. With straightforward matrix multiplication algorithms, each of the $\ell$ matrix products has a computation complexity of $\mathcal{O}\left(\lambda\kappa\mu\right)$ for a total of $\mathcal{O}\left(\ell\lambda\kappa\mu\right)$, which upon normalization by $L=\ell K_c$, yields a complexity of $\mathcal{C}_s=\mathcal{O}\left(\lambda\kappa\mu/K_c\right)$ per server. The responses from any $R=(\ell+1)K_c-1$ servers provide $R$ observations to the user, each comprised of linear combinations of various product matrices, including both desired products and undesired products (interference). Interpreting the $R$ observations as occupying an $R$-dimensional vector space, the $L$ desired matrix products $({\bf A}_l{\bf B}_l)_{l\in[L]}$ occupy $L=\ell K_c$ of these $R$ dimensions, leaving only $R-L=K_c-1$ dimensions for interference. Remarkably, while there are a total of $\ell K_c(K_c-1)$ undesired matrix products, ${\bf A}_l{\bf B}_{l'}, l\neq l'$ that appear in the responses from the servers, they collectively occupy only a total of $K_c-1$ dimensions. This is because of \emph{cross-subspace alignment} \cite{Jia_Sun_Jafar_XSTPIR}, facilitated by the specialized Cauchy structure of the encoding. Since $L=\ell K_c$ desired matrix products are recovered from a total of $R$ that are downloaded, the normalized download cost is $\frac{R}{L}=\frac{(\ell+1)K_c-1}{\ell K_c}$. Note that the decoding operation involves inverting a Cauchy-Vandermonde matrix, where the Cauchy part spans the dimensions carrying desired signals while the Vandermonde part spans the dimensions carrying interference. Fast algorithms for inverting such matrices are also known \cite{Finck_FastCV}, which limits the decoding complexity to $\widetilde{\mathcal{O}}(\lambda\mu\log^2R)$.

\subsection{Observations}\label{sec:csaobs}
In this section we present some observations to place CSA Codes into perspective. In particular we would like to compare CSA codes which generalize and improve upon the state of art of batch processing approaches (LCC codes), against EP codes which represent the state of art for matrix-partitioning approaches.
\begin{enumerate}
\item From the conditions of Theorem \ref{thm:csacodes}, the field size $|\mathbb{F}|$ must be at least equal to $S+L$. However, it is possible to reduce the field size requirement to $|\mathbb{F}|\geq S$ by constructing a \emph{systematic} version of the code (see Section \ref{sec:systematic}).
\item To estimate the complexity of computation at each server we use only straightforward matrix multiplication algorithms that require $\widetilde{\mathcal{O}}(\lambda\mu\kappa)$ arithmetic operations over $\mathbb{F}$ in order to compute the product of a $\lambda\times\mu$ matrix with a $\mu\times\kappa$ matrix. It is well known that this complexity can be improved upon by using more sophisticated\footnote{Notably, for $\lambda=\mu=\kappa$  the best known algorithms \cite{LeGall} thus far have computation complexity that is still super-quadratic (more than $\mathcal{O}(\lambda^{2.3})$), and are not considered practical \cite{SCostas} due to large hidden constants in the $\mathcal{O}$ notation.} algorithms \cite{Strassen, Coppersmith_Winograd, LeGall}. Such improvements do not constitute  a relative advantage because they can be applied similarly to other  codes, such as Entangled Polynomial codes as well.
\item We are primarily interested in balanced settings, e.g., $\lambda=\mu=\kappa$, that are typically studied for complexity analysis. While the achievability claims of Theorem \ref{thm:csacodes} are also applicable to unbalanced settings, it is not difficult to improve upon Theorem \ref{thm:csacodes} in certain aspects  in highly unbalanced settings. For example, as shown recently in \cite{Jia_Jafar_SDMM}, when $\kappa\ll \min(\lambda,\mu)$, it may be significantly beneficial for the user  in terms of download cost to retrieve the ${\bf A}, {\bf B}$ matrices separately from the distributed servers and do the computation locally. 
\item First let us compare CSA codes with LCC codes, both of which are based on batch processing. Remarkably, setting $\ell=1$ in CSA codes recovers the LCC code for CDBMM, i.e., LCC codes are a special case of CSA codes. The parameter $\ell$ in CSA codes is mainly\footnote{$\ell$ may be also useful for parallel processing within each server because the computation at each server is naturally split into $\ell$ independent computations.} useful to reduce download cost (by choosing large $\ell$). On the one hand, note that the download cost in \eqref{eq:dcost} is always bounded between $1$ and $2$, so even the worst case choice of $\ell$ will at most double the download cost.  So if the download cost is only important in the $\mathcal{O}$ sense (as a function of $R$), then it is desirable to set $\ell=1$ and $K_c=L$ which reduces the number of parameters for the coding scheme. On the other hand, for settings where the download cost is the dominating concern, the generalization to $\ell>1$ is important. For example, suppose for some application due to latency concerns there is a hard threshold that the download from each server cannot exceed  the equivalent of one matrix multiplication, i.e., no more than $\lambda^2$ elements of $\mathbb{F}$. Then for large batch sizes $L$, the lower download cost of CSA codes translates into a smaller recovery threshold by up to a factor of $2$ relative to LCC codes (albeit at the cost of increased upload and server computation).
\item Next, let us compare the performance of CSA codes with Entangled Polynomial\footnote{Entangled Polynomial codes generalize MatDot codes and Polynomial codes, improve upon PolyDot codes, and have similar performance as Generalized PolyDot codes, so it suffices to compare {CSA} codes with Entangled Polynomial codes.} codes\cite{Yu_Maddah-Ali_Avestimehr}. For this comparison we will only focus on $\ell=1$, so this also applies equivalently to LCC codes instead of CSA codes. In order to compute a batch of matrix products $\left({\bf A}_l{\bf B}_l\right)_{l\in[L]}$,  we will show that joint/batch processing of all $L$ products with {CSA codes} achieves significantly better communication (upload-download) costs than separate  application of Entangled Polynomial codes for each $l\in[L]$, under the same recovery-threshold-computational-complexity-trade-off. It is proved in \cite{Yu_Maddah-Ali_Avestimehr} that for any positive integers $(p, m, n)$, Entangled Polynomial codes  achieve
\begin{align}
   \text{Recovery threshold:}&& R&=pmn+p-1,\\
   \text{Upload cost:}&& (U_A, U_B)&=(S/pm, S/pn),\\
   \text{Download cost:}&& D&=\frac{pmn+p-1}{mn}.
\end{align}
To simplify the order analysis, let us assume that $\lambda=\kappa=\mu$, and to balance the upload costs $(U_A, U_B)$ let us choose $m=n$. Let us regard the recovery threshold $R$ as a variable, and consider the upload cost $U_A, U_B$ and the download cost $D$ as functions of $R$. So for the Entangled Polynomial codes \cite{Yu_Maddah-Ali_Avestimehr}, we have
\begin{align}
    U_A&=U_B=U=\frac{mS}{pm^2}\geq m\left(\frac{S}{R}\right),&&&    D&=\frac{R}{m^2}.
\end{align}
A tradeoff is evident. For example, if we want download cost of $\mathcal{O}(1)$, then we need $m= \Theta(\sqrt{R})$ which yields upload cost of $\mathcal{O}(S/\sqrt{R})$. On the other hand, if we want upload cost of $\mathcal{O}(S/R)$, then we should set $m=\Theta(1)$ which yields download cost of $\mathcal{O}(R)$. If $S/R$ is held constant, then to best balance the upload and download cost, we need $m=\Theta(R^{1/3})$, which yields both upload cost  and download cost of $\mathcal{O}(R^{1/3})$. Evidently, it is not possible to achieve both upload and download cost of $\mathcal{O}(1)$ with Entagled Polynomial codes. However, with CSA codes, setting $\ell=1$, we have upload cost of $\mathcal{O}(1)$ and download cost of $\mathcal{O}(S/R)$. In particular, if $S/R$ is a constant, then both upload and download costs are $\mathcal{O}(1)$. Note that   {CSA codes}  have the same server computational complexity of $\mathcal{O}(\lambda^3/R)$ \emph{normalized by batch size} as EP codes. %

\item Continuing with the comparison between CSA codes and EP codes, Figure \ref{fig:compareuda}, \ref{fig:compareudb} and \ref{fig:compareudc} show lower convex hulls of achievable (balanced upload cost, download cost) pairs of Entangled Polynomial codes and {CSA codes} given the number of servers and the recovery threshold $(S=30, R\leq 25)$, $(S=300, R\leq 250)$ and $(S=3000, R\leq 2500)$ respectively. Each value of $(S,R)$ produces an achievable region in the $(U,D)$ plane (including all possible choices of $m,n,p$ parameters for Entagled Polynomial codes, and all choices of $\ell, K_c$ parameters for CSA codes). What is shown in the figure is the union of these regions for each case, e.g., in the first figure the union is over all $(S,R)$ with $(S=30, R\leq 25)$.  Evidently, the advantage of CSA codes over Entangled Polynomial codes in terms of communication cost is significant and grows stronger for larger $(S, R)$ values.
\begin{figure}[htbp]
\begin{subfigure}{0.33\textwidth}
\centering
\includegraphics[scale=0.65]{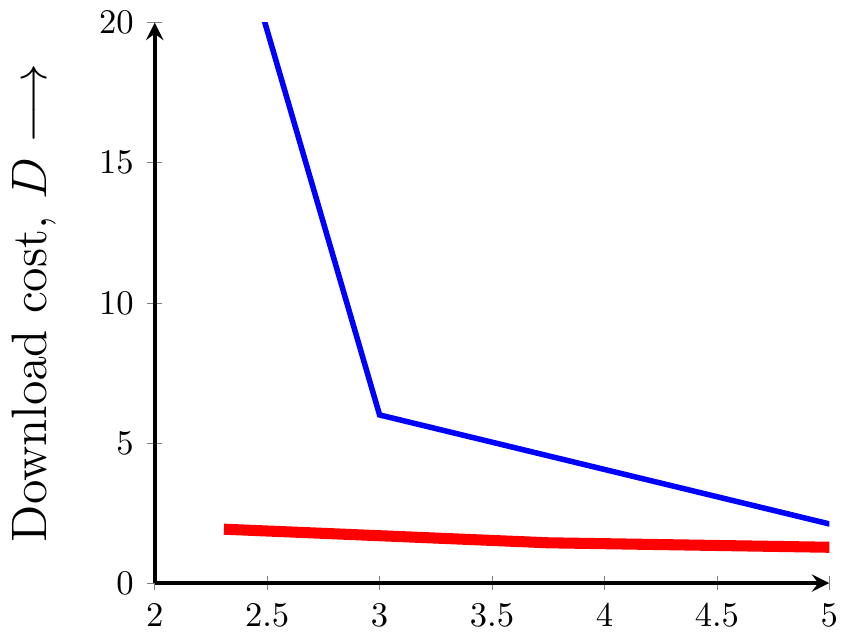}
\vspace{0cm}
\caption{$S=30, R\leq 25$}\label{fig:compareuda}
\end{subfigure}
\begin{subfigure}{0.33\textwidth}
\centering
\includegraphics[scale=0.65]{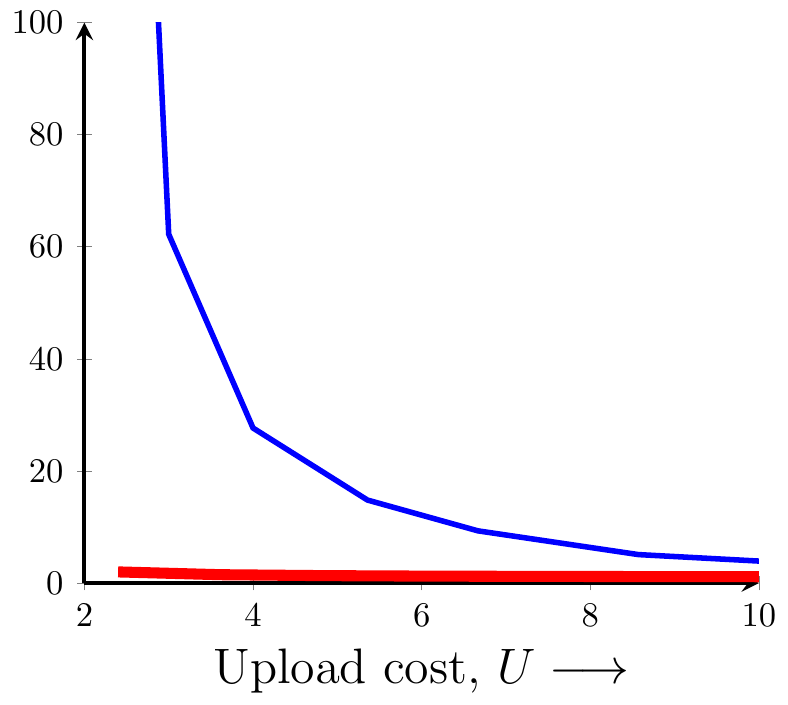}
\caption{$S=300, R\leq 250$} 
\label{fig:compareudb}
\end{subfigure}
\begin{subfigure}{0.33\textwidth}
\centering
\includegraphics[scale=0.65]{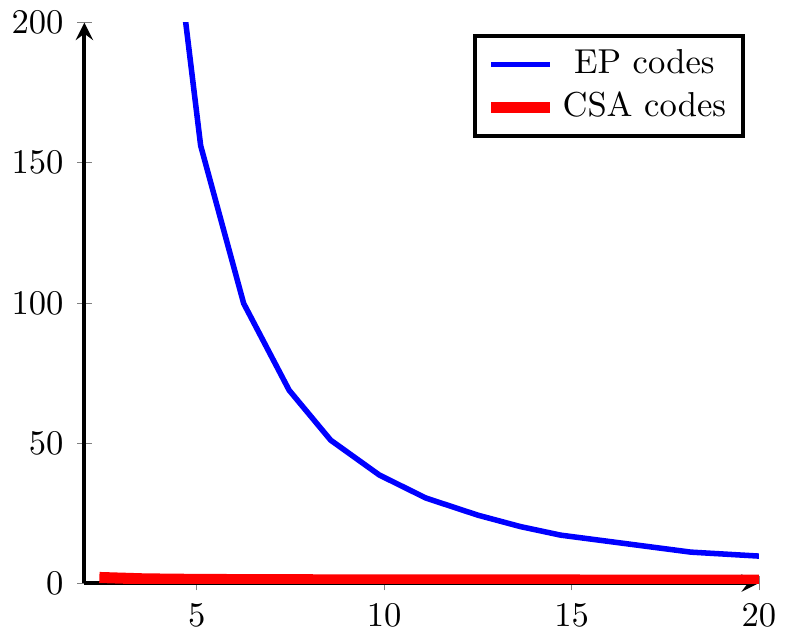}
\vspace{0.4cm}
\caption{$S=3000, R\leq 2500$} 
\label{fig:compareudc}
\end{subfigure}
   
    \caption{\it Lower convex hulls of achievable (balanced upload cost, download cost) pairs $(U,D)$ of Entangled Polynomial codes (EP codes) and \emph{cross subspace alignment codes} (CSA codes) given (a) $(S=30, R\leq 25)$, (b) $(S=300, R\leq 250)$ and (c) $(S=3000, R\leq 2500)$. }
    \label{fig:compareud}
\end{figure}

\item  CSA codes show a similar advantage over Entangled Polynomial codes in terms of the tradeoff between encoding complexity and decoding complexity normalized by batch size. For example, consider the balanced setting of $m=n$, $\lambda=\mu=\kappa$, and constant $S/R$. The encoding  complexity of Entangled Polynomial codes is $\widetilde{\mathcal{O}}(\lambda^2 U\log^2S)$, and the decoding  complexity is $\widetilde{\mathcal{O}}(\lambda^2D\log^2R)$, where $U=U_A=U_B$ is the balanced upload  cost, and $D$ is the download cost. For CSA codes, the encoding  complexity  is $\widetilde{\mathcal{O}}(\lambda^2\log^2S)$, and the decoding  complexity  is $\widetilde{\mathcal{O}}(\lambda^2\log^2R)$, which corresponds to $U=D=\mathcal{O}(1)$. Thus, the communication cost advantage of CSA codes over Entangled Polynomial codes is further manifested in the improved tradeoff between encoding and decoding complexity.

\item Finally, let us place CSA codes in perspective with previous applications of cross-subspace alignment. The idea of cross-subspace alignment was introduced in the context of $X$-secure $T$-private information retrieval (XSTPIR) \cite{Jia_Sun_Jafar_XSTPIR}. The goal of XSTPIR is to allow a user to retrieve, as efficiently as possible, a desired message $W_\theta$ out of $K$ messages, $W_1, W_2, \cdots, W_K$ that are `secret-shared' across $S$ servers in an $X$-secure fashion,  without revealing any information about the index $\theta$ to any group of up to $T$ colluding servers. According to the scheme proposed in \cite{Jia_Sun_Jafar_XSTPIR} the $\ell^{th}$ symbol of each message is stored in the $1\times K$ vector ${\bf W}_\ell$. The query vector ${\bf Q}_\theta$ is the $\theta^{th}$ column of a $K\times K$ identity matrix, so that retrieving the product ${\bf W}_\ell{\bf Q}_\theta$ retrieves the $\ell^{th}$ symbol of the desired message $W_\theta$. In order to guarantee security of data and privacy of queries, the ${\bf W}_\ell$ and ${\bf Q}_\theta$ vectors are mixed with independent noise terms. Intuitively, by replacing ${\bf W}_\ell$ and ${\bf Q}_\theta$ with matrices ${\bf A}$ and ${\bf B}$, and eliminating the corresponding noise terms if the privacy and/or security constraints are relaxed, cross-subspace alignment schemes can be used to retrieve arbitrary matrix products ${\bf AB}$. This intuition helps with some of the achievable schemes\footnote{Notably, batch processing is used in \cite{Jia_Jafar_SDMM} while matrix partitioning is used in \cite{Kakar_Ebadifar_Sezgin_CSA}. The achievable schemes in \cite{Jia_Jafar_SDMM, Kakar_Ebadifar_Sezgin_CSA} can be regarded as special cases of $X$-secure CSA codes presented in this work with $K_c=1$. See Appendix \ref{sec:symxs} for details.} in \cite{Jia_Jafar_SDMM, Kakar_Ebadifar_Sezgin_CSA}.
However, in \cite{Jia_Sun_Jafar_XSTPIR}, the ${\bf W}_\ell$ vectors are not jointly encoded. Each ${\bf W}_\ell$ vector is separately mixed with noise. Similarly, in \cite{Jia_Jafar_SDMM, Kakar_Ebadifar_Sezgin_CSA}  the matrices are separately mixed with noise for security, and not jointly encoded. Joint encoding of messages arises in PIR when instead of replicated storage \cite{Sun_Jafar_PIR, Sun_Jafar_TPIR}, coded storage is assumed \cite{Banawan_Ulukus, Tajeddine_Rouayheb, FREIJ_HOLLANTI,Sun_Jafar_MDSTPIR,  Zhang_Ge_Variant, Wang_Skoglund_TSPIR}. PIR with MDS-coded storage, $X$-secure data and $T$-private queries is studied in \cite{Jia_Jafar_MDSXSTPIR} and indeed a generalized cross-subspace alignment scheme is the key contribution of \cite{Jia_Jafar_MDSXSTPIR}. However, since there is only one query vector ${\bf Q}_\theta$, applications of this cross-subspace alignment scheme are useful primarily for  matrix multiplications of the form ${\bf A}_1{\bf B}, {\bf A}_2{\bf B}, \cdots, {\bf A}_L{\bf B}$, where we have only one ${\bf B}$ matrix to be multiplied with each ${\bf A}$ matrix. This is indeed how the scheme is applied in the context of private secure distributed matrix multiplication (PSDMM) in \cite{Jia_Jafar_MDSXSTPIR}. Batch multiplications of the form ${\bf A}_1{\bf B}_1, {\bf A}_2{\bf B}_2, \cdots, {\bf A}_L{\bf B}_L$, that are studied in this work, present a significantly greater challenge in that joint coding is now to be applied both among ${\bf A}_1, {\bf A}_2, \cdots, {\bf A}_L$ and among ${\bf B}_1, {\bf B}_2, \cdots, {\bf B}_L$ matrices, which introduces new interference terms ${\bf A}_l{\bf B}_{l'}, l\neq l'$. A central technical challenge behind this work is to determine if and how these terms can be aligned. The CSA codes introduced in this work present a solution to this challenge.

\end{enumerate}

\subsection{Proof of Theorem \ref{thm:csacodes}}\label{sec:proof}
In this section, we present the construction of {CSA codes}. Let $L=\ell K_c$. Recall Lemma 1 in \cite{Jia_Jafar_MDSXSTPIR}, which is also a standard result for Cauchy-Vandermonde matrices \cite{Gasca_Martinez_Muhlbach}, replicated here for the sake of completeness. 
\begin{lemma}\label{lemma:csa}
If $f_{1,1}, f_{1,2}, \cdots, f_{\ell,K_c}, \alpha_1,\alpha_2,\cdots,\alpha_R$ are $R+L$ distinct elements of $\mathbb{F}$, with $1\leq \ell K_c=L\leq R-1$ and $|\mathbb{F}|\geq R+L$, then the following $R\times R$ Cauchy-Vandermonde matrix is invertible over $\mathbb{F}$.
\begin{align}
{\bf V}_{\ell,K_c,R}&\triangleq\left[
\begin{matrix}
\frac{1}{f_{1,1}-\alpha_1}&\frac{1}{f_{1,2}-\alpha_1}&\cdots&\frac{1}{f_{\ell,K_c}-\alpha_1}&1&\alpha_1&\cdots&\alpha_1^{R-L-1}\\
\frac{1}{f_{1,1}-\alpha_2}&\frac{1}{f_{1,2}-\alpha_2}&\cdots&\frac{1}{f_{\ell,K_c}-\alpha_2}&1&\alpha_2&\cdots&\alpha_2^{R-L-1}\\
\vdots&\vdots&\vdots&\vdots&\vdots&\vdots&\vdots&\vdots\\
\frac{1}{f_{1,1}-\alpha_R}&\frac{1}{f_{1,2}-\alpha_R}&\cdots&\frac{1}{f_{\ell,K_c}-\alpha_R}&1&\alpha_R&\cdots&\alpha_R^{R-L-1}\\
\end{matrix}
\right]
\end{align}
\end{lemma}

\noindent Before presenting the general code construction let us start with some illustrative examples.
\subsubsection{$\ell=1$, $K_c=2$, $L=2$}
Let $f_{1,1}, f_{1,2}, \alpha_{1}, \alpha_{2},\dots,\alpha_{S}$ represent $(S+2)$ distinct elements from $\mathbb{F}$. For all $s\in[S]$, define,
\begin{equation}
    \Delta_{s}^{1,2}=(f_{1,1}-\alpha_s)(f_{1,2}-\alpha_s).
\end{equation}
 Shares of matrices $\mathbf{A}$ are constructed as follows.
\begin{align}
    \widetilde{A}^s&=\Delta_{s}^{1,2}\left(\frac{1}{f_{1,1}-\alpha_s}\mathbf{A}_{1,1}+\frac{1}{f_{1,2}-\alpha_s}\mathbf{A}_{1,2}\right)\\
    &=(f_{1,2}-\alpha_s)\mathbf{A}_{1,1}+(f_{1,1}-\alpha_s)\mathbf{A}_{1,2},
\end{align}
where we set $\mathbf{A}_{1,1}=\mathbf{A}_{1}$ and  $\mathbf{A}_{1,2}=\mathbf{A}_{2}$. Shares of matrices $\mathbf{B}$ are constructed as follows.
\begin{align}
    \widetilde{B}^s=\frac{1}{f_{1,1}-\alpha_s}\mathbf{B}_{1,1}+\frac{1}{f_{1,2}-\alpha_s}\mathbf{B}_{1,2},
\end{align}
where we similarly set that $\mathbf{B}_{1,1}=\mathbf{B}_{1}$ and $\mathbf{B}_{1,2}=\mathbf{B}_{2}$. The answer from the $s^{th}$ server,  $Y_s$ is constructed as follows.
\begin{align}
    Y_s&=\widetilde{A}^s\widetilde{B}^s\\
    &=\frac{f_{1,2}-\alpha_s}{f_{1,1}-\alpha_s}\mathbf{A}_{1,1}\mathbf{B}_{1,1}+\frac{f_{1,1}-\alpha_s}{f_{1,2}-\alpha_s}\mathbf{A}_{1,2}\mathbf{B}_{1,2}+(\mathbf{A}_{1,1}\mathbf{B}_{1,2}+\mathbf{A}_{1,2}\mathbf{B}_{1,1}).
\end{align}
Now let us see how the user is able to recover the desired matrix products $(\mathbf{A}_{1}\mathbf{B}_{1},~\mathbf{A}_{2}\mathbf{B}_{2})=(\mathbf{A}_{1,1}\mathbf{B}_{1,1},~\mathbf{A}_{1,2}\mathbf{B}_{1,2})$ with recovery threshold $R=(\ell+1)K_c-1=2\times2-1=3$. We can rewrite $Y_s$ as follows.
\begin{align}
    Y_s&=\frac{f_{1,2}-\alpha_s}{f_{1,1}-\alpha_s}\mathbf{A}_{1,1}\mathbf{B}_{1,1}+\frac{f_{1,1}-\alpha_s}{f_{1,2}-\alpha_s}\mathbf{A}_{1,2}\mathbf{B}_{1,2}+(\mathbf{A}_{1,1}\mathbf{B}_{1,2}+\mathbf{A}_{1,2}\mathbf{B}_{1,1})\\
    &=\frac{f_{1,1}-\alpha_s+(f_{1,2}-f_{1,1})}{f_{1,1}-\alpha_s}\mathbf{A}_{1,1}\mathbf{B}_{1,1}+\frac{f_{1,2}-\alpha_s+(f_{1,1}-f_{1,2})}{f_{1,2}-\alpha_s}\mathbf{A}_{1,2}\mathbf{B}_{1,2}\notag\\
    &\quad\quad+(\mathbf{A}_{1,1}\mathbf{B}_{1,2}+\mathbf{A}_{1,2}\mathbf{B}_{1,1})\\
    &=\frac{f_{1,2}-f_{1,1}}{f_{1,1}-\alpha_s}\mathbf{A}_{1,1}\mathbf{B}_{1,1}+\frac{f_{1,1}-f_{1,2}}{f_{1,2}-\alpha_s}\mathbf{A}_{1,2}\mathbf{B}_{1,2}\notag\\
    &\quad\quad+\underbrace{(\mathbf{A}_{1,1}\mathbf{B}_{1,1}+\mathbf{A}_{1,2}\mathbf{B}_{1,2}+\mathbf{A}_{1,1}\mathbf{B}_{1,2}+\mathbf{A}_{1,2}\mathbf{B}_{1,1})}_{I_1},
\end{align}
where $I_1$ represents \emph{interference} due to undesired terms.
For any $R=3$ servers, whose indices are denoted as $s_1,s_2,s_3$, we can represent their answers in the following matrix form.
\begin{equation}
    \begin{bmatrix}
    Y_{s_1}\\
    Y_{s_2}\\
    Y_{s_3}
    \end{bmatrix}=\underbrace{\begin{bmatrix}
    \frac{1}{f_{1,1}-\alpha_{s_1}}&\frac{1}{f_{1,2}-\alpha_{s_1}}&1\\
    \frac{1}{f_{1,1}-\alpha_{s_2}}&\frac{1}{f_{1,2}-\alpha_{s_2}}&1\\
    \frac{1}{f_{1,1}-\alpha_{s_3}}&\frac{1}{f_{1,2}-\alpha_{s_3}}&1
    \end{bmatrix}}_{\mathbf{V}_{1,2,3}}\underbrace{\begin{bmatrix}
    f_{1,2}-f_{1,1}&&\\
    &f_{1,1}-f_{1,2}&\\
    &&1
    \end{bmatrix}}_{\mathbf{V}'_{1,2,3}}\otimes\mathbf{I}_{\lambda}\begin{bmatrix}
    \mathbf{A}_{1,1}\mathbf{B}_{1,1}\\
    \mathbf{A}_{1,2}\mathbf{B}_{1,2}\\
    I_1
    \end{bmatrix}
\end{equation}
It follows from Lemma \ref{lemma:csa} that the $3\times 3$ matrix $\mathbf{V}_{1,2,3}$ is  invertible. By definition, $f_{1,1}\neq f_{1,2}$, thus the matrix $\mathbf{V}_{1,2,3}\mathbf{V}'_{1,2,3}$ is  invertible. Since the  Kronecker product of non-singular matrices is non-singular, the $3\lambda\times 3\lambda$ matrix $(\mathbf{V}_{1,2,3}\mathbf{V}'_{1,2,3})\otimes\mathbf{I}_{\lambda}$ is also invertible. Thus the user is able to recover desired products $(\mathbf{A}_{1,1}\mathbf{B}_{1,1},\mathbf{A}_{1,2}\mathbf{B}_{1,2})$ by inverting the matrix $(\mathbf{V}_{1,2,3}\mathbf{V}'_{1,2,3})\otimes\mathbf{I}_{\lambda}$. This completes the proof of recovery threshold $R=3$. Now let us calculate the upload cost and download cost of the code. Since  $\widetilde{A}^s$ consists of $\lambda\kappa$ $q$-ary symbols, and $\widetilde{B}^s$ consists of $\kappa\mu$ $q$-ary symbols, we have $U_A=U_B=S/2=S/K_c$. On the other hand, for all $s\in[S]$, $Y_s$ consists of $\lambda\mu$ $q$-ary symbols, and from any $R=3$ answers, the user is able to recover the two desired matrix products. So we have normalized download cost $D=3/2$. 

\subsubsection{$\ell=2, K_c=2, L=4$}
Let $f_{1,1},f_{1,2},f_{2,1},f_{2,2},\alpha_1,\dots,\alpha_S$ represent $(S+\ell K_c)=(S+4)$ distinct elements from $\mathbb{F}$. For all $s\in[S]$, let us define
\begin{align}
    \Delta_s^{1,2}&=(f_{1,1}-\alpha_s)(f_{1,2}-\alpha_s),\\
    \Delta_s^{2,2}&=(f_{2,1}-\alpha_s)(f_{2,2}-\alpha_s).
\end{align}
Let us set $\mathbf{A}_{l,k}=\mathbf{A}_{K_c(l-1)+k}$ and $\mathbf{B}_{lk}=\mathbf{B}_{K_c(l-1)+k}$ for all $l\in[2],k\in[2]$. Coded shares of matrices $\mathbf{A}$ are constructed as follows.
\begin{equation}
    \widetilde{A}^s=(\widetilde{A}_{1}^s,\widetilde{A}_{2}^s),
\end{equation}
where
\begin{align}
    \widetilde{A}_{1}^s&=\Delta_s^{1,2}\left(\frac{1}{f_{1,1}-\alpha_s}\mathbf{A}_{1,1}+\frac{1}{f_{1,2}-\alpha_s}\mathbf{A}_{1,2}\right)\\
    &=(f_{1,2}-\alpha_s)\mathbf{A}_{1,1}+(f_{1,1}-\alpha_s)\mathbf{A}_{1,2},\\
    \widetilde{A}_{2}^s&=\Delta_s^{2,2}\left(\frac{1}{f_{2,1}-\alpha_s}\mathbf{A}_{2,1}+\frac{1}{f_{2,2}-\alpha_s}\mathbf{A}_{2,2}\right)\\
    &=(f_{2,2}-\alpha_s)\mathbf{A}_{2,1}+(f_{2,1}-\alpha_s)\mathbf{A}_{2,2}.
\end{align}
Coded shares of matrices $\mathbf{B}$ are constructed as follows.
\begin{equation}
    \widetilde{B}^s=(\widetilde{B}_{1}^s,\widetilde{B}_{2}^s),
\end{equation}
where
\begin{align}
    \widetilde{B}_{1}^s&=\frac{1}{f_{1,1}-\alpha_s}\mathbf{B}_{1,1}+\frac{1}{f_{1,2}-\alpha_s}\mathbf{B}_{1,2},\\
    \widetilde{B}_{2}^s&=\frac{1}{f_{2,1}-\alpha_s}\mathbf{B}_{2,1}+\frac{1}{f_{2,2}-\alpha_s}\mathbf{B}_{2,2}.
\end{align}
The answer provided by the $s^{th}$ server to the user is constructed as follows.
\begin{equation}
    Y_s=\widetilde{A}_{1}^s\widetilde{B}_{1}^s+\widetilde{A}_{2}^s\widetilde{B}_{2}^s.
\end{equation}
To see why the $R=(\ell+1)K_c-1=5$ recovery threshold holds, we rewrite $Y_s$ as follows.
\begin{align}
    Y_s&=\widetilde{A}_{1}^s\widetilde{B}_{1}^s+\widetilde{A}_{2}^s\widetilde{B}_{2}^s\\
    &=\frac{f_{1,2}-\alpha_s}{f_{1,1}-\alpha_s}\mathbf{A}_{1,1}\mathbf{B}_{1,1}+\frac{f_{1,1}-\alpha_s}{f_{1,2}-\alpha_s}\mathbf{A}_{1,2}\mathbf{B}_{1,2}+\frac{f_{2,2}-\alpha_s}{f_{2,1}-\alpha_s}\mathbf{A}_{2,1}\mathbf{B}_{2,1}+\frac{f_{2,1}-\alpha_s}{f_{2,2}-\alpha_s}\mathbf{A}_{2,2}\mathbf{B}_{2,2}\notag\\
    &\quad\quad+(\mathbf{A}_{1,1}\mathbf{B}_{1,2}+\mathbf{A}_{1,2}\mathbf{B}_{1,1}+\mathbf{A}_{2,1}\mathbf{B}_{2,2}+\mathbf{A}_{2,2}\mathbf{B}_{2,1})\\
    &=\frac{f_{1,1}-\alpha_s+(f_{1,2}-f_{1,1})}{f_{1,1}-\alpha_s}\mathbf{A}_{1,1}\mathbf{B}_{1,1}+\frac{f_{1,2}-\alpha_s+(f_{1,1}-f_{1,2})}{f_{1,2}-\alpha_s}\mathbf{A}_{1,2}\mathbf{B}_{1,2}\notag\\
    &\quad\quad+\frac{f_{2,1}-\alpha_s+(f_{2,2}-f_{2,1})}{f_{2,1}-\alpha_s}\mathbf{A}_{2,1}\mathbf{B}_{2,1}+\frac{f_{2,2}-\alpha_s+(f_{2,1}-f_{2,2})}{f_{2,2}-\alpha_s}\mathbf{A}_{2,2}\mathbf{B}_{2,2}\notag\\
    &\quad\quad+(\mathbf{A}_{1,1}\mathbf{B}_{1,2}+\mathbf{A}_{1,2}\mathbf{B}_{1,1}+\mathbf{A}_{2,1}\mathbf{B}_{2,2}+\mathbf{A}_{2,2}\mathbf{B}_{2,1})\\
    &=\frac{f_{1,2}-f_{1,1}}{f_{1,1}-\alpha_s}\mathbf{A}_{1,1}\mathbf{B}_{1,1}+\frac{f_{1,1}-f_{1,2}}{f_{1,2}-\alpha_s}\mathbf{A}_{1,2}\mathbf{B}_{1,2}\notag\\
    &\quad\quad+\frac{f_{2,2}-f_{2,1}}{f_{2,1}-\alpha_s}\mathbf{A}_{2,1}\mathbf{B}_{2,1}+\frac{f_{2,1}-f_{2,2}}{f_{2,2}-\alpha_s}\mathbf{A}_{2,2}\mathbf{B}_{2,2}\notag\\
    &\quad\quad\quad\quad\underbrace{\begin{aligned}&+(\mathbf{A}_{1,1}\mathbf{B}_{1,1}+\mathbf{A}_{1,2}\mathbf{B}_{1,2}+\mathbf{A}_{2,1}\mathbf{B}_{2,1}+\mathbf{A}_{2,2}\mathbf{B}_{2,2}\\&\quad\quad+\mathbf{A}_{1,1}\mathbf{B}_{1,2}+\mathbf{A}_{1,2}\mathbf{B}_{1,1}+\mathbf{A}_{2,1}\mathbf{B}_{2,2}+\mathbf{A}_{2,2}\mathbf{B}_{2,1}).\end{aligned}}_{I_1}
\end{align}
Therefore, for any $R=5$ servers, whose indices are denoted as $s_1,s_2,\dots,s_5$, we can represent their answers in the following matrix form.
\begin{equation}
    \begin{bmatrix}
    Y_{s_1}\\
    Y_{s_2}\\
    Y_{s_3}\\
    Y_{s_4}\\
    Y_{s_5}\\
    \end{bmatrix}=\underbrace{\begin{bmatrix}
    \frac{1}{f_{1,1}-\alpha_{s_1}}&\frac{1}{f_{1,2}-\alpha_{s_1}}&\frac{1}{f_{2,1}-\alpha_{s_1}}&\frac{1}{f_{2,2}-\alpha_{s_1}}&1\\
    \frac{1}{f_{1,1}-\alpha_{s_2}}&\frac{1}{f_{1,2}-\alpha_{s_2}}&\frac{1}{f_{2,1}-\alpha_{s_2}}&\frac{1}{f_{2,2}-\alpha_{s_2}}&1\\
    \frac{1}{f_{1,1}-\alpha_{s_3}}&\frac{1}{f_{1,2}-\alpha_{s_3}}&\frac{1}{f_{2,1}-\alpha_{s_3}}&\frac{1}{f_{2,2}-\alpha_{s_3}}&1\\
    \frac{1}{f_{1,1}-\alpha_{s_4}}&\frac{1}{f_{1,2}-\alpha_{s_4}}&\frac{1}{f_{2,1}-\alpha_{s_4}}&\frac{1}{f_{2,2}-\alpha_{s_4}}&1\\
    \frac{1}{f_{1,1}-\alpha_{s_5}}&\frac{1}{f_{1,2}-\alpha_{s_5}}&\frac{1}{f_{2,1}-\alpha_{s_5}}&\frac{1}{f_{2,2}-\alpha_{s_5}}&1
    \end{bmatrix}}_{\mathbf{V}_{2,2,5}}\underbrace{\begin{bmatrix}
    c_{1,1}&&&&\\
    &c_{1,2}&&&\\
    &&c_{2,1}&&\\
    &&&c_{2,2}&\\
    &&&&1
    \end{bmatrix}}_{\mathbf{V}'_{2,2,5}}\otimes\mathbf{I}_{\lambda}\begin{bmatrix}
    \mathbf{A}_{1,1}\mathbf{B}_{1,1}\\
    \mathbf{A}_{1,2}\mathbf{B}_{1,2}\\
    \mathbf{A}_{2,1}\mathbf{B}_{2,1}\\
    \mathbf{A}_{2,2}\mathbf{B}_{2,2}\\
    I_1
    \end{bmatrix},
\end{equation}
where $c_{1,1}=f_{1,2}-f_{1,1}$, $c_{2,1}=f_{2,2}-f_{2,1}$, $c_{1,2}=-c_{1,1}$, $c_{2,2}=-c_{2,1}$. Since $f_{1,1}, f_{1,2}, f_{2,1}, f_{2,2}$ are distinct elements from $\mathbb{F}$, the constants $c_{1,1}, c_{1,2}, c_{2,1}, c_{2,2}$ take non-zero values. Guaranteed by Lemma \ref{lemma:csa} and the fact that Kronecker product of non-singular matrices is non-singular, the $5\lambda\times 5\lambda$ matrix $(\mathbf{V}_{2,2,5}\mathbf{V}'_{2,2,5})\otimes\mathbf{I}_{\lambda}$ is invertible, and the user is able to recover desired products $(\mathbf{A}_{1}\mathbf{B}_{1},\dots,\mathbf{A}_{4}\mathbf{B}_{4})=(\mathbf{A}_{l,k}\mathbf{B}_{l,k})_{l\in[2],k\in[2]}$ from the answers received from  any $R=5$ servers. This completes the proof of the $R=5$ recovery threshold. Finally, note that the upload cost  is $U_A=U_B=S/2=S/K_c$ and the download cost is $D=4/5$ because a total of $R=5$   matrix products, each of dimension $\lambda\times\mu$, are downloaded (one from each server) from which the $4$ desired matrix products, also each of dimension $\lambda\times \mu$, are recovered.

\subsubsection{$\ell=1, K_c=3, L=3$}
Let $f_{1,1}, f_{1,2}, f_{1,3}, \alpha_1,\alpha_2,\dots,\alpha_S$ represent $(S+\ell K_c)=(S+3)$ distinct elements from $\mathbb{F}$. For all $s\in[S]$, let us define
\begin{equation}
    \Delta_s^{1,3}=(f_{1,1}-\alpha_s)(f_{1,2}-\alpha_s)(f_{1,3}-\alpha_s).
\end{equation}
Shares of $\mathbf{A}$ and $\mathbf{B}$ at the $s^{th}$ server are constructed as follows.
\begin{align}
    \widetilde{A}^s&=\Delta_s^{1,3}\left(\frac{1}{f_{1,1}-\alpha_s}\mathbf{A}_{1,1}+\frac{1}{f_{1,2}-\alpha_s}\mathbf{A}_{1,2}+\frac{1}{f_{1,3}-\alpha_s}\mathbf{A}_{1,3}\right),\\
    \widetilde{B}^s&=\frac{1}{f_{1,1}-\alpha_s}\mathbf{B}_{1,1}+\frac{1}{f_{1,2}-\alpha_s}\mathbf{B}_{1,2}+\frac{1}{f_{1,3}-\alpha_s}\mathbf{B}_{1,3},
\end{align}
where we set $\mathbf{A}_{1,k}=\mathbf{A}_{k}$ and $\mathbf{B}_{1,k}=\mathbf{B}_{k}$ for $k\in[3]$. The answer returned by the $s^{th}$ server to the user is
\begin{equation}
    Y_s=\widetilde{A}^s\widetilde{B}^s.
\end{equation}
Now let us prove that the user is able to recover desired products $(\mathbf{A}_{l}\mathbf{B}_{l})_{l\in[3]}=(\mathbf{A}_{1,k}\mathbf{B}_{1,k})_{k\in[3]}$ with recovery threshold $R=(\ell+1)K_c-1=2\times 3-1=5$. Let us rewrite $Y_s$ as follows.
\begin{align}
    Y_s&=\widetilde{A}^s\widetilde{B}^s\\
    &=\frac{(f_{1,2}-\alpha_s)(f_{1,3}-\alpha_s)}{f_{1,1}-\alpha_s}\mathbf{A}_{1,1}\mathbf{B}_{1,1}+\frac{(f_{1,1}-\alpha_s)(f_{1,3}-\alpha_s)}{f_{1,2}-\alpha_s}\mathbf{A}_{1,2}\mathbf{B}_{1,2}\notag\\
    &\quad\quad+\frac{(f_{1,1}-\alpha_s)(f_{1,2}-\alpha_s)}{f_{1,3}-\alpha_s}\mathbf{A}_{1,3}\mathbf{B}_{1,3}+(f_{1,1}-\alpha_s)(\mathbf{A}_{1,2}\mathbf{B}_{1,3}+\mathbf{A}_{1,3}\mathbf{B}_{1,2})\notag\\
    &\quad\quad+(f_{1,2}-\alpha_s)(\mathbf{A}_{1,1}\mathbf{B}_{1,3}+\mathbf{A}_{1,3}\mathbf{B}_{1,1})+(f_{1,3}-\alpha_s)(\mathbf{A}_{1,1}\mathbf{B}_{1,2}+\mathbf{A}_{1,2}\mathbf{B}_{1,1}).\label{eq:ex3ans}
\end{align}
Next let us manipulate the first term on the RHS. By  long division of polynomials (regard numerator and denominator as polynomials of $\alpha_s$), we have
\begin{align}
    &\frac{(f_{1,2}-\alpha_s)(f_{1,3}-\alpha_s)}{f_{1,1}-\alpha_s}\mathbf{A}_{1,1}\mathbf{B}_{1,1}\\
    &=\left(-\alpha_s+(f_{1,2}+f_{1,3}-f_{1,1})+\frac{(f_{1,2}-f_{1,1})(f_{1,3}-f_{1,1})}{f_{1,1}-\alpha_s}\right)\mathbf{A}_{1,1}\mathbf{B}_{1,1}.
\end{align}
Now it is obvious  that the scaling factor of $\mathbf{A}_{1,1}\mathbf{B}_{1,1}$ can be expanded into weighted sums of the terms $(f_{1,1}-\alpha_s)^{-1},1$ and $\alpha_s$. For the second and third terms in \eqref{eq:ex3ans}, by the long division of polynomials, we can similarly show that the second term can be expanded into weighted sums of the terms $(f_{1,2}-\alpha_s)^{-1},1$ and $\alpha_s$ and that the third term can be expanded into weighted sums of the terms $(f_{1,3}-\alpha_s)^{-1},1$ and $\alpha_s$. Note that the last three terms in \eqref{eq:ex3ans} can be expanded into weighted sums of the terms $1,\alpha_s$. Now, consider any $R=5$ servers, whose indices are denoted as $s_i, i\in[5]$, and we can represent their answers in the following matrix notation.
\begin{equation}
    \begin{bmatrix}
    Y_{s_1}\\
    Y_{s_2}\\
    Y_{s_3}\\
    Y_{s_4}\\
    Y_{s_5}
    \end{bmatrix}=\underbrace{\begin{bmatrix}
    \frac{1}{f_{1,1}-\alpha_{s_1}}&\frac{1}{f_{1,2}-\alpha_{s_1}}&\frac{1}{f_{1,3}-\alpha_{s_1}}&1&\alpha_1\\
    \frac{1}{f_{1,1}-\alpha_{s_2}}&\frac{1}{f_{1,2}-\alpha_{s_2}}&\frac{1}{f_{1,3}-\alpha_{s_2}}&1&\alpha_2\\
    \frac{1}{f_{1,1}-\alpha_{s_3}}&\frac{1}{f_{1,2}-\alpha_{s_3}}&\frac{1}{f_{1,3}-\alpha_{s_3}}&1&\alpha_3\\
    \frac{1}{f_{1,1}-\alpha_{s_4}}&\frac{1}{f_{1,2}-\alpha_{s_4}}&\frac{1}{f_{1,3}-\alpha_{s_4}}&1&\alpha_4\\
    \frac{1}{f_{1,1}-\alpha_{s_5}}&\frac{1}{f_{1,2}-\alpha_{s_5}}&\frac{1}{f_{1,3}-\alpha_{s_5}}&1&\alpha_5
    \end{bmatrix}}_{\mathbf{V}_{1,3,5}}\underbrace{\begin{bmatrix}
    c_{1,1}&&&&\\
    &c_{1,2}&&&\\
    &&c_{1,3}&&\\
    &&&1&\\
    &&&&1
    \end{bmatrix}}_{\mathbf{V}'_{1,3,5}}\otimes\mathbf{I}_{\lambda}\begin{bmatrix}
    \mathbf{A}_{1,1}\mathbf{B}_{1,1}\\
    \mathbf{A}_{1,2}\mathbf{B}_{1,2}\\
    \mathbf{A}_{1,3}\mathbf{B}_{1,3}\\
    *\\
    *
    \end{bmatrix},
\end{equation}
where we have used $*$ to represent various combinations of interference symbols that can be found explicitly by expanding \eqref{eq:ex3ans}, since those forms are not important. We have $c_{1,1}=(f_{1,2}-f_{1,1})(f_{1,3}-f_{1,1})$, $c_{1,2}=(f_{1,1}-f_{1,2})(f_{1,3}-f_{1,2})$ and  $c_{1,3}=(f_{1,1}-f_{1,3})(f_{1,2}-f_{1,3})$. Since $f_{1,1}, f_{1,2}$ and $f_{1,3}$ are distinct by definition, it follows that $c_{1,1}$, $c_{1,2}$ and $c_{1,3}$ are non-zero values. Therefore, the matrix $(\mathbf{V}_{1,3,5}\mathbf{V}'_{1,3,5})\otimes\mathbf{I}_{\lambda}$ is invertible according to Lemma \ref{lemma:csa} and the properties of Kronecker products. Thus, the user is able to recover the desired matrix products by inverting the matrix $(\mathbf{V}_{1,3,5}\mathbf{V}'_{1,3,5})\otimes\mathbf{I}_{\lambda}$. This completes the proof of $R=5$ recovery threshold. Similarly, we can compute the upload cost and download cost of the code as follows, $U_A=U_B=S/K_c=S/3$, and $D=5/3$, which achieves desired costs.

\subsubsection{Arbitrary $\ell, K_c$ and $L=\ell K_c$}
Now let us present the general code construction. 
Let $f_{1,1}, f_{1,2},\cdots,f_{\ell,K_c}, \alpha_1,\alpha_2,\cdots,\alpha_S$ represent $(S+L)$ distinct elements from $\mathbb{F}$. For all $l\in[\ell], s\in[S]$, let us define
\begin{equation}
    \Delta_s^{\l,K_c}=\prod_{k\in[K_c]}(f_{l,k}-\alpha_s).
\end{equation}
Let us also define
\begin{align}
    \mathbf{A}_{l,k}&=\mathbf{A}_{K_c(l-1)+k},\\
    \mathbf{B}_{l,k}&=\mathbf{B}_{K_c(l-1)+k},
\end{align}
for all $l\in[\ell],k\in[K_c]$. Note that by this definition, desired products can be represented as follows.
\begin{align}
    \left(\begin{matrix}
    \mathbf{A}_{1,1}\mathbf{B}_{1,1}&\cdots&\mathbf{A}_{1,K_c}\mathbf{B}_{1,K_c}\\
    \mathbf{A}_{2,1}\mathbf{B}_{2,1}&\cdots&\mathbf{A}_{2, K_c}\mathbf{B}_{2,K_c}\\
    \vdots&\vdots&\vdots\\
    \mathbf{A}_{\ell,1}\mathbf{B}_{\ell,1}&\cdots&\mathbf{A}_{\ell, K_c}\mathbf{B}_{\ell, K_c}\\
    \end{matrix}\right)&=   
     \left(\begin{matrix}
    \mathbf{A}_{1}\mathbf{B}_{1}&\cdots&\mathbf{A}_{K_c}\mathbf{B}_{K_c}\\
    \mathbf{A}_{K_c+1}\mathbf{B}_{K_c+1}&\cdots&\mathbf{A}_{2K_c}\mathbf{B}_{2K_c}\\
    \vdots&\vdots&\vdots\\
    \mathbf{A}_{(\ell-1)K_c+1}\mathbf{B}_{(\ell-1)K_c+1}&\cdots&\mathbf{A}_{\ell K_c}\mathbf{B}_{\ell K_c}\\
    \end{matrix}\right).
\end{align}
Now we are ready to construct the \emph{CSA code} with arbitrary parameters $(\ell,K_c)$. For all $s\in[S]$, let us construct shares of matrices $\mathbf{A}$ and $\mathbf{B}$ at the $s^{th}$ server as follows.
\begin{align}
    \widetilde{A}^s&=(\widetilde{A}_{1}^s,\widetilde{A}_{2}^s,\dots,\widetilde{A}_{\ell}^s),\label{eq:An}\\
    \widetilde{B}^s&=(\widetilde{B}_{1}^s,\widetilde{B}_{2}^s,\dots,\widetilde{B}_{\ell}^s)\label{eq:Bn},
\end{align}
where for $l\in[\ell]$, let us set
\begin{align}
    \widetilde{A}_{l}^s&=\Delta_s^{l,K_c}\sum_{k\in[K_c]}\frac{1}{f_{l,k}-\alpha_s}\mathbf{A}_{l,k},\label{eq:Anl}\\
    \widetilde{B}_{l}^s&=\sum_{k\in[K_c]}\frac{1}{f_{l,k}-\alpha_s}\mathbf{B}_{l,k}.\label{eq:Bnl}
\end{align}
The answer returned by the $s^{th}$ server to the user is constructed as follows.
\begin{align}
    Y_s&=\sum_{l\in[\ell]}\widetilde{A}^s_{l}\widetilde{B}^s_{l}\\
    &=\widetilde{A}^s_{1}\widetilde{B}^s_{1}+\widetilde{A}^s_{2}\widetilde{B}^s_{2}+\dots+\widetilde{A}^s_{\ell}\widetilde{B}^s_{\ell}.
\end{align}
Now let us see why the $R=(\ell+1)K_c-1$ recovery threshold holds. First, let us rewrite $Y_s$ as follows.
\begin{align}
    Y_s&=\widetilde{A}^s_{1}\widetilde{B}^s_{1}+\widetilde{A}^s_{2}\widetilde{B}^s_{2}+\dots+\widetilde{A}^s_{\ell}\widetilde{B}^s_{\ell}\\
    &=\sum_{l\in[\ell]}\Delta_s^{l,K_c}\left(\sum_{k\in[K_c]}\frac{1}{f_{l,k}-\alpha_s}\mathbf{A}_{l,k}\right)\left(\sum_{k\in[K_c]}\frac{1}{f_{l,k}-\alpha_s}\mathbf{B}_{l,k}\right)\\
    &=\sum_{l\in[\ell]}\Delta_s^{l,K_c}\left(\sum_{k\in[K_c]}\sum_{k'\in[K_c]}\frac{\mathbf{A}_{l,k}\mathbf{B}_{l,k'}}{(f_{l,k}-\alpha_s)(f_{l,k'}-\alpha_s)}\right)\\
    &=\sum_{l\in[\ell]}\sum_{k\in[K_c]}\frac{\prod_{k'\in[K_c]\setminus\{k\}}(f_{l,k'}-\alpha_s)}{(f_{l,k}-\alpha_s)}\mathbf{A}_{l,k}\mathbf{B}_{l,k}\notag\\
    &\quad\quad+\sum_{l\in[\ell]}\sum_{\substack{k,k'\in[K_c]\\k\neq k'}}\left(\prod_{k''\in[K_c]\setminus\{k,k'\}}(f_{l,k''}-\alpha_s)\right)\mathbf{A}_{l,k}\mathbf{B}_{l,k'}\label{eq:anscorr},
\end{align}
where in the last step, we split the summation into two parts depending on whether or not $k=k'$. 

Let us consider the first term in \eqref{eq:anscorr}. If we regard both numerator and denominator as polynomials of $\alpha_s$, then by  long division of polynomials, for each $l\in[\ell], k\in[K_c]$, the following term
\begin{equation}
    \frac{\prod_{k'\in[K_c]\setminus\{k\}}(f_{l,k'}-\alpha_s)}{(f_{l,k}-\alpha_s)}\mathbf{A}_{l,k}\mathbf{B}_{l,k},
\end{equation}
can be expanded into weighted sums of the terms $(f_{l,k}-\alpha_s)^{-1},1,\alpha_s,\cdots,\alpha_s^{K_c-2}$, i.e., it can be rewritten as
\begin{equation}
    \Big(c_{-1}(f_{l,k}-\alpha_s)^{-1}+c_0+c_1\alpha_s+\dots+c_{K_c-2}\alpha_s^{K_c-2}\Big)\mathbf{A}_{l,k}\mathbf{B}_{l,k}.
\end{equation}
Now note that the numerator polynomial $\prod_{k'\in[K_c]\setminus\{k\}}(f_{l,k'}-\alpha_s)$ has no root $f_{l,k}$, while $f_{l,k}$ is the only root of the denominator polynomial. Since $(f_{l,k})_{l\in[\ell],k\in[K_c]}$ are distinct elements from $\mathbb{F}$ by definition, by the polynomial remainder theorem, $c_{-1}=\prod_{k'\in[K_c]\setminus\{k\}}(f_{l,k'}-f_{l,k})\neq 0$. 

Next we note that the second term in \eqref{eq:anscorr} can be expanded\footnote{ When $K_c=1$, the second term in \eqref{eq:anscorr} equal zero, thus the Vandermonde terms do not appear and the matrix form representation only involves Cauchy matrices, i.e., Cauchy-Vandermonde matrices without Vandermonde part. } into weighted sums of the terms $1,\alpha_s,\cdots,\alpha_s^{K_c-2}$, so in the matrix form, answers from any $R=(\ell+1)K_c-1$ servers, whose indices are denoted as $s_1,s_2,\cdots,s_R$, can be written as follows.
\begin{align}
    \begin{bmatrix}
    Y_{s_1}\\
    Y_{s_2}\\
    \vdots\\
    Y_{s_R}
    \end{bmatrix}&=\underbrace{\begin{bmatrix}
\frac{1}{f_{1,1}-\alpha_{s_1}}&\frac{1}{f_{1,2}-\alpha_{s_1}}&\cdots&\frac{1}{f_{\ell,K_c}-\alpha_{s_1}}&1&\alpha_{s_1}&\cdots&\alpha_{s_1}^{R-L-1}\\
\frac{1}{f_{1,1}-\alpha_{s_2}}&\frac{1}{f_{1,2}-\alpha_{s_2}}&\cdots&\frac{1}{f_{\ell,K_c}-\alpha_{s_2}}&1&\alpha_{s_2}&\cdots&\alpha_{s_2}^{R-L-1}\\
\vdots&\vdots&\vdots&\vdots&\vdots&\vdots&\vdots&\vdots\\
\frac{1}{f_{1,1}-\alpha_{s_R}}&\frac{1}{f_{1,2}-\alpha_{s_R}}&\cdots&\frac{1}{f_{\ell,K_c}-\alpha_{s_R}}&1&\alpha_{s_R}&\cdots&\alpha_{s_R}^{R-L-1}\\
    \end{bmatrix}}_{\mathbf{V}_{\ell,K_c,R}}\notag\\
    &\quad\quad\underbrace{\begin{bmatrix}
    c_{1,1}&&&&&&\\
    &c_{1,2}&&&&&\\
    &&\ddots&&&&\\
    &&&c_{\ell,K_c}&&&\\
    &&&&1&&\\
    &&&&&\ddots&\\
    &&&&&&1
    \end{bmatrix}}_{\mathbf{V}'_{\ell,K_c,R}}\otimes\mathbf{I}_{\lambda}\begin{bmatrix}
    \mathbf{A}_{1,1}\mathbf{B}_{1,1}\\
    \mathbf{A}_{1,2}\mathbf{B}_{1,2}\\
    \vdots\\
    \mathbf{A}_{\ell, K_c}\mathbf{B}_{\ell, K_c}\\
    *\\
    \vdots\\
    *
    \end{bmatrix},
\end{align}
where we have used $*$ to represent various combinations of interference symbols that can be found explicitly by expanding \eqref{eq:anscorr}, whose exact forms are irrelevant. We note that $R-L-1=(\ell+1)K_c-1-\ell K_c-1=K_c-2$. And we also note that for all $l\in[\ell]$ and $k\in[K_c]$, $c_{l,k}=\prod_{k'\in[K_c]\setminus\{k\}}(f_{l,k'}-f_{l,k})\neq 0$. Therefore, guaranteed by Lemma \ref{lemma:csa} and the fact that the Kronecker product of non-singular matrices is non-singular, the matrix $(\mathbf{V}_{\ell,K_c,R}\mathbf{V}'_{\ell,K_c,R})\otimes\mathbf{I}_{\lambda}$ is invertible. Therefore, the user is able to recover desired products $(\mathbf{A}_{l,k}\mathbf{B}_{l,k})_{l\in[\ell],k\in[K_c]}$ by inverting the matrix. This completes the proof of $R=(\ell+1)K_c-1$ recovery threshold. For the upload costs, it is easy to see that we have $U_A=U_B=(\ell S)/L=S/K_c$. The download cost is $D=R/L= \left((\ell+1)K_c-1\right)/(\ell K_c)$. The computational complexity at each server is $\mathcal{O}(\lambda\kappa\mu/K_c)$ if we assume  straightforward matrix multiplication algorithms. 

Finally, let us consider the encoding   and decoding complexity. Recall the encoding functions \eqref{eq:An}, \eqref{eq:Bn}, \eqref{eq:Anl}, \eqref{eq:Bnl}. Note that  each of the $\widetilde{A}^s_{\ell}$ can be regarded as products of an $S\times K_c$ Cauchy matrix with a total of $\lambda\kappa$ column vectors of length $K_c$. Similarly, each of the $\widetilde{B}^s_{\ell}$ can be considered as products of an $S\times K_c$ Cauchy matrix by a total of $\kappa\mu$ column vectors of length $K_c$. Remarkably, the problem of efficiently multiplying an $S\times S$ Cauchy matrix with a column vector is known as Trummer's problem\cite{Golub_Trummer}. Fast algorithms exist \cite{Gerasoulis_Grigoriadis_Trummer,Gerasoulis_Trummer,Pan_Tabanjeh_Trummer} that solve Trummer's problem with  computational complexity as low as $\widetilde{\mathcal{O}}(S\log^2S)$, in contrast to  straightforward algorithms that have computational complexity of $\mathcal{O}(S^2)$. Similarly, with fast algorithms the computational complexity of multiplying a $S\times K_c$ Cauchy matrix with a column vector is at most $\widetilde{\mathcal{O}}(S\log^2S)$, so the encoding  complexity of $\widetilde{A}^{[S]}$ and $\widetilde{B}^{[S]}$ is at most $\widetilde{\mathcal{O}}\left(\left(\lambda\kappa  S\log^2S\right)/K_c\right)$ and $\widetilde{\mathcal{O}}\left(\left(\kappa\mu  S\log^2S\right)/K_c\right)$, respectively. On the other hand, consider the decoding procedure of \emph{CSA codes}, which can be regarded as solving a total of $\lambda\mu$ linear systems defined by an $R\times R$ coefficient matrix. Indeed, this coefficient matrix is a Cauchy-Vandermonde matrix. There is a  large body of literature studying fast algorithms for solving linear systems defined by $R\times R$ Cauchy-Vandermonde matrices, and the best known computational complexity is $\widetilde{\mathcal{O}}(R\log^2 R)$, see, e.g., \cite{Finck_FastCV}\footnote{The fast algorithm of solving Cauchy-Vandermonde type linear systems here takes inputs of only parameters of a Cauchy-Vandermonde matrix $\mathbf{V}$, i.e, $(\alpha_{s_1},\alpha_{s_2},\cdots,\alpha_{s_R},f_{1,1},f_{1,2},\cdots,f_{\ell,K_c})$ and a column vector $\mathbf{y}$, and outputs the column vector $\mathbf{x}$ such that $\mathbf{Vx}=\mathbf{y}$ with the computational complexity of at most $\mathcal{O}(R\log^2R)$. Therefore, it is not necessary for the user (decoder) to store extra information beyond $\alpha_{[S]}$ and $(f_{l,k})_{l\in[\ell],k\in[K_c]}$.}. Therefore,  the decoding  complexity of $\widetilde{\mathcal{O}}\left((\lambda\mu R\log^2R)/L\right)=\widetilde{\mathcal{O}}(\lambda\mu\log^2R)$ is achievable. This completes the proof of Theorem \ref{thm:csacodes}.

\subsection{Systematic Construction of CSA Codes}\label{sec:systematic}
In this section, we present a systematic construction of CSA codes. Instead of uploading coded version of matrices $\mathbf{A}$ and $\mathbf{B}$ to all of the $S$ servers, the systematic construction of CSA codes uploads uncoded constituent matrix pair $(\mathbf{A}_s,\mathbf{B}_s)$ directly to the $s^{th}$ server for the first $L$ servers, i.e., for all $s\in[L]$. For the remaining $S-L$ servers, coded shares are uploaded following the same construction that was presented in Section \ref{sec:proof}. We will see that the systematic construction of CSA codes works on a smaller field $\mathbb{F}$, compared to the construction presented in Section \ref{sec:proof}. The systematic construction of CSA codes requires less encoding complexity. Its decoding complexity decreases as more of the first $L$ servers respond. In fact, if all of the first $L$ servers respond, then no  computation is required at all for decoding. The systematic construction also preserves backward compatibility to current systems that apply straightforward parallelization strategies. Formally, we have
\begin{align}
    \widetilde{A}^s&=\mathbf{A}_s,\\
    \widetilde{B}^s&=\mathbf{B}_s
\end{align}
for all $s\in[L]$ and
\begin{align}
    \widetilde{A}^s&=(\widetilde{A}_{1}^s,\widetilde{A}_{2}^s,\dots,\widetilde{A}_{\ell}^s),\\
    \widetilde{B}^s&=(\widetilde{B}_{1}^s,\widetilde{B}_{2}^s,\dots,\widetilde{B}_{\ell}^s)
\end{align}
for all $s\in\{L+1,\cdots,S\}$, where $\widetilde{A}_{[\ell]}^s$ and $\widetilde{B}_{[\ell]}^s$ are defined in \eqref{eq:Anl} and \eqref{eq:Bnl} respectively. Similarly, the answer returned by the $s^{th}$ server is constructed as follows.
\begin{equation}
    Y_s=\widetilde{A}^s\widetilde{B}^s
\end{equation}
for all $s\in[L]$ and
\begin{equation}
    Y_s=\sum_{l\in[\ell]}\widetilde{A}^s_l\widetilde{B}^s_l
\end{equation}
for all $s\in\{L+1,\cdots,S\}$. Note that since coded shares are used only for $S-L$ servers, we no longer need distinct values $\alpha_1, \alpha_2, \cdots, \alpha_L$, so the field size required is only $|\mathbb{F}|\geq S$. 

Now, let us prove that the recovery threshold $R$ is not affected by the systematic construction, i.e., the desired products are still recoverable from the answers of any $R=L+K_c-1$ servers. Denote the set of responsive servers as $\mathcal{R}$, $|\mathcal{R}|=R$. Note that if $[L]\subset\mathcal{R}$, then the desired products $\mathbf{A}\mathbf{B}$ can be directly recovered from answers of the first $L$ servers. On the other hand, if $[L]\cap\mathcal{R}=\emptyset$, then we can recover the desired products $\mathbf{A}\mathbf{B}$ following the same argument that was presented in Section \ref{sec:proof}. When $[L]\cap\mathcal{R}\neq\emptyset$, denote the elements in the set $\mathcal{R}\setminus[L]$ as $(s_1, s_2, \cdots, s_{R'})$. The answers from these $R'$ servers can be written in the following matrix form. \begin{align}
    \begin{bmatrix}
    Y_{s_1}\\
    Y_{s_2}\\
    \vdots\\
    Y_{s_{R'}}
    \end{bmatrix}&=\begin{bmatrix}
\frac{c_{1,1}}{f_{1,1}-\alpha_{s_1}}&\frac{c_{1,2}}{f_{1,2}-\alpha_{s_1}}&\cdots&\frac{c_{\ell,K_c}}{f_{\ell,K_c}-\alpha_{s_1}}&1&\alpha_{s_1}&\cdots&\alpha_{s_1}^{R-L-1}\\
\frac{c_{1,1}}{f_{1,1}-\alpha_{s_2}}&\frac{c_{1,2}}{f_{1,2}-\alpha_{s_2}}&\cdots&\frac{c_{\ell,K_c}}{f_{\ell,K_c}-\alpha_{s_2}}&1&\alpha_{s_2}&\cdots&\alpha_{s_2}^{R-L-1}\\
\vdots&\vdots&\vdots&\vdots&\vdots&\vdots&\vdots&\vdots\\
\frac{c_{1,1}}{f_{1,1}-\alpha_{s_{R'}}}&\frac{c_{1,2}}{f_{1,2}-\alpha_{s_{R'}}}&\cdots&\frac{c_{\ell,K_c}}{f_{\ell,K_c}-\alpha_{s_{R'}}}&1&\alpha_{s_R}&\cdots&\alpha_{s_{R'}}^{R-L-1}\\
    \end{bmatrix}\otimes\mathbf{I}_{\lambda}\begin{bmatrix}
    \mathbf{A}_{1,1}\mathbf{B}_{1,1}\\
    \mathbf{A}_{1,2}\mathbf{B}_{1,2}\\
    \vdots\\
    \mathbf{A}_{\ell, K_c}\mathbf{B}_{\ell, K_c}\\
    *\\
    \vdots\\
    *
    \end{bmatrix},
\end{align}
Note that the dimension of the first matrix on the RHS, or the decoding matrix, is $(R'\times R)$, thus it appears to be not invertible. However, from answers of the servers in the set $\mathcal{R}\cap[L]$, we can directly recover $|\mathcal{R}\cap[L]|$ desired products. Note that the desired products appear along the dimension spanned by the Cauchy part. By subtracting these known products from the answers, we obtain the decoding matrix of dimension $(R'\times R')$, which is invertible by Lemma \ref{lemma:csa}. This completes the proof of recovery threshold $R=L+K_c-1$. It is easy to see that the upload and download costs are also not affected by the systematic construction. For the encoding complexity, the systematic construction requires less arithmetic operations because no computation is needed to obtain $\widetilde{A}^{[L]}$ and $\widetilde{B}^{[L]}$. For the decoding complexity, when $[L]\subset\mathcal{R}$, no computation is needed, and when $[L]\cap\mathcal{R}=\emptyset$, it follows the same argument presented in Section \ref{sec:proof}. When $[L]\cap\mathcal{R}\neq\emptyset$, the user (decoder) eliminates all products obtained from answers of the servers in the set $\mathcal{R}\cap[L]$, and then decodes the remaining products according to the fast decoding algorithm. Thus the decoding complexity is not increased.

\section{Generalized Cross-Subspace Alignment (GCSA) Codes: Combining Batch Processing and Matrix-Partitioning}\label{sec:gcsa}
In this section, we present Generalized CSA codes (GCSA codes), which combine the batch processing of CSA codes with the matrix-partitioning approach of EP codes. Although we have shown that batch processing with CSA codes significantly improves the tradeoff between upload-download costs, evidently CSA codes require at least one matrix multiplication of dimensions $\lambda\times\kappa$ and $\kappa\times\mu$ at each server. For a computation-latency limited setting, partitioning may be necessary to reduce the computation load per server. A naive approach to combine the batch processing of CSA codes and the partitioning of EP codes is to first (separately for each $l\in[L]$) apply EP codes for each pair of matrices, ${\bf A}_l, {\bf B}_l$. Next, since only matrix multiplication is involved in obtaining the  answers for EP codes,  we can then apply CSA codes for these matrix multiplications. Specifically, for positive integers $\ell',K_c',S',p,m,n$ such that $L=\ell' K_c'$, $m\mid\lambda$, $n\mid\mu$, $p\mid\kappa$ and $S'\geq pmn+p-1$,  apply EP codes of parameter $p,m,n$ with $S'$ servers for each matrix multiplication. This yields a total of $S'\ell'K_c'$ matrix multiplications. For these matrix multiplications, we can further apply CSA codes. Now we can see that by this construction, if we choose CSA codes parameters $\ell=\ell', K_c=K_c'S'$, we can achieve the upload cost $(U_A,U_B)=(S/(K_c'pm),S/(K_c'pn))$. It is also easy to see that for this simple combination of EP and CSA codes,  the recovery threshold achieved is $R=\ell' K_c'S'+K_c'S'-1$, and the  download cost is $D=(\ell' K_c'S'+K_c'S'-1)/(\ell' K_c'mn)$. However, we will see that under the same upload cost, GCSA codes can improve the recovery threshold to $R=pmn(\ell'K'_c+K_c'-1)+p-1$ and the download cost to $D=(pmn(\ell'K'_c+K_c'-1)+p-1)/(\ell' K_c'mn)$. This result is better than the naive construction because $S'\geq pmn+p-1\geq 1$. On the other hand, note that the Lagrange Coded Computation (LCC) codes in \cite{Yu_Lagrange} can be regarded as a special case of CSA codes with parameter $\ell=1$, so the naive approach of combining LCC codes with EP  codes achieves the recovery threshold of $R=2LS'-1$. With GCSA codes of parameter $\ell=1$, the recovery threshold is improved to $R=2Lpmn+p-1$.

\subsection{GCSA Codes: Main Result}\label{sec:gcsamain}
Our main result for GCSA codes appears in the following theorem.
\begin{theorem}\label{thm:gcsa}
For CDBMM over a field $\mathbb{F}$ with $S$ servers, and positive integers $(\ell, K_c, p, m, n)$ such that $m\mid\lambda$, $n\mid\mu$, $p\mid\kappa$ and $L=\ell K_c\leq|\mathbb{F}|-S$, the GCSA codes presented in this work achieve
\begin{align}
   \text{Recovery Threshold:}&& R&=pmn((\ell+1)K_c-1)+p-1,\\
     \text{Upload Cost for $\widetilde{A}^{[S]},\widetilde{B}^{[S]}$:}&&(U_A, U_B)&=\left(\frac{S}{K_cpm},\frac{S}{K_cpn}\right),\\
    \text{Download Cost:} &&D&=\frac{pmn((\ell+1)K_c-1)+p-1}{mn\ell K_c},\\
    \text{Server Computation Complexity:}&& \mathcal{C}_s&=\mathcal{O}\left(\frac{\lambda\kappa\mu}{K_cpmn}\right),\\
    \text{Encoding Complexity for $\widetilde{A}^{[S]},\widetilde{B}^{[S]}$:}&&(\mathcal{C}_{eA},\mathcal{C}_{eB})&=\left(\widetilde{\mathcal{O}}\left(\frac{\lambda\kappa S\log^2S}{K_cpm}\right),  \widetilde{\mathcal{O}}\left(\frac{\kappa\mu S\log^2S}{K_cpn}\right)\right),\\
     \text{Decoding Complexity:}&&\mathcal{C}_d&=\widetilde{\mathcal{O}}\left(\lambda\mu p\log^2R\right).
\end{align}
\end{theorem}

\subsection{Observations}
\begin{enumerate}
\item GCSA codes generalize almost all state of art approaches for coded distributed batch matrix multiplication. Setting $m=n=p$ reduces GCSA codes to CSA codes. Further setting $\ell=1$ recovers LCC codes. Setting $\ell=K_c=1$ reduces GCSA codes to EP codes. Further setting $p=1$ recovers Polynomial codes, while setting $m=n=1$ recovers MatDot codes.
\item Let us explain why GCSA codes, which include CSA codes, LCC codes and EP codes as special cases, are capable of achieving more than what each of these codes can achieve in general. Consider a finite horizon setting, where\footnote{We make a distinction between batch size and job size, in that the job size is fixed as part of the problem specification while the batch size may be chosen arbitrarily by a coding scheme, e.g., to partition the job into smaller jobs.} the \emph{job size} $J$, i.e., the number of matrices to be multiplied, is fixed. So we need to compute $J$ matrix multiplications, ${\bf A}_1{\bf B}_1, \cdots, {\bf A}_{J}{\bf B}_{J}$, where each ${\bf A}_j, {\bf B}_j, j\in[J]$ is a $\lambda\times\lambda$ matrix. Suppose each scalar multiplication takes $T_m$ seconds, and  it takes $T_c$ seconds to communicate one scalar over any communication channel. For simplicity let us assume that multiplying two $\lambda\times\lambda$ matrices requires  $\lambda^3 T_m$ seconds of computation time. There is a required latency constraint for this job, such that the total computation time at each server cannot  exceed $\lambda^3T_m/K$, where $K>1$ is a given parameter that determines the server latency constraint. Note that this latency constraint immediately rules out LCC codes, and even CSA codes because they need at least $\lambda^3T_m$ seconds of computation time at each server which violates the given constraint. Now consider EP codes which can partition the matrices to reduce the size of computation task at each server, but need to repeat the process $J$ times because each ${\bf A}_j{\bf B}_j$ is computed separately.  EP codes need computation time $J \lambda^3T_m/(pmn)$ at each server, so they can satisfy the latency constraint by choosing $pmn\geq JK$. On the other hand, GCSA codes with, say $\ell=1$ and $K_c=J$, need computation time $\lambda^3T_m/(p''m''n'')$ at each server. So GCSA codes can satisfy the latency constraint by choosing $p''m''n''\geq K$, i.e., with less partitioning than needed for EP codes. Note that GCSA codes need less partitioning than EP codes to satisfy the same latency constraint, because they make up some of the computation time by batch processing of the $J$ multiplications that must be carried out separately by EP codes. It turns out that this allows GCSA codes to have lower communication cost. EP codes require  a total  download time of $\frac{JR_{EP}\lambda^2T_c}{mn}=\frac{J(pmn+p-1)\lambda^2T_c}{mn}$, and an upload time of $\frac{JS\lambda^2T_c}{pm}+\frac{JS\lambda^2T_c}{pn}$ seconds, where $S$ is the number of servers utilized by the scheme. For straggler tolerance, suppose $R_{EP}/S=\eta<1$, so that the upload time is expressed as $\frac{J(pmn+p-1)\lambda^2T_c}{\eta p}(\frac{1}{m}+\frac{1}{n}$).  For balanced upload and download times we need $m=n$ and $\eta pm/2\approx m^2$, so that the balanced upload/download time for EP codes is $\approx \frac{J(2m^3/\eta+2m/\eta-1)\lambda^2T_c}{m^2}$. Given the latency constraint which forces $2m^3/\eta\geq JK$, we find that the optimal balanced upload/download time for EP codes is achieved with $m\approx\left(\frac{\eta JK}{2}\right)^{1/3}$. On the other hand, now consider  GCSA codes, which need total download time of $\frac{R_{GCSA}\lambda^2T_c}{m''n''}=\frac{(p''m''n''(2J-1)+p''-1)\lambda^2T_c}{m''n''}$, and total upload time of $\frac{S''\lambda^2T_c}{p''m''}+\frac{S''\lambda^2T_c}{p''n''}$. For similar straggler robustness, let $R_{GCSA}/S''=\eta$ be the same as for EP codes. For balanced costs, similarly set $m''=n''$ and $\eta p''m''/2\approx m''^2$. Thus GCSA codes achieve balanced upload/download time of $\approx\frac{((2m''^3/\eta)(2J-1)+2m''/\eta-1)\lambda^2T_c}{m''^2}$, respectively. Combined with the latency constraint which forces $2m''^3/\eta\geq K$, we find that the optimal balanced upload/download time for this particular construction of GCSA codes is achieved with $m''\approx\left(\frac{\eta K}{2}\right)^{1/3}$. For a quick comparison of approximately optimal balanced upload/download time, note that for EP codes it is lower bounded by $\frac{2Jm\lambda^2T_c}{\eta}$, and for GCSA codes it is upper bounded by $\frac{8Jm''\lambda^2T_c}{\eta}$, so EP codes need more balanced communication time by a factor of at least $\frac{m}{4m''}\approx  \frac{J^{1/3}}{4}$ which can be significantly  larger than $1$ for large job sizes $J$. To complement the approximate analysis, Figure \ref{fig:job} explicitly compares the balanced upload/download time (the maximum of upload and download times) versus the job latency constraint parameter $K$. The values shown for EP codes are precisely lowerbounds, i.e., EP codes cannot do any better, while those for GCSA codes are strictly achievable. Thus, GCSA codes can achieve more than what can be achieved by CSA, LCC or EP codes.

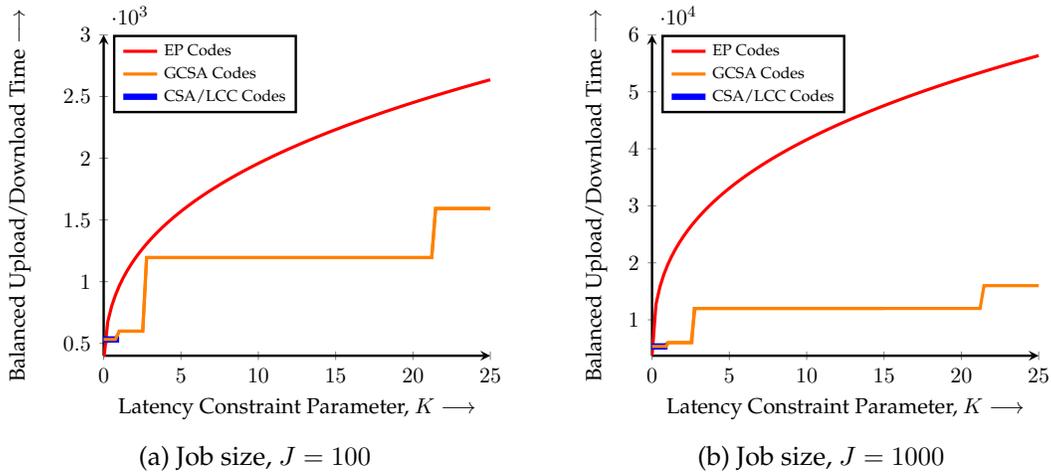
\begin{figure}[htbp]
\centering
\begin{subfigure}{0.45\textwidth}
\centering
\begin{tikzpicture}[scale=0.75]
\begin{axis}[legend cell align={left},
    axis lines = left,
    ylabel style={align=center},
    ylabel shift = 0.3 cm,
    scaled y ticks=base 10:-3,
    xlabel = {Latency Constraint Parameter, $K\longrightarrow$},
	ylabel = {Balanced Upload/Download Time $ \longrightarrow$},
    xmin=0,
    xmax=25,
    ymin=400,
    ymax=3000,
    very thick,
     legend style={at={(0.5,1)}}
]

\addplot[domain=1/\Ja:25, red, ultra thick, samples=100] {\Ja*((x*\Ja*0.75/2)^(2/3)*2*(x*\Ja*0.75/2)^(1/3)/0.75+2*(x*\Ja*0.75/2)^(1/3)/0.75-1)/(x*\Ja*0.75/2)^(2/3)};
\addlegendentry{\scriptsize EP Codes}

\addplot[domain=0:25, orange, ultra thick, samples=100] {(x>1)*(ceil((x*0.75/2)^(1/3))*ceil((x*0.75/2)^(1/3))*ceil(2*ceil((x*0.75/2)^(1/3))/0.75)*(2*\Ja-1)+ceil(2*ceil((x*0.75/2)^(1/3))/0.75)-1)/ceil((x*0.75/2)^(1/3))/ceil((x*0.75/2)^(1/3))+(x<1)*2*ceil((2*\Ja-1)/0.75)};
\addlegendentry{\scriptsize GCSA Codes}

\addplot[domain=-1:1, blue, line width=3.2pt, samples=100] {2*ceil((2*\Ja-1)/0.75)};
\addlegendentry{\scriptsize CSA/LCC Codes}

\addplot[domain=0:25, orange, ultra thick, samples=100] {(x>1)*(ceil((x*0.75/2)^(1/3))*ceil((x*0.75/2)^(1/3))*ceil(2*ceil((x*0.75/2)^(1/3))/0.75)*(2*\Ja-1)+ceil(2*ceil((x*0.75/2)^(1/3))/0.75)-1)/ceil((x*0.75/2)^(1/3))/ceil((x*0.75/2)^(1/3))+(x<1)*2*ceil((2*\Ja-1)/0.75)};

\end{axis}
\end{tikzpicture}
\caption{Job size, $J=\Ja$}\label{fig:j1}
\end{subfigure}
\begin{subfigure}{0.45\textwidth}
\centering
\begin{tikzpicture}[scale=0.75]
\begin{axis}[legend cell align={left}, ylabel shift = -20 pt,
    axis lines = left,
    ylabel style={align=center},
    ylabel shift = 0.1 cm,
    xlabel = {Latency Constraint Parameter, $K\longrightarrow$},
    ylabel = {Balanced Upload/Download Time $ \longrightarrow$},
    xmin=0,
    xmax=25,
    ymax=60000,
    very thick,
    legend style={at={(0.5,1)}}
]

\addplot[domain=1/\Jb:25, red, ultra thick, samples=100] {\Jb*((x*\Jb*0.75/2)^(2/3)*2*(x*\Jb*0.75/2)^(1/3)/0.75+2*(x*\Jb*0.75/2)^(1/3)/0.75-1)/(x*\Jb*0.75/2)^(2/3)};
\addlegendentry{\scriptsize EP Codes}

\addplot[domain=0:25, orange, ultra thick, samples=100] {(x>1)*(ceil((x*0.75/2)^(1/3))*ceil((x*0.75/2)^(1/3))*ceil(2*ceil((x*0.75/2)^(1/3))/0.75)*(2*\Jb-1)+ceil(2*ceil((x*0.75/2)^(1/3))/0.75)-1)/ceil((x*0.75/2)^(1/3))/ceil((x*0.75/2)^(1/3))+(x<1)*2*ceil((2*\Jb-1)/0.75)};
\addlegendentry{\scriptsize GCSA Codes}

\addplot[domain=-1:1, blue, line width=3.2pt, samples=100] {2*ceil((2*\Jb-1)/0.75)};
\addlegendentry{\scriptsize CSA/LCC Codes}

\addplot[domain=0:15, orange, ultra thick, samples=100] {(x>1)*(ceil((x*0.75/2)^(1/3))*ceil((x*0.75/2)^(1/3))*ceil(2*ceil((x*0.75/2)^(1/3))/0.75)*(2*\Jb-1)+ceil(2*ceil((x*0.75/2)^(1/3))/0.75)-1)/ceil((x*0.75/2)^(1/3))/ceil((x*0.75/2)^(1/3))+(x<1)*2*ceil((2*\Jb-1)/0.75)};

\end{axis}
\end{tikzpicture}
\vspace{0.0cm}
\caption{\small Job size, $J=\Jb$} 
\label{fig:j2}
\end{subfigure}
\caption{\small \it Balanced upload/download time vs the value of the latency constraint parameter $K$ for EP codes, CSA/LCC codes and GCSA codes (normalized by $\lambda^3 T_c$). CSA/LCC codes are not feasible for $K>1$. The values for EP codes are lower bounds while those for GCSA codes are upper bounds, showing that GCSA codes strictly outperform both batch processing (CSA/LCC) and matrix-partitioning (EP) codes.}\label{fig:job}
\end{figure}

\item The finite horizon, i.e., fixed job size and fixed latency constraint for each job is important in the previous discussion. If instead of the absolute value of server latency for a fixed job size, we only insisted  on normalized server latency \emph{per job}, where each job is still comprised of $J$ matrix multiplications, then we could jointly process $K$ jobs codes with an absolute latency of $\lambda^3T_c$ which allows LCC and CSA codes, while still achieving the latency \emph{per job} of $\lambda^3/K$. Since batch processing approaches like LCC codes and CSA codes have already been shown to achieve better communication costs than any matrix partitioning approach, neither EP codes nor GCSA codes would be needed in that case. On the other hand, it is also worth noting that if we go to the other extreme and require a fixed server latency  of less than $\lambda^3T_c/K$ for each matrix multiplication, i.e., set $J=1$, then it can be seen that batch processing cannot help, i.e., matrix partitioning alone is enough. In other words, if latency constraints are imposed on each matrix multiplication $(J=1)$ , then EP codes suffice, and neither LCC codes, nor CSA or GCSA codes are needed.

\item Figure \ref{fig:job} shows the advantage of GCSA codes in terms of communication cost over exclusively batch processing (LCC) and matrix-partitioning (EP) codes under absolute server latency constraints, even when GCSA codes are restricted to the choice $\ell=1$. However, the advantage of GCSA codes over LCC and EP codes can be seen even without absolute latency constraints. This is  illustrated in Figure \ref{fig:gcsa} which only constrains the recovery threshold $R$ and the number of servers $S$. The figure shows lower convex hulls of achievable (balanced upload cost, download cost) pairs of GCSA codes for various bounds on the matrix partitioning parameters $pmn$, given that the number of servers $S=300$ and the overall recovery threshold $R\leq 250$. Each value of $(S,R,pmn)$ produces an achievable region in the $(U,D)$ plane (including all possible choices of parameters $m,n,p,\ell,K_c$). What is shown in the figure is the union of these regions. The larger the value of $pmn$, the more the GCSA code construction shifts toward EP codes, generally with the benefit of reduced latency of computation at each server that comes with matrix partitioning. On the other hand, the smaller the value of $pmn$, the more the GCSA code construction shifts toward CSA codes, with the benefit of improved communication costs that come with batch processing. As noted previously, when no matrix partitioning is allowed,  LCC codes can be recovered as a special case of GCSA codes by setting $\ell=1$. The figure also shows how GCSA codes are capable of improving upon LCC codes in terms of download cost by choosing $\ell>1$.

\begin{figure}[htbp]
\centering
\includegraphics[width=3in]{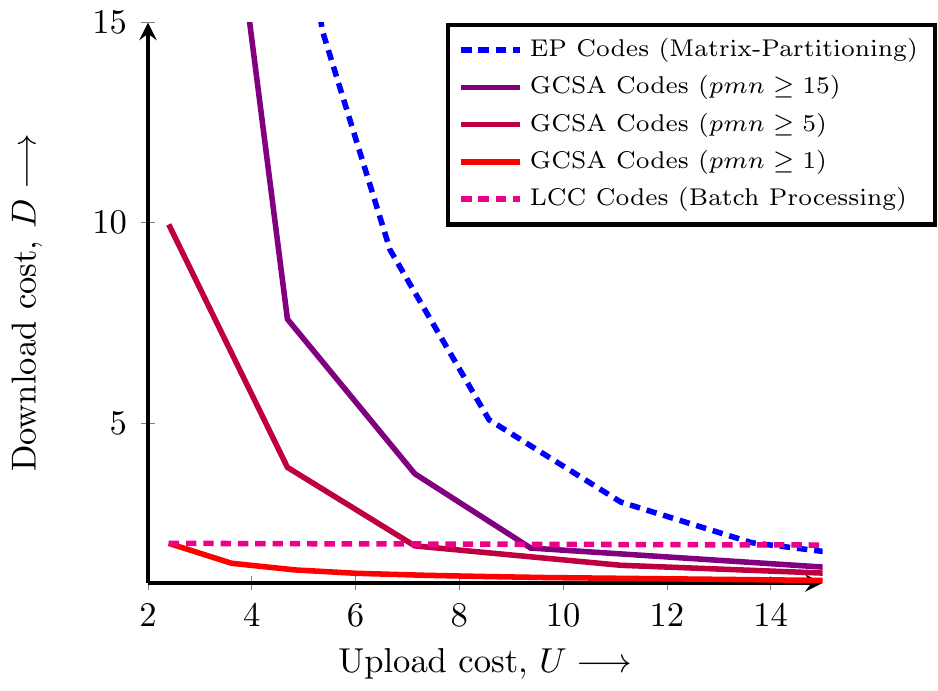}
\caption{\it \small Lower convex hulls of achievable (balanced upload cost, download cost) pairs $(U,D)$ of GCSA codes for various bounds on $pmn$, given that $S=300$ and  the overall recovery threshold $R\leq 250$. Note that EP codes and LCC codes are also special cases of GCSA codes, obtained by setting $\ell=K_c=1$, and $\ell=m=n=p=1$, respectively. CSA codes are obtained by setting $m=n=p=1$.}
\label{fig:gcsa}
\end{figure}

\end{enumerate}

\subsection{Proof of Theorem \ref{thm:gcsa}}
Let us recall the standard result for Confluent Cauchy-Vandermonde matrices \cite{Gasca_Martinez_Muhlbach}, reproduced here for the sake of completeness.
\begin{lemma}\label{lemma:ccv}
If $f_{1,1}, f_{1,2}, \cdots, f_{\ell,K_c}, \alpha_1,\alpha_2,\cdots,\alpha_R$ are $R+L$ distinct elements of $\mathbb{F}$, with $|\mathbb{F}|\geq R+L$ and $L=\ell K_c$, then the following $R\times R$ Confluent Cauchy-Vandermonde matrix is invertible over $\mathbb{F}$.
\begin{align}
\hat{{\bf V}}_{\ell,K_c,{R'},R}&\triangleq\left[
\begin{matrix}
\frac{1}{(f_{1,1}-\alpha_1)^{R'}}&\cdots&\frac{1}{f_{1,1}-\alpha_1}&\cdots&\frac{1}{(f_{\ell,K_c}-\alpha_1)^{R'}}&\cdots&\frac{1}{f_{\ell,K_c}-\alpha_1}&1&\cdots&\alpha_1^{R-{R'}L-1}\\
\frac{1}{(f_{1,1}-\alpha_2)^{R'}}&\cdots&\frac{1}{f_{1,1}-\alpha_2}&\cdots&\frac{1}{(f_{\ell,K_c}-\alpha_2)^{R'}}&\cdots&\frac{1}{f_{\ell,K_c}-\alpha_2}&1&\cdots&\alpha_2^{R-{R'}L-1}\\
\vdots&\vdots&\vdots&\vdots&\vdots&\vdots&\vdots&\vdots&\vdots&\vdots\\
\frac{1}{(f_{1,1}-\alpha_R)^{R'}}&\cdots&\frac{1}{f_{1,1}-\alpha_R}&\cdots&\frac{1}{(f_{\ell,K_c}-\alpha_R)^{R'}}&\cdots&\frac{1}{f_{\ell,K_c}-\alpha_R}&1&\cdots&\alpha_R^{R-{R'}L-1}\\
\end{matrix}
\right]
\end{align}
\end{lemma}

Before presenting the generalized CSA codes construction let us start with some illustrative examples.
\subsubsection{$\ell=1, K_c=2, L=2, p=1, m=n=2$}
Let $f_{1,1}, f_{1,2}, \alpha_{1}, \alpha_{2},\dots,\alpha_{S}$ represent $(S+2)$ distinct elements from $\mathbb{F}$. For all $s\in[S]$, define,
\begin{equation}
    \Delta_{s}^{1,2}=(f_{1,1}-\alpha_s)^4(f_{1,2}-\alpha_s)^4.
\end{equation}
We set $\mathbf{A}_{1,1}=\mathbf{A}_1$,~ $\mathbf{A}_{1,2}=\mathbf{A}_2$,~ $\mathbf{B}_{1,1}=\mathbf{B}_1$ and $\mathbf{B}_{1,2}=\mathbf{B}_2$. Besides, we partition each of the matrices $\mathbf{A}_{1,1}$ and $\mathbf{A}_{1,2}$ into $2\times1$ blocks, denoted as $\mathbf{A}_{1,1}^{1,1}$, $\mathbf{A}_{1,1}^{2,1}$ and $\mathbf{A}_{1,2}^{1,1}$, $\mathbf{A}_{1,2}^{2,1}$ respectively. Similarly, we partition each of the matrices $\mathbf{B}_{1,1}$ and $\mathbf{B}_{1,2}$ into $1\times2$ blocks, denoted as $\mathbf{B}_{1,1}^{1,1}$, $\mathbf{B}_{1,1}^{1,2}$ and $\mathbf{B}_{1,2}^{1,1}$, $\mathbf{B}_{1,2}^{1,2}$ respectively. Note that the desired products $\mathbf{A}_{1,1}\mathbf{B}_{1,1},\mathbf{A}_{1,2}\mathbf{B}_{1,2}$ corresponds to the products $(\mathbf{A}_{1,k}^{i,1}\mathbf{B}_{1,k}^{1,j})_{i\in[2],j\in[2],k\in[2]}$.
 Shares of matrices $\mathbf{A}$ are constructed as follows.
\begin{align}
    \widetilde{A}^s&=\Delta_{s}^{1,2}\left(\frac{1}{(f_{1,1}-\alpha_s)^4}\left(\mathbf{A}_{1,1}^{1,1}+(f_{1,1}-\alpha_s)\mathbf{A}_{1,1}^{2,1}\right)+\frac{1}{(f_{1,2}-\alpha_s)^4}\left(\mathbf{A}_{1,2}^{1,1}+(f_{1,2}-\alpha_s)\mathbf{A}_{1,2}^{2,1}\right)\right)\\
    &=(f_{1,2}-\alpha_s)^4\underbrace{\left(\mathbf{A}_{1,1}^{1,1}+(f_{1,1}-\alpha_s)\mathbf{A}_{1,1}^{2,1}\right)}_{P_s^{1,1}}+(f_{1,1}-\alpha_s)^4\underbrace{\left(\mathbf{A}_{1,2}^{1,1}+(f_{1,2}-\alpha_s)\mathbf{A}_{1,2}^{2,1}\right)}_{P_s^{1,2}}.
\end{align}
Note that the term $P_s^{1,1}$ follows the construction of Entangled Polynomial codes of parameter $m=n=2, p=1$. In the original construction of Entangled Polynomial codes, the term $P_s^{1,1}$ is a polynomial of $\alpha_s$, however in the construction of generalized CSA codes, it is a polynomial of $(f_{1,1}-\alpha_s)$. Similarly, the term $P_s^{1,2}$ follows the construction of Entangled Polynomial codes, and it is a polynomial of $(f_{1,2}-\alpha_s)$. Shares of matrices $\mathbf{B}$ are constructed as follows.
\begin{align}
    \widetilde{B}^s=\frac{1}{(f_{1,1}-\alpha_s)^4}\underbrace{\left(\mathbf{B}_{1,1}^{1,1}+(f_{1,1}-\alpha_s)^2\mathbf{B}_{1,1}^{1,2}\right)}_{Q_s^{1,1}}+\frac{1}{(f_{1,2}-\alpha_s)^4}\underbrace{\left(\mathbf{B}_{1,2}^{1,1}+(f_{1,2}-\alpha_s)^2\mathbf{B}_{1,2}^{1,2}\right)}_{Q_s^{1,2}}.
\end{align}
The term $Q_s^{1,1}$ and $Q_s^{1,2}$ follows the construction of Entangled Polynomial codes of given parameter, and they are polynomials of $(f_{1,1}-\alpha_s)$ and $(f_{1,2}-\alpha_s)$ respectively.

The answer from the $s^{th}$ server,  $Y_s$ is constructed as $Y_s=\widetilde{A}^s\widetilde{B}^s$. To see why it is possible to recover the desired products from the answers of any $R=12$ servers, let us rewrite $Y_s$ as follows.
\begin{align}
    Y_s&=\widetilde{A}^s\widetilde{B}^s\\
    &=\frac{(f_{1,2}-\alpha_s)^4}{(f_{1,1}-\alpha_s)^4}P_s^{1,1}Q_s^{1,1}+\frac{(f_{1,1}-\alpha_s)^4}{(f_{1,2}-\alpha_s)^4}P_s^{1,2}Q_s^{1,2}+(P_s^{1,1}Q_s^{1,2}+P_s^{1,2}Q_s^{1,1})\\
    &=\frac{((f_{1,1}-\alpha_s)+(f_{1,2}-f_{1,1}))^4}{(f_{1,1}-\alpha_s)^4}P_s^{1,1}Q_s^{1,1}+\frac{((f_{1,2}-\alpha_s)+(f_{1,1}-f_{1,2}))^4}{(f_{1,2}-\alpha_s)^4}P_s^{1,2}Q_s^{1,2}\notag\\
    &\quad\quad+(P_s^{1,1}Q_s^{1,2}+P_s^{1,2}Q_s^{1,1})\\
    &=\left(\frac{c_{1,1,0}}{(f_{1,1}-\alpha_s)^4}+\frac{c_{1,1,1}}{(f_{1,1}-\alpha_s)^3}+\frac{c_{1,1,2}}{(f_{1,1}-\alpha_s)^2}+\frac{c_{1,1,3}}{f_{1,1}-\alpha_s}\right)P_s^{1,1}Q_s^{1,1}\notag\\
    &\quad\quad+\left(\frac{c_{1,2,0}}{(f_{1,2}-\alpha_s)^4}+\frac{c_{1,2,1}}{(f_{1,2}-\alpha_s)^3}+\frac{c_{1,2,2}}{(f_{1,2}-\alpha_s)^2}+\frac{c_{1,2,3}}{f_{1,2}-\alpha_s}\right)P_s^{1,2}Q_s^{1,2}\notag\\
    &\quad\quad+(P_s^{1,1}Q_s^{1,1}+P_s^{1,2}Q_s^{1,2}+P_s^{1,1}Q_s^{1,2}+P_s^{1,2}Q_s^{1,1}),\label{eq:vcsaex11}
\end{align}
where in the last step, we perform binomial expansion for numerator polynomials. According to the Binomial theorem, $(c_{1,k,i})_{k\in[2],i\in\{0,1,2,3\}}$ are non-zero. Note that
\begin{align}
    P_s^{1,1}Q_s^{1,1}&=\mathbf{A}_{1,1}^{1,1}\mathbf{B}_{1,1}^{1,1}+(f_{1,1}-\alpha_s)\mathbf{A}_{1,1}^{2,1}\mathbf{B}_{1,1}^{1,1}+(f_{1,1}-\alpha_s)^2\mathbf{A}_{1,1}^{1,1}\mathbf{B}_{1,1}^{1,2}+(f_{1,1}-\alpha_s)^3\mathbf{A}_{1,1}^{2,1}\mathbf{B}_{1,1}^{1,2},\\
    P_s^{1,2}Q_s^{1,2}&=\mathbf{A}_{1,2}^{1,1}\mathbf{B}_{1,2}^{1,1}+(f_{1,2}-\alpha_s)\mathbf{A}_{1,2}^{2,1}\mathbf{B}_{1,2}^{1,1}+(f_{1,2}-\alpha_s)^2\mathbf{A}_{1,2}^{1,1}\mathbf{B}_{1,2}^{1,2}+(f_{1,2}-\alpha_s)^3\mathbf{A}_{1,2}^{2,1}\mathbf{B}_{1,2}^{1,2}.
\end{align}
Therefore, we can further rewrite the first term in \eqref{eq:vcsaex11} as follows.
\begin{align}
    &\frac{c_{1,1,0}\mathbf{A}_{1,1}^{1,1}\mathbf{B}_{1,1}^{1,1}}{(f_{1,1}-\alpha_s)^4}+\frac{c_{1,1,1}\mathbf{A}_{1,1}^{1,1}\mathbf{B}_{1,1}^{1,1}+c_{1,1,0}\mathbf{A}_{1,1}^{2,1}\mathbf{B}_{1,1}^{1,1}}{(f_{1,1}-\alpha_s)^3}+\frac{c_{1,1,2}\mathbf{A}_{1,1}^{1,1}\mathbf{B}_{1,1}^{1,1}+c_{1,1,1}\mathbf{A}_{1,1}^{2,1}\mathbf{B}_{1,1}^{1,1}+c_{1,1,0}\mathbf{A}_{1,1}^{1,1}\mathbf{B}_{1,1}^{1,2}}{(f_{1,1}-\alpha_s)^2}\notag\\
    &\quad\quad+\frac{c_{1,1,3}\mathbf{A}_{1,1}^{1,1}\mathbf{B}_{1,1}^{1,1}+c_{1,1,2}\mathbf{A}_{1,1}^{2,1}\mathbf{B}_{1,1}^{1,1}+c_{1,1,1}\mathbf{A}_{1,1}^{1,1}\mathbf{B}_{1,1}^{1,2}+c_{1,1,0}\mathbf{A}_{1,1}^{2,1}\mathbf{B}_{1,1}^{1,2}}{f_{1,1}-\alpha_s}\notag\\
    &\quad\quad+(c_{1,1,1}\mathbf{A}_{1,1}^{2,1}\mathbf{B}_{1,1}^{1,2}+c_{1,1,2}\mathbf{A}_{1,1}^{1,1}\mathbf{B}_{1,1}^{1,2}+c_{1,1,3}\mathbf{A}_{1,1}^{2,1}\mathbf{B}_{1,1}^{1,1})\notag\\
    &\quad\quad+(f_{1,1}-\alpha_s)(c_{1,1,2}\mathbf{A}_{1,1}^{2,1}\mathbf{B}_{1,1}^{1,2}+c_{1,1,3}\mathbf{A}_{1,1}^{1,1}\mathbf{B}_{1,1}^{1,2})+(f_{1,1}-\alpha_s)^2(c_{1,1,3}\mathbf{A}_{1,1}^{2,1}\mathbf{B}_{1,1}^{1,2}).\label{eq:vcsaex12}
\end{align}
The second term in \eqref{eq:vcsaex11} can be similarly rewritten. Note that the third term in \eqref{eq:vcsaex11} and the last three terms in \eqref{eq:vcsaex12} can be expanded into weighted sums of the terms $1,\alpha_s,\alpha_s^2,\alpha_s^3$, so in the matrix form, answers from any $12$ servers, whose indices are denoted as $s_1,s_2,\cdots,s_{12}$, can be written as follows.
\begin{align}
    \begin{bmatrix}
    Y_{s_1}\\
    Y_{s_2}\\
    \vdots\\
    Y_{s_{12}}
    \end{bmatrix}&=\underbrace{\left[\begin{array}{ccc;{4pt/4pt}ccc;{4pt/4pt}cccc}
\frac{1}{(f_{1,1}-\alpha_{s_1})^4}&\cdots&\frac{1}{f_{1,1}-\alpha_{s_1}}&\frac{1}{(f_{1,2}-\alpha_{s_1})^4}&\cdots&\frac{1}{f_{1,2}-\alpha_{s_1}}&1&\alpha_{s_1}&\cdots&\alpha_{s_1}^3\\
\frac{1}{(f_{1,1}-\alpha_{s_2})^4}&\cdots&\frac{1}{f_{1,1}-\alpha_{s_2}}&\frac{1}{(f_{1,2}-\alpha_{s_2})^4}&\cdots&\frac{1}{f_{1,2}-\alpha_{s_2}}&1&\alpha_{s_2}&\cdots&\alpha_{s_2}^3\\
\vdots&\vdots&\vdots&\vdots&\vdots&\vdots&\vdots&\vdots&\vdots&\vdots\\
\frac{1}{(f_{1,1}-\alpha_{s_{12}})^4}&\cdots&\frac{1}{f_{1,1}-\alpha_{s_{12}}}&\frac{1}{(f_{1,2}-\alpha_{s_{12}})^4}&\cdots&\frac{1}{f_{1,2}-\alpha_{s_{12}}}&1&\alpha_{s_{12}}&\cdots&\alpha_{s_{12}}^3\\
    \end{array}\right]}_{\hat{\mathbf{V}}_{1,2,4,12}}\notag\\
    &\underbrace{\left[\begin{array}{c;{4pt/4pt}c;{4pt/4pt}c}
    \mathbf{T}(c_{1,1,0},\cdots,c_{1,1,3})&&\\\hdashline[4pt/4pt]
    &\mathbf{T}(c_{1,2,0},\cdots,c_{1,2,3})&\\\hdashline[4pt/4pt]
    &&\mathbf{I}_4\\
    \end{array}\right]}_{\hat{\mathbf{V}}'_{1,2,4,12}}\otimes\mathbf{I}_{\lambda/m}\left[\begin{array}{c}
    \mathbf{A}_{1,1}^{1,1}\mathbf{B}_{1,1}^{1,1}\\
    \mathbf{A}_{1,1}^{2,1}\mathbf{B}_{1,1}^{1,1}\\
    \mathbf{A}_{1,1}^{1,1}\mathbf{B}_{1,1}^{1,2}\\
    \mathbf{A}_{1,1}^{2,1}\mathbf{B}_{1,1}^{1,2}\\\hdashline[4pt/4pt]
    \mathbf{A}_{1,2}^{1,1}\mathbf{B}_{1,2}^{1,1}\\
    \mathbf{A}_{1,2}^{2,1}\mathbf{B}_{1,2}^{1,1}\\
    \mathbf{A}_{1,2}^{1,1}\mathbf{B}_{1,2}^{1,2}\\
    \mathbf{A}_{1,2}^{2,1}\mathbf{B}_{1,2}^{1,2}\\\hdashline[4pt/4pt]
    *\\
    \vdots\\
    *
    \end{array}\right],
\end{align}
where we have used $*$ to represent various combinations of interference symbols that can be found explicity by exapnding \eqref{eq:vcsaex11}, whose exact forms are irrelevant. Note that the matrix $\hat{\mathbf{V}}'_{1,2,4,12}$ is a block diagonal matrix composed with two lower triangular toeplitz matrices and an identity matrix, thus is invertible, and the matrix $\hat{\mathbf{V}}_{1,2,4,12}\hat{\mathbf{V}}'_{1,2,4,12}\otimes\mathbf{I}_{\lambda/m}$ is then invertible from Lemma \ref{lemma:ccv} and the fact that the Kronecker product of invertible matrices is invertible. Therefore, the user is able to recover desired products from the answers of any $12$ servers by inverting the matrix. This completes the proof of recovery threshold $R=12$. Finally, it is straightforward to verify that the upload cost is $U_A=S/4=S/(K_cpm)$,  $U_B=S/4=S/(K_cpn)$, and the download cost is $D=12/8=3/2$, which matches Theorem \ref{thm:gcsa}.

\subsubsection{$\ell=1, K_c=2, L=2, p=2, m=n=1$}
Let $f_{1,1}, f_{1,2}, \alpha_{1}, \alpha_{2},\dots,\alpha_{S}$ represent $(S+2)$ distinct elements from $\mathbb{F}$. For all $s\in[S]$, define,
\begin{equation}
    \Delta_{s}^{1,2}=(f_{1,1}-\alpha_s)^2(f_{1,2}-\alpha_s)^2.
\end{equation}
We set $\mathbf{A}_{1,1}=\mathbf{A}_1$,~ $\mathbf{A}_{1,2}=\mathbf{A}_2$,~ $\mathbf{B}_{1,1}=\mathbf{B}_1$ and $\mathbf{B}_{1,2}=\mathbf{B}_2$. Besides, we partition each of the matrices $\mathbf{A}_{1,1}$ and $\mathbf{A}_{1,2}$ into $1\times2$ blocks, denoted as $\mathbf{A}_{1,1}^{1,1}$, $\mathbf{A}_{1,1}^{1,2}$ and $\mathbf{A}_{1,2}^{1,1}$, $\mathbf{A}_{1,2}^{1,2}$ respectively. Similarly, we partition each of the matrices $\mathbf{B}_{1,1}$ and $\mathbf{B}_{1,2}$ into $2\times1$ blocks, denoted as $\mathbf{B}_{1,1}^{1,1}$, $\mathbf{B}_{1,1}^{2,1}$ and $\mathbf{B}_{1,2}^{1,1}$, $\mathbf{B}_{1,2}^{2,1}$ respectively. Note that the desired products $\mathbf{A}_{1,1}\mathbf{B}_{1,1},\mathbf{A}_{1,2}\mathbf{B}_{1,2}$ can be written as follows.
\begin{align}
\mathbf{A}_{1,1}\mathbf{B}_{1,1}&=\mathbf{A}_{1,1}^{1,1}\mathbf{B}_{1,1}^{1,1}+\mathbf{A}_{1,1}^{1,2}\mathbf{B}_{1,1}^{2,1},\\
\mathbf{A}_{1,2}\mathbf{B}_{1,2}&=\mathbf{A}_{1,2}^{1,1}\mathbf{B}_{1,2}^{1,1}+\mathbf{A}_{1,2}^{1,2}\mathbf{B}_{1,2}^{2,1}.
\end{align}
 Shares of matrices $\mathbf{A}$ are constructed as follows.
\begin{align}
    \widetilde{A}^s&=\Delta_{s}^{1,2}\left(\frac{1}{(f_{1,1}-\alpha_s)^2}\left(\mathbf{A}_{1,1}^{1,1}+(f_{1,1}-\alpha_s)\mathbf{A}_{1,1}^{1,2}\right)+\frac{1}{(f_{1,2}-\alpha_s)^2}\left(\mathbf{A}_{1,2}^{1,1}+(f_{1,2}-\alpha_s)\mathbf{A}_{1,2}^{1,2}\right)\right)\\
    &=(f_{1,2}-\alpha_s)^2\underbrace{\left(\mathbf{A}_{1,1}^{1,1}+(f_{1,1}-\alpha_s)\mathbf{A}_{1,1}^{1,2}\right)}_{P_s^{1,1}}+(f_{1,1}-\alpha_s)^2\underbrace{\left(\mathbf{A}_{1,2}^{1,1}+(f_{1,2}-\alpha_s)\mathbf{A}_{1,2}^{1,2}\right)}_{P_s^{1,2}}.
\end{align}
Note that now the term $P_s^{1,1}$ follows the construction of Entangled Polynomial codes of parameter $m=n=1, p=2$, and it is a polynomial of $(f_{1,1}-\alpha_s)$. Similarly, the term $P_s^{1,2}$ follows the construction of Entangled Polynomial codes, and it is a polynomial of $(f_{1,2}-\alpha_s)$. Shares of matrices $\mathbf{B}$ are constructed as follows.
\begin{align}
    \widetilde{B}^s=\frac{1}{(f_{1,1}-\alpha_s)^2}\underbrace{\left((f_{1,1}-\alpha_s)\mathbf{B}_{1,1}^{1,1}+\mathbf{B}_{1,1}^{2,1}\right)}_{Q_s^{1,1}}+\frac{1}{(f_{1,2}-\alpha_s)^2}\underbrace{\left((f_{1,2}-\alpha_s)\mathbf{B}_{1,2}^{1,1}+\mathbf{B}_{1,2}^{2,1}\right)}_{Q_s^{1,2}}.
\end{align}
The terms $Q_s^{1,1}$ and $Q_s^{1,2}$ also follow the construction of EP codes for the given parameter values $p,m,n$, and they are polynomials of $(f_{1,1}-\alpha_s)$ and $(f_{1,2}-\alpha_s)$ respectively.

The answer from the $s^{th}$ server,  $Y_s$ is constructed as $Y_s=\widetilde{A}^s\widetilde{B}^s$. To see why it is possible to recover the desired products from the answers of any $R=7$ servers, let us rewrite $Y_s$ as follows.
\begin{align}
    Y_s&=\widetilde{A}^s\widetilde{B}^s\\
    &=\frac{(f_{1,2}-\alpha_s)^2}{(f_{1,1}-\alpha_s)^2}P_s^{1,1}Q_s^{1,1}+\frac{(f_{1,1}-\alpha_s)^2}{(f_{1,2}-\alpha_s)^2}P_s^{1,2}Q_s^{1,2}+(P_s^{1,1}Q_s^{1,2}+P_s^{1,2}Q_s^{1,1})\\
    &=\frac{((f_{1,1}-\alpha_s)+(f_{1,2}-f_{1,1}))^2}{(f_{1,1}-\alpha_s)^2}P_s^{1,1}Q_s^{1,1}+\frac{((f_{1,2}-\alpha_s)+(f_{1,1}-f_{1,2}))^2}{(f_{1,2}-\alpha_s)^2}P_s^{1,2}Q_s^{1,2}\notag\\
    &\quad\quad+(P_s^{1,1}Q_s^{1,2}+P_s^{1,2}Q_s^{1,1})\\
    &=\left(\frac{c_{1,1,0}}{(f_{1,1}-\alpha_s)^2}+\frac{c_{1,1,1}}{f_{1,1}-\alpha_s}\right)P_s^{1,1}Q_s^{1,1}+\left(\frac{c_{1,2,0}}{(f_{1,2}-\alpha_s)^2}+\frac{c_{1,2,1}}{f_{1,2}-\alpha_s}\right)P_s^{1,2}Q_s^{1,2}\notag\\
    &\quad\quad+(P_s^{1,1}Q_s^{1,1}+P_s^{1,2}Q_s^{1,2}+P_s^{1,1}Q_s^{1,2}+P_s^{1,2}Q_s^{1,1}),\label{eq:vcsaex21}
\end{align}
where in the last step, we perform binomial expansion for numerator polynomials. According to the Binomial Theorem, $(c_{1,k,i})_{k\in[2],i\in\{0,1\}}$ are non-zero. Note that
\begin{align}
    P_s^{1,1}Q_s^{1,1}&=\mathbf{A}_{1,1}^{1,1}\mathbf{B}_{1,1}^{2,1}+(f_{1,1}-\alpha_s)(\mathbf{A}_{1,1}^{1,1}\mathbf{B}_{1,1}^{1,1}+\mathbf{A}_{1,1}^{1,2}\mathbf{B}_{1,1}^{2,1})+(f_{1,1}-\alpha_s)^2\mathbf{A}_{1,1}^{1,2}\mathbf{B}_{1,1}^{1,1},\\
    P_s^{1,2}Q_s^{1,2}&=\mathbf{A}_{1,2}^{1,1}\mathbf{B}_{1,2}^{2,1}+(f_{1,2}-\alpha_s)(\mathbf{A}_{1,2}^{1,1}\mathbf{B}_{1,2}^{1,1}+\mathbf{A}_{1,2}^{1,2}\mathbf{B}_{1,2}^{2,1})+(f_{1,2}-\alpha_s)^2\mathbf{A}_{1,2}^{1,2}\mathbf{B}_{1,2}^{1,1}.
\end{align}
Therefore, we can further rewrite the first term in \eqref{eq:vcsaex21} as follows.
\begin{align}
    &\frac{c_{1,1,0}\mathbf{A}_{1,1}^{1,1}\mathbf{B}_{1,1}^{2,1}}{(f_{1,1}-\alpha_s)^2}+\frac{c_{1,1,1}\mathbf{A}_{1,1}^{1,1}\mathbf{B}_{1,1}^{2,1}+c_{1,1,0}(\mathbf{A}_{1,1}^{1,1}\mathbf{B}_{1,1}^{1,1}+\mathbf{A}_{1,1}^{1,2}\mathbf{B}_{1,1}^{2,1})}{f_{1,1}-\alpha_s}\notag\\
    &\quad\quad+(c_{1,1,0}\mathbf{A}_{1,1}^{1,2}\mathbf{B}_{1,1}^{1,1}+c_{1,1,1}(\mathbf{A}_{1,1}^{1,1}\mathbf{B}_{1,1}^{1,1}+\mathbf{A}_{1,1}^{1,2}\mathbf{B}_{1,1}^{2,1}))\notag\\
    &\quad\quad+(f_{1,1}-\alpha_s)(c_{1,1,1}\mathbf{A}_{1,1}^{1,2}\mathbf{B}_{1,1}^{1,1}).\label{eq:vcsaex22}
\end{align}
The second term in \eqref{eq:vcsaex21} can be similarly rewritten. Note that the third term in \eqref{eq:vcsaex21} and the last two terms in \eqref{eq:vcsaex22} can be expanded into weighted sums of the terms $1,\alpha_s,\alpha_s^2$, so in the matrix form, answers from any $7$ servers, whose indices are denoted as $s_1,s_2,\cdots,s_{7}$, can be written as follows.
\begin{align}
    \begin{bmatrix}
    Y_{s_1}\\
    Y_{s_2}\\
    \vdots\\
    Y_{s_{7}}
    \end{bmatrix}&=\underbrace{\left[\begin{array}{cc;{4pt/4pt}cc;{4pt/4pt}ccc}
\frac{1}{(f_{1,1}-\alpha_{s_1})^2}&\frac{1}{f_{1,1}-\alpha_{s_1}}&\frac{1}{(f_{1,2}-\alpha_{s_1})^2}&\frac{1}{f_{1,2}-\alpha_{s_1}}&1&\alpha_{s_1}&\alpha_{s_1}^2\\
\frac{1}{(f_{1,1}-\alpha_{s_2})^2}&\frac{1}{f_{1,1}-\alpha_{s_2}}&\frac{1}{(f_{1,2}-\alpha_{s_2})^2}&\frac{1}{f_{1,2}-\alpha_{s_2}}&1&\alpha_{s_2}&\alpha_{s_2}^2\\
\vdots&\vdots&\vdots&\vdots&\vdots&\vdots&\vdots\\
\frac{1}{(f_{1,1}-\alpha_{s_{7}})^2}&\frac{1}{f_{1,1}-\alpha_{s_{7}}}&\frac{1}{(f_{1,2}-\alpha_{s_{7}})^2}&\frac{1}{f_{1,2}-\alpha_{s_{7}}}&1&\alpha_{s_{7}}&\alpha_{s_{7}}^2\\
    \end{array}\right]}_{\hat{\mathbf{V}}_{1,2,2,7}}\notag\\
    &\underbrace{\left[\begin{array}{c;{4pt/4pt}c;{4pt/4pt}c}
    \mathbf{T}(c_{1,1,0},c_{1,1,1})&&\\\hdashline[4pt/4pt]
    &\mathbf{T}(c_{1,2,0},c_{1,2,1})&\\\hdashline[4pt/4pt]
    &&\mathbf{I}_3\\
    \end{array}\right]}_{\hat{\mathbf{V}}'_{1,2,2,7}}\otimes\mathbf{I}_{\lambda/m}\left[\begin{array}{c}
    \mathbf{A}_{1,1}^{1,1}\mathbf{B}_{1,1}^{2,1}\\
    \mathbf{A}_{1,1}^{1,1}\mathbf{B}_{1,1}^{1,1}+\mathbf{A}_{1,1}^{1,2}\mathbf{B}_{1,1}^{2,1}\\\hdashline[4pt/4pt]
    \mathbf{A}_{1,2}^{1,1}\mathbf{B}_{1,2}^{2,1}\\
    \mathbf{A}_{1,2}^{1,1}\mathbf{B}_{1,2}^{1,1}+\mathbf{A}_{1,2}^{1,2}\mathbf{B}_{1,2}^{2,1}\\\hdashline[4pt/4pt]
    *\\
    *\\
    *
    \end{array}\right],
\end{align}
where we have used $*$ to represent various combinations of interference symbols that can be found explicity by exapnding \eqref{eq:vcsaex21}, whose exact forms are irrelevant. Note that the matrix $\hat{\mathbf{V}}'_{1,2,2,7}$ is a block diagonal matrix composed with two lower triangular toeplitz matrices and an identity matrix, thus is invertible, and the matrix $\hat{\mathbf{V}}_{1,2,2,7}\hat{\mathbf{V}}'_{1,2,2,7}\otimes\mathbf{I}_{\lambda/m}$ is then invertible from Lemma \ref{lemma:ccv} and the fact that the Kronecker product of invertible matrices is invertible. Therefore, the user is able to recover desired products, i.e., $(\mathbf{A}_{1,1}^{1,1}\mathbf{B}_{1,1}^{1,1}+\mathbf{A}_{1,1}^{1,2}\mathbf{B}_{1,1}^{2,1})$ and $(\mathbf{A}_{1,2}^{1,1}\mathbf{B}_{1,2}^{1,1}+\mathbf{A}_{1,2}^{1,2}\mathbf{B}_{1,2}^{2,1})$, from the answers of any $7$ servers by inverting the matrix. This completes the proof of recovery threshold $R=7$. Finally, it is straightforward to verify that the upload cost is $U_A=S/4=S/(K_cpm)$,  $U_B=S/4=S/(K_cpn)$, and the download cost is $D=7/2$, which matches Theorem \ref{thm:gcsa}.

\subsubsection{Arbitrary $(\ell,K_c,p,m,n)$ and $L=\ell K_c$}
Define $R'=pmn$. Let $f_{1,1}$, $f_{1,2}$, $\cdots$, $f_{\ell,K_c}$, $\alpha_1$, $\alpha_2$, $\cdots$, $\alpha_S$ be $(S+L)$ distinct elements from the field $\mathbb{F}$. For all $l\in[\ell],k\in[K_c]$, we define $c_{l,k,i}, i\in\{0,1,\cdots,R'(K_c-1)\}$ to be the coefficients satisfying
\begin{align}
    \Psi_{l,k}(\alpha)=\prod_{k'\in[K_c]\setminus\{k\}}\left(\alpha+(f_{l,k'}-f_{l,k})\right)^{R'}=\sum_{i=0}^{R'(K_c-1)}c_{l,k,i}\alpha^i,
\end{align}
i.e., they are the coefficients of the polynomial $\Psi_{l,k}(\alpha)=\prod_{k'\in[K_c]\setminus\{k\}}\left(\alpha+(f_{l,k'}-f_{l,k})\right)^{R'}$, which is defined here by its roots. Now for all $l\in[\ell], s\in[S]$, let us define
\begin{equation}
    \Delta_s^{\l,K_c}=\prod_{k\in[K_c]}(f_{l,k}-\alpha_s)^{R'}.
\end{equation}
Let us also split the $L=\ell K_c$ instances of $\mathbf{A}$ and $\mathbf{B}$ matrices into $\ell$ groups, i.e.,
\begin{align}
    \mathbf{A}_{l,k}&=\mathbf{A}_{K_c(l-1)+k},\\
    \mathbf{B}_{l,k}&=\mathbf{B}_{K_c(l-1)+k}
\end{align}
for all $l\in[\ell],k\in[K_c]$. Further, for each matrix $\mathbf{A}_{l,k}$, we partition it into $m\times p$ blocks, denoted as $\mathbf{A}_{l,k}^{1,1},\mathbf{A}_{l,k}^{1,2},\cdots,\mathbf{A}_{l,k}^{m,p}$. Similarly, for each matrix $\mathbf{B}_{l,k}$, we partition it into $p\times n$ blocks, denoted as $\mathbf{B}_{l,k}^{1,1},\mathbf{B}_{l,k}^{1,2},\cdots,\mathbf{B}_{l,k}^{p,n}$. Now, for all $l\in[\ell],k\in[K_c]$, let us define
\begin{align}
    P_s^{l,k}&=\sum_{m'\in[m]}\sum_{p'\in[p]}\mathbf{A}_{l,k}^{m',p'}(f_{l,k}-\alpha_s)^{p'-1+p(m'-1)},\\
    Q_s^{l,k}&=\sum_{p'\in[p]}\sum_{n'\in[n]}\mathbf{B}_{l,k}^{p',n'}(f_{l,k}-\alpha_s)^{p-p'+pm(n'-1)},
\end{align}
i.e., we apply EP codes for each $\mathbf{A}_{l,k}$ and  $\mathbf{B}_{l,k}$. Note that the original EP codes can be regarded as polynomials of $\alpha_s$, and here for each $(l,k)$, we construct the EP codes as polynomials of $(f_{l,k}-\alpha_s)$. Now recall that by the construction of EP codes, the product $P_s^{l,k}Q_s^{l,k}$ can be written as weighted sums of the terms $1,(f_{l,k}-\alpha_s),\cdots,(f_{l,k}-\alpha_s)^{R'+p-2}$, i.e.,
\begin{align}
    P_s^{l,k}Q_s^{l,k}=\sum_{i=0}^{R'+p-2}\mathbf{C}_{l,k}^{(i+1)}(f_{l,k}-\alpha_s)^i,
\end{align}
where $\mathbf{C}_{l,k}^{(1)},\mathbf{C}_{l,k}^{(2)},\cdots,\mathbf{C}_{l,k}^{(R'+p-1)}$ are various linear combinations of products of blocks of $\mathbf{A}_{l,k}$ and blocks of $\mathbf{B}_{l,k}$. In particular, the desired product $\mathbf{A}_{l,k}\mathbf{B}_{l,k}$ can be obtained from $\mathbf{C}_{l,k}^{(1)},\cdots,\mathbf{C}_{l,k}^{(R')}$. Now we are ready to formally present the construction of generalized CSA codes. For all $s\in[S]$, let us construct shares of matrices $\mathbf{A}$ and $\mathbf{B}$ at the $s^{th}$ server as follows.
\begin{align}
    \widetilde{A}^s&=(\widetilde{A}_{1}^s,\widetilde{A}_{2}^s,\dots,\widetilde{A}_{\ell}^s),\\
    \widetilde{B}^s&=(\widetilde{B}_{1}^s,\widetilde{B}_{2}^s,\dots,\widetilde{B}_{\ell}^s),
\end{align}
where for $l\in[\ell]$, let us set
\begin{align}
    \widetilde{A}_{l}^s&=\Delta_s^{l,K_c}\sum_{k\in[K_c]}\frac{1}{(f_{l,k}-\alpha_s)^{R'}}P_s^{l,k},\\
    \widetilde{B}_{l}^s&=\sum_{k\in[K_c]}\frac{1}{(f_{l,k}-\alpha_s)^{R'}}Q_s^{l,k}.
\end{align}
The answer returned by the $s^{th}$ server to the user is constructed as follows.
\begin{align}
    Y_s&=\sum_{l\in[\ell]}\widetilde{A}^s_{l}\widetilde{B}^s_{l}.
\end{align}
Now let us prove that the generalized CSA codes are $R=pmn((\ell+1)K_c-1)+p-1$ recoverable. Let us rewrite $Y_s$ as follows.
\begin{align}
    Y_s&=\widetilde{A}^s_{1}\widetilde{B}^s_{1}+\widetilde{A}^s_{2}\widetilde{B}^s_{2}+\dots+\widetilde{A}^s_{\ell}\widetilde{B}^s_{\ell}\\
    &=\sum_{l\in[\ell]}\Delta_s^{l,K_c}\left(\sum_{k\in[K_c]}\frac{1}{(f_{l,k}-\alpha_s)^{R'}}P_s^{l,k}\right)\left(\sum_{k\in[K_c]}\frac{1}{(f_{l,k}-\alpha_s)^{R'}}Q_s^{l,k}\right)\\
    &=\sum_{l\in[\ell]}\sum_{k\in[K_c]}\frac{\prod_{k'\in[K_c]\setminus\{k\}}(f_{l,k'}-\alpha_s)^{R'}}{(f_{l,k}-\alpha_s)^{R'}}P_s^{l,k}Q_s^{l,k}\notag\\
    &\quad\quad+\sum_{l\in[\ell]}\sum_{\substack{k,k'\in[K_c]\\k\neq k'}}\left(\prod_{k''\in[K_c]\setminus\{k,k'\}}(f_{l,k''}-\alpha_s)^{R'}\right)P_s^{l,k}Q_s^{l,k'}.\label{eq:anscorrv}
\end{align}
Note that in the last step, we split the summation into two parts depending on whether or not $k=k'$. 

Let us consider the first term in \eqref{eq:anscorrv}. For each $l\in[\ell], k\in[K_c]$, we have
\begin{align}
    &\frac{\prod_{k'\in[K_c]\setminus\{k\}}(f_{l,k'}-\alpha_s)^{R'}}{(f_{l,k}-\alpha_s)^{R'}}P_s^{l,k}Q_s^{l,k}\\
    =&\frac{\prod_{k'\in[K_c]\setminus\{k\}}\left((f_{l,k}-\alpha_s)+(f_{l,k'}-f_{l,k})\right)^{R'}}{(f_{l,k}-\alpha_s)^{R'}}P_s^{l,k}Q_s^{l,k}\\
    =&\frac{\Psi_{l,k}(f_{l,k}-\alpha_s)}{(f_{l,k}-\alpha_s)^{R'}}P_s^{l,k}Q_s^{l,k}\label{eq:vcsagalpha}\\
    =&\left(\frac{c_{l,k,0}}{(f_{l,k}-\alpha_s)^{R'}}+\frac{c_{l,k,1}}{(f_{l,k}-\alpha_s)^{R'-1}}+\cdots+\frac{c_{l,k,R'-1}}{f_{l,k}-\alpha_s}\right)P_s^{l,k}Q_s^{l,k}\notag\\
    &\quad\quad+\left(\sum_{i=R'}^{R'(K_c-1)}c_{l,k,i}(f_{l,k}-\alpha_s)^{i-R'}\right)P_s^{l,k}Q_s^{l,k},\label{eq:anscorrv1}
\end{align}
where in \eqref{eq:vcsagalpha}, we used the definition of $\Psi_{l,k}(\cdot)$, and in the next step, we rewrite the polynomial $\Psi_{l,k}(f_{l,k}-\alpha_s)$ in terms of its coefficients. Let us consider the first term in \eqref{eq:anscorrv1}.
\begin{align}
    &\left(\frac{c_{l,k,0}}{(f_{l,k}-\alpha_s)^{R'}}+\frac{c_{l,k,1}}{(f_{l,k}-\alpha_s)^{R'-1}}+\cdots+\frac{c_{l,k,R'-1}}{f_{l,k}-\alpha_s}\right)P_s^{l,k}Q_s^{l,k}\\
    =&\left(\frac{c_{l,k,0}}{(f_{l,k}-\alpha_s)^{R'}}+\frac{c_{l,k,1}}{(f_{l,k}-\alpha_s)^{R'-1}}+\cdots+\frac{c_{l,k,R'-1}}{f_{l,k}-\alpha_s}\right)\sum_{i=0}^{R'+p-2}\mathbf{C}_{l,k}^{(i+1)}(f_{l,k}-\alpha_s)^i\\
    =&\sum_{i=0}^{R'-1}\frac{\sum_{i'=0}^{i}c_{l,k,i-i'}\mathbf{C}_{l,k}^{(i'+1)}}{(f_{l,k}-\alpha_s)^{R'-i}}+\sum_{i=0}^{p-2}(f_{l,k}-\alpha_s)^i\left(\sum_{i'=i+1}^{R'+i}c_{l,k,R'-i'+i}\mathbf{C}_{l,k}^{(i'+1)}\right)\notag\\
    &\quad\quad+\sum_{i=p-1}^{R'+p-3}(f_{l,k}-\alpha_s)^i\left(\sum_{i'=i+1}^{R'+p-2}c_{l,k,R'-i'+i}\mathbf{C}_{l,k}^{(i'+1)}\right)\label{eq:anscorrv2}.
\end{align}
We further note that when $K_c=1$, for all $i\neq 0, c_{l,k,i}=0$, thus the second term in \eqref{eq:anscorrv}, the second term in \eqref{eq:anscorrv1} and the third term in \eqref{eq:anscorrv2} equal zero. The second term in \eqref{eq:anscorrv2} can be expanded\footnote{ When $K_c=p=1$, the second term in \eqref{eq:anscorrv2} is zero, thus the Vandermonde terms do not appear. The matrix form representation now involves only confluent Cauchy matrices, i.e., confluent Cauchy-Vandermonde matrices without Vandermonde part. } into weighted sums of the terms $1,\alpha_s,\cdots,\alpha_s^{p-2}$. Since $K_c=1$, we can equivalently write these terms as $1,\alpha_s,\cdots,\alpha_s^{R'(K_c-1)+p-2}$. On the other hand, when $K_c>1$, the second term in \eqref{eq:anscorrv}, the second term in \eqref{eq:anscorrv1}, the second and the third terms in \eqref{eq:anscorrv2} can also be expanded into weighted sums of the terms $1,\alpha_s,\cdots,\alpha_s^{R'(K_c-1)+p-2}$. Because  $R'(K_c-1)+p-2=R-R' L-1$, in the matrix form, answers from any $R=pmn((\ell+1)K_c-1)+p-1$ servers, whose indices are denoted as $s_1,s_2,\cdots,s_R$, can be written as follows.
\begin{align}
    \begin{bmatrix}
    Y_{s_1}\\
    Y_{s_2}\\
    \vdots\\
    Y_{s_R}
    \end{bmatrix}&=\underbrace{\left[
\begin{array}{ccc;{4pt/4pt}c;{4pt/4pt}ccc;{4pt/4pt}ccc}
\frac{1}{(f_{1,1}-\alpha_{s_1})^{R'}}&\cdots&\frac{1}{f_{1,1}-\alpha_{s_1}}&\cdots&\frac{1}{(f_{\ell,K_c}-\alpha_{s_1})^{R'}}&\cdots&\frac{1}{f_{\ell,K_c}-\alpha_{s_1}}&1&\cdots&\alpha_{s_1}^{R-{R'}L-1}\\
\frac{1}{(f_{1,1}-\alpha_{s_2})^{R'}}&\cdots&\frac{1}{f_{1,1}-\alpha_{s_2}}&\cdots&\frac{1}{(f_{\ell,K_c}-\alpha_{s_2})^{R'}}&\cdots&\frac{1}{f_{\ell,K_c}-\alpha_{s_2}}&1&\cdots&\alpha_{s_2}^{R-{R'}L-1}\\
\vdots&\vdots&\vdots&\vdots&\vdots&\vdots&\vdots&\vdots&\vdots&\vdots\\
\frac{1}{(f_{1,1}-\alpha_{s_R})^{R'}}&\cdots&\frac{1}{f_{1,1}-\alpha_{s_R}}&\cdots&\frac{1}{(f_{\ell,K_c}-\alpha_{s_R})^{R'}}&\cdots&\frac{1}{f_{\ell,K_c}-\alpha_{s_R}}&1&\cdots&\alpha_{s_R}^{R-{R'}L-1}\\
\end{array}
\right]}_{\hat{\mathbf{V}}_{\ell,K_c,R',R}}\notag\\
    &\underbrace{\left[\begin{array}{c;{4pt/4pt}c;{4pt/4pt}c;{4pt/4pt}c}
    \mathbf{T}(c_{1,1,0},\cdots,c_{1,1,R'-1})&&&\\\hdashline[4pt/4pt]
    &\ddots&&\\\hdashline[4pt/4pt]
    &&\mathbf{T}(c_{\ell,K_c,0},\cdots,c_{\ell,K_c,R'-1})&\\\hdashline[4pt/4pt]
    &&&\mathbf{I}_{R-R' L}
    \end{array}\right]}_{\hat{\mathbf{V}}'_{\ell,K_c,R',R}}\otimes\mathbf{I}_{\lambda/m}\left[\begin{array}{c}
    \mathbf{C}_{1,1}^{(1)}\\
    \vdots\\
    \mathbf{C}_{1,1}^{(R')}\\\hdashline[4pt/4pt]
    \vdots\\\hdashline[4pt/4pt]
    \mathbf{C}_{\ell,K_c}^{(1)}\\
    \vdots\\
    \mathbf{C}_{\ell,K_c}^{(R')}\\\hdashline[4pt/4pt]
    *\\
    \vdots\\
    *
    \end{array}\right],
\end{align}
We have used $*$ to represent various combinations of interference symbols that can be found explicitly by expanding \eqref{eq:anscorrv}, whose exact forms are irrelevant. Now since $f_{1,1}, f_{1,2}, \cdots, f_{\ell,K_c}$ are distinct, for all $l\in[\ell],k\in[K_c]$, we must have 
\begin{equation}
    c_{l,k,0}=\prod_{k'\in[K_c]\setminus\{k\}}(f_{l,k'}-f_{l,k})^{R'}
\end{equation}
are non-zero. Hence, the lower triangular toeplitz matrices $\mathbf{T}(c_{1,1,0},c_{1,1,1},\cdots,c_{1,1,R'-1}),\cdots,\\\mathbf{T}(c_{\ell,K_c,0},c_{\ell,K_c,1},\cdots,c_{\ell,K_c,R'-1})$ are non-singular, and the block diagonal matrix $\hat{\mathbf{V}}'_{\ell,K_c,R',R}$ is invertible. Guaranteed by Lemma \ref{lemma:ccv} and the fact that the Kronecker product of non-singular matrices is non-singular, the matrix $(\hat{\mathbf{V}}_{\ell,K_c,R',R}\hat{\mathbf{V}}'_{\ell,K_c,R',R})\otimes\mathbf{I}_{\lambda/m}$ is invertible. Therefore, the user is able to recover $(\mathbf{C}_{l,k}^{(i)})_{l\in[\ell],k\in[K_c],i\in[R']}$ by inverting the matrix. And the desired products $(\mathbf{A}_l\mathbf{B}_l)_{l\in[L]}$ are recoverable from $(\mathbf{C}_{l,k}^{(i)})_{l\in[\ell],k\in[K_c],i\in[R']}$, guaranteed by the construction of Entangled Polynomial codes. This completes the proof of recovery threshold $R=pmn((\ell+1)K_c-1)+p-1$. It is also easy to see that the upload cost $U_A=S/(K_cpm)$ and $U_B=S/(K_cpn)$. Note that we are able to recover $Lmn$ desired symbols from $R$ downloaded answers, so the download cost is $D=\frac{R}{Lmn}=\frac{pmn((\ell+1)K_c-1)+p-1}{mn\ell K_c}$. Thus the desired costs are achievable. Note that the encoding procedure can be considered as products of Confluent Cauchy matrices by vectors. By fast algorithms \cite{Olshevsky_Shokrollahi}, the encoding complexity of $(\mathcal{C}_{eA},\mathcal{C}_{eB})=\left(\widetilde{\mathcal{O}}\left(\frac{\lambda\kappa S\log^2S}{K_cpm}\right),  \widetilde{\mathcal{O}}\left(\frac{\kappa\mu S\log^2S}{K_cpn}\right)\right)$ is achievable. Now let us consider the decoding complexity. Note that the decoding procedure involves matrix-vector multiplications of inverse of Toeplitz matrix and inverse of confluent Cauchy-Vandermonde matrix. From the inverse formula of confluent Cauchy-Vandermonde matrix presented in \cite{Yang_Hu_GCV}, the matrix-vector multiplication of the inverse of confluent Cauchy-Vandermonde matrix $\hat{\mathbf{V}}_{\ell,K_c,R',R}$ can be decomposed into a series of structured matrix-vector multiplications including confluent Cauchy matrix, transpose of Vandermonde matrix, Hankel matrix and Toeplitz matrix. By fast algorithms \cite{Olshevsky_Shokrollahi,Gohberg_Olshevsky_Fast}, the complexity of decoding is at most $\widetilde{\mathcal{O}}(\lambda\mu p \log^2R)$. With straightforward matrix multiplication algorithms, the server computation complexity is $\mathcal{C}_s=(\lambda\kappa\mu)/(K_cpmn)$.
This completes the proof of Theorem \ref{thm:gcsa}.

\section{$N$-CSA Codes for $N$-linear Coded Distributed Batch Computation ($N$-CDBC)}\label{sec:ncsa}
\subsection{$N$-CSA Codes: Main Result}
In this section, let us generalize CSA codes for $N$-CDBC. The generalization, called $N$-CSA codes, is presented in the following theorem.
\begin{theorem}\label{thm:nlcsa}
For $N$-CDBC over a field $\mathbb{F}$ with $S$ servers, and positive integers $\ell, K_c$ such that $L=\ell K_c\leq|\mathbb{F}|-S$, the $N$-CSA codes introduced in this section achieve
\begin{align}
   \text{Recovery Threshold:}&& R&=K_c(N+\ell-1)-N+1,\\
     \text{Upload Cost for $\widetilde{X^{(n)}}^{[S]}, n\in[N]$:}&&U_{X^{(n)}}&=\frac{S}{K_c},\\
    \text{Download Cost:} &&D&=\frac{K_c(N+\ell-1)-N+1}{\ell K_c},\\
    \text{Server Computation Complexity:}&& \mathcal{C}_s&=\mathcal{O}(\omega/K_c),\\
    \text{Encoding complexity for $\widetilde{X^{(n)}}^{[S]}$, $n\in[N]$:} &&\mathcal{C}_{eX^{(n)}}&=\widetilde{\mathcal{O}}\left(\frac{\dim (V_n)S\log^2 S}{K_c}\right),\\
    \text{Decoding complexity:} &&\mathcal{C}_{d}&=\widetilde{\mathcal{O}}\left(\frac{\ell+N-1}{\ell}\dim(W)R\log^2R\right),
\end{align}
where $\omega$ is the number of arithmetic operations required to compute the $N$-linear map $\Omega(\cdot)$.
\end{theorem}
\subsection{Proof of Theorem \ref{thm:nlcsa}}
Now let us present the construction of $N$-CSA codes for $N$-CDBC. Let $f_{1,1}, f_{1,2},\cdots,f_{\ell,K_c},\alpha_1,\alpha_2,\\\cdots,\alpha_S$ represent $(S+L)$ distinct elements from $\mathbb{F}$. For all $l\in[\ell], s\in[S]$, let us define
\begin{equation}
    \Delta_s^{\l,K_c}=\prod_{k\in[K_c]}(f_{l,k}-\alpha_s).
\end{equation}
For all $n\in[N],l\in[\ell],k\in[K_c]$, we define
\begin{align}
    x_{l,k}^{(n)}=x^{(n)}_{K_c(l-1)+k}.
\end{align}
For all $s\in[S],n\in[N]$, we construct $\widetilde{X^{(n)}}^{s}$ as follows.
\begin{align}
    \widetilde{X^{(n)}}^{s}=(\widetilde{X^{(n)}}^{s}_1,\widetilde{X^{(n)}}^{s}_2,\cdots,\widetilde{X^{(n)}}^{s}_{\ell}),
\end{align}
where for $l\in[\ell]$, let us set
\begin{align}
    \widetilde{X^{(n)}}^{s}_l=\Delta_s^{\l,K_c}\sum_{k\in[K_c]}\frac{1}{f_{l,k}-\alpha_s}x^{(n)}_{l,k}.
\end{align}
The answer returned by the $s^{th}$ server is constructed as follows.
\begin{align}
    Y_s=\sum_{l\in[\ell]}\frac{1}{\Delta_s^{\l,K_c}}\Omega(\widetilde{X^{(1)}}^{s}_l,\widetilde{X^{(2)}}^{s}_l,\cdots,\widetilde{X^{(N)}}^{s}_l).
\end{align}
To prove that the code is $R$-recoverable, let us rewrite $Y_s$ as follows.
\begin{align}
    Y_s&=\sum_{l\in[\ell]}\frac{1}{\Delta_s^{\l,K_c}}\Omega(\widetilde{X^{(1)}}^{s}_l,\widetilde{X^{(2)}}^{s}_l,\cdots,\widetilde{X^{(N)}}^{s}_l)\\
    &=\sum_{l\in[\ell]}\frac{1}{\Delta_s^{\l,K_c}}\Omega\left(\Delta_s^{\l,K_c}\sum_{k\in[K_c]}\frac{1}{f_{l,k}-\alpha_s}x^{(1)}_{l,k},\cdots,\Delta_s^{\l,K_c}\sum_{k\in[K_c]}\frac{1}{f_{l,k}-\alpha_s}x^{(N)}_{l,k}\right)\\
    &=\sum_{l\in[\ell]}(\Delta_s^{\l,K_c})^{N-1}\left(\sum_{k_1\in[K_c]}\frac{1}{f_{l,k_1}-\alpha_s}\cdots\sum_{k_N\in[K_c]}\frac{1}{f_{l,k_N}-\alpha_s}\left(\Omega(x^{(1)}_{l,k_1},\cdots,x^{(N)}_{l,k_N})\right)\right)\\
    &=\sum_{l\in[\ell]}\sum_{k\in[K_c]}\frac{\prod_{k'\in[K_c]\setminus\{k\}}(f_{l,k'}-\alpha_s)^{N-1}}{(f_{l,k}-\alpha_s)}\Omega(x^{(1)}_{l,k},\cdots,x^{(N)}_{l,k})\notag\\
    &\quad\quad+\sum_{l\in[\ell]}\sum_{\substack{k_1,\cdots,k_N\in[K_c],\\\neg(k_1=\cdots=k_N)}}\left(\frac{(\Delta_s^{\l,K_c})^{N-1}}{(f_{l,k_1}-\alpha_s)\cdots(f_{l,k_N}-\alpha_s)}\Omega(x^{(1)}_{l,k_1},\cdots,x^{(N)}_{l,k_N})\right),\label{eq:anscorrnl}
\end{align}
where in \eqref{eq:anscorrnl}, we split the summation depending on whether or not $k_1=k_2=\cdots=k_N$. Following the same argument presented in Section \ref{sec:proof}, by performing long division of polynomials for the first term in \eqref{eq:anscorrnl}, and noting that the second term in \eqref{eq:anscorrnl} can be expanded to weighted sums of the terms $1,\alpha_s,\alpha_s^2,\cdots,\alpha_s^{K_c(N-1)-N}$, the presented code is $(R=K_c(N+\ell-1)-N+1)$-recoverable as long as the following matrix is non-singular.
\begin{align}
    \underbrace{\begin{bmatrix}
\frac{1}{f_{1,1}-\alpha_{s_1}}&\frac{1}{f_{1,2}-\alpha_{s_1}}&\cdots&\frac{1}{f_{\ell,K_c}-\alpha_{s_1}}&1&\alpha_{s_1}&\cdots&\alpha_{s_1}^{R-L-1}\\
\frac{1}{f_{1,1}-\alpha_{s_2}}&\frac{1}{f_{1,2}-\alpha_{s_2}}&\cdots&\frac{1}{f_{\ell,K_c}-\alpha_{s_2}}&1&\alpha_{s_2}&\cdots&\alpha_{s_2}^{R-L-1}\\
\vdots&\vdots&\vdots&\vdots&\vdots&\vdots&\vdots&\vdots\\
\frac{1}{f_{1,1}-\alpha_{s_R}}&\frac{1}{f_{1,2}-\alpha_{s_R}}&\cdots&\frac{1}{f_{\ell,K_c}-\alpha_{s_R}}&1&\alpha_{s_R}&\cdots&\alpha_{s_R}^{R-L-1}\\
    \end{bmatrix}}_{\mathbf{V}_{\ell,K_c,R}}\underbrace{\begin{bmatrix}
    c_{1,1}&&&&&&\\
    &c_{1,2}&&&&&\\
    &&\ddots&&&&\\
    &&&c_{\ell,K_c}&&&\\
    &&&&1&&\\
    &&&&&\ddots&\\
    &&&&&&1
    \end{bmatrix}}_{\mathbf{V}'_{\ell,K_c,R}},\label{eq:dmCDBC}
\end{align}
where for all $l\in[\ell],k\in[K_c]$, $c_{l,k}=\prod_{k'\in[K_c]\setminus\{k\}}(f_{l,k'}-f_{l,k})^{N-1}$. The indices of any $R$ responsive servers are denoted as $s_1,s_2,\cdots,s_R$. Since $f_{1,1},f_{1,2},\cdots,f_{l,k}$ are distinct elements from $\mathbb{F}$, $(c_{l,k})_{l\in[\ell],k\in[K_c]}$ are non-zero, and $R-L-1=K_c(N-1)-N$, the matrix $\mathbf{V}_{\ell,K_c,R}\mathbf{V}'_{\ell,K_c,R}$ is invertible guaranteed by Lemma \ref{lemma:csa}. This completes the proof of recovery threshold. The upload cost for $\widetilde{X^{(n)}}^{[S]}, n\in[N]$ is readily verified to be $S/K_c$, and the download cost is $D=R/L=\frac{K_c(N+\ell-1)-N+1}{\ell K_c}$. By fast algorithms discussed in Section \ref{sec:proof}, we can achieve the encoding/decoding complexity as presented in Theorem \ref{thm:nlcsa}. The computational complexity at each server is $\mathcal{O}(\ell\omega/L)=\mathcal{O}(\omega/K_c)$, where $\omega$ is the number of arithmetic operations required to compute $\Omega(\cdot)$. This completes the proof of Theorem \ref{thm:nlcsa}.

{\it Remark 1: }Let us regard a multivariate polynomial of total degree $N$ as a linear combination of various restricted evaluations of $N$-linear maps. Note that the construction for $\widetilde{X^{(n)}}^s$ is symmetric across all $n\in[N]$. $N$-CSA codes can also be transformed to evaluate a multivariate polynomial at $L$ points as follows. For each server $s\in[S]$, the answer is computed for each $N$-linear map according to $N$-CSA codes, and each server returns the user with the linear combination of the answers. It is easy to see that the user is able to recover the evaluation of the multivariate polynomial of total degree $N$ at the given $L$ points from answers of any $R=K_c(N+\ell-1)-N+1$ servers. The LCC codes in \cite{Yu_Lagrange}, which achieve the recovery threshold $R=K_cN-N+1$, are a special case of this construction, where $\ell=1$.

{\it Remark 2: }The systematic construction presented in Section \ref{sec:systematic} can be also applied directly to $N$-CSA codes for $N$-CDBC, i.e., for all $s\in[L]$, uncoded variables $(x^{(n)}_s)_{n\in[N]}$ are uploaded to the $s^{th}$ server, and coded shares are uploaded to the remaining $S-L$ servers, according to the same coding scheme. Similarly, the recovery threshold is not affected by the systematic construction.

{\it Remark 3: }The Lagrange codes presented in \cite{Yu_Lagrange} for $N$-CDBC and can be considered as a special case of $N$-CSA codes obtained by setting the parameter $\ell=1$. Note that the download cost can be written as $D=1+\left(\frac{N-1}{\ell}\right)\left(\frac{K_c-1}{K_c}\right)$. The parameter $\ell$ plays an important role in improving the download cost, which may be of interest when $N$ is large and the down-link is costly. For example, let us assume that $R/S$ is held constant, then the order of the download cost achieved is $\mathcal{O}(1+(N-1)/\ell)$ and the order of the upload cost for $\widetilde{X^{(n)}}^{[S]}, n\in[N]$ achieved is $\mathcal{O}(\ell+(N-1))$, which offers flexible trade-off between the upload cost and download cost.

\section{Conclusion}\label{sec:conc}
The main contribution of this work is a class of codes, based on the idea of Cross Subspace Alignment (CSA) that originated in private information retrieval (PIR) literature. These codes are shown to unify, generalize and improve upon existing algorithms for coded distributed batch matrix multiplication, $N$-linear batch computation, and multivariate batch polynomial evaluation, such as  Polynomial, MatDot and PolyDot codes, Generalized PolyDot and Entangled Polynomial (EP) codes, and Lagrange Coded Computing (LCC). CSA codes for coded distributed batch matrix multiplication, which include LCC codes as a special case, improve significantly upon state of art matrix-partitioning approaches (EP codes)  in terms of communication cost, and upon LCC codes in download-constrained settings. Generalized CSA (GCSA) codes bridge the extremes of matrix partitioning based approaches (EP codes) and batch processing approaches (CSA codes, LCC codes), and allow a tradeoff between server computation complexity, which is improved by emphasizing the matrix partitioning aspect, and communication costs, which are improved by emphasizing the batch processing aspect. $N$-CSA codes for $N$-linear batch computations and multivariate polynomial evaluations similarly generalize LCC codes, offering advantages especially in download constrained settings. As a final observation, note that LCC codes in \cite{Yu_Lagrange} also allow settings with $X$-secure data and $B$-byzantine servers. Given that cross-subspace alignment schemes originated in PIR with $X$-security constraints \cite{Jia_Sun_Jafar_XSTPIR} and have also been applied to $B$-byzantine settings in \cite{Jia_Jafar_MDSXSTPIR}, extensions of CSA codes, GCSA codes and $N$-CSA codes to $X$-secure and $B$-byzantine settings are relatively straightforward, as shown in Appendix \ref{sec:symxs}. An interesting direction for future work is the possibility of task partitioning (similar to matrix partitioning) for $N$-CSA codes to reduce the computation cost per server in settings where latency constraints prevent any server from fully computing the $N$-linear map, or the multivariate polynomial evaluation by itself.

\appendix

\section{$N$-CSA Codes for $X$-secure $B$-byzantine $N$-linear Coded Distributed Batch Computation}\label{sec:symxs}
Let us consider the problem of $X$-secure $B$-byzantine $N$-linear coded distributed batch computation (XSBNCDBC) over a finite field $\mathbb{F}_q$, where the shares $\widetilde{X^{(n)}}^{[S]},n\in[N]$ are coded in an $X$-secure fashion, i.e.,
any $X$ colluding servers learn nothing about the data, $x^{(n)}_{[L]}$. Formally, we have
\begin{equation}
    I\left(\widetilde{X^{(n)}}^{\mathcal{X}};x^{(n)}_{[L]}\right)=0,\quad \forall\mathcal{X}\subset[S],|\mathcal{X}|=X,n\in[N].
\end{equation}
Furthermore, we assume that there exists a set of servers $\mathcal{B}$, $\mathcal{B}\subset[S]$, $|\mathcal{B}|
\leq B$, known as Byzantine servers. The user knows
the number of Byzantine servers $B$ but the realization of the set $\mathcal{B}$ is not known to the user \emph{apriori}. The Byzantine
servers respond to the user arbitrarily, possibly introducing errors. However, the remaining servers, i.e., all servers $s\in[S]\setminus\mathcal{B}$, if they respond at all, respond  truthfully with the function $h_s$. We will follow the problem statement and definitions of $N$-CDBC in all other aspects. The goal in this section is to present a generalized $N$-CSA codes construction for XSBNCDBC, which achieves the recovery threshold $R=K_c(N+\ell-1)+N(X-1)+2B+1$. To construct $N$-CSA codes for XSBNCDBC, let  $f_{1,1}, f_{1,2},\cdots,f_{\ell,K_c}$ and $\alpha_1,\alpha_2,\cdots,\alpha_S$ be $(S+L)$ distinct elements from $\mathbb{F}_q$, where $q\geq S+L$. For all $n\in[N]$, let $(z^{(n)}_{l,k,x})_{l\in[\ell],k\in[K_c],x\in[X]}$ be independent uniformly random noise vectors from $V_n$, that are used to guarantee the security. The independence between data and random noise symbols is specified as follows.
\begin{align}
    H(\mathbf{x}_{[L]},(z^{(n)}_{l,k,x})_{n\in[N],l\in[\ell],k\in[K_c],x\in[X]})=H(\mathbf{x}_{[L]})+\sum_{\substack{n\in[N],l\in[\ell],\\k\in[K_c],x\in[X]}}H(z^{(n)}_{l,k,x}).\label{eq:symind}
\end{align}
For all $l\in[\ell], s\in[S]$, let us define
\begin{equation}
    \Delta_s^{\l,K_c}=\prod_{k\in[K_c]}(f_{l,k}-\alpha_s).
\end{equation}
For all $n\in[N],l\in[\ell],k\in[K_c]$, we define
\begin{align}
    x_{l,k}^{(n)}=x^{(n)}_{K_c(l-1)+k}.
\end{align}
For all $s\in[S],n\in[N]$, we construct $\widetilde{X^{(n)}}^{s}$ as follows.
\begin{align}
    \widetilde{X^{(n)}}^{s}=(\widetilde{X^{(n)}}^{s}_1,\widetilde{X^{(n)}}^{s}_2,\cdots,\widetilde{X^{(n)}}^{s}_{\ell}),
\end{align}
where for $l\in[\ell]$, let us set
\begin{align}
    \widetilde{X^{(n)}}^{s}_l=\Delta_s^{\l,K_c}\left(\sum_{k\in[K_c]}\frac{1}{f_{l,k}-\alpha_s}x^{(n)}_{l,k}+\sum_{x\in[X]}\alpha_s^{x-1}z^{(n)}_{l,k,x}\right).
\end{align}
Now it is readily seen that the $X$-security of data is guaranteed by the i.i.d. and  uniformly distributed noise terms, i.e., $(z^{(n)}_{l,k,x})_{n\in[N],l\in[\ell],k\in[K_c],x\in[X]}$ that are coded according to an MDS$(X,S)$ code (a Reed-Solomon code). The answer returned by the $s^{th}$ server is constructed as follows.
\begin{align}
    Y_s=\sum_{l\in[\ell]}\frac{1}{\Delta_s^{\l,K_c}}\Omega(\widetilde{X^{(1)}}^{s}_l,\widetilde{X^{(2)}}^{s}_l,\cdots,\widetilde{X^{(N)}}^{s}_l).
\end{align}
Now let us see why it is possible to recover the desired evaluations from the answers of any $R=K_c(N+\ell-1)+N(X-1)+1$ servers. Note that $Y_s$ can be rewritten as follows.
\begin{align}
    Y_s&=\sum_{l\in[\ell]}\frac{1}{\Delta_s^{\l,K_c}}\Omega(\widetilde{X^{(1)}}^{s}_l,\widetilde{X^{(2)}}^{s}_l,\cdots,\widetilde{X^{(N)}}^{s}_l)\\
    &=\sum_{l\in[\ell]}(\Delta_s^{\l,K_c})^{N-1}\Omega\left(\sum_{k\in[K_c]}\frac{1}{f_{l,k}-\alpha_s}x^{(1)}_{l,k}+\sum_{x\in[X]}\alpha_s^{x-1}z^{(1)}_{l,k,x},\cdots,\right.\notag\\
    &\hspace{5cm},\cdots,\left.\sum_{k\in[K_c]}\frac{1}{f_{l,k}-\alpha_s}x^{(N)}_{l,k}+\sum_{x\in[X]}\alpha_s^{x-1}z^{(N)}_{l,k,x}\right)\label{eq:symans1}\\
    &=\sum_{l\in[\ell]}\sum_{k\in[K_c]}\frac{\prod_{k'\in[K_c]\setminus\{k\}}(f_{l,k'}-\alpha_s)^{N-1}}{(f_{l,k}-\alpha_s)}\Omega(x^{(1)}_{l,k},\cdots,x^{(N)}_{l,k})+\sum_{i\in[(K_c-1)(N-1)+NX]}\alpha_s^{i-1}I_i\label{eq:symans2}.
\end{align}
In \eqref{eq:symans2}, we rewrite \eqref{eq:symans1} following the same argument that we used in Section \ref{sec:proof}. Note that $I_i, i\in[(K_c-1)(N-1)+NX]$ represent various linear combinations of $\Omega(\cdot)$, which can be found explicitly by expanding  \eqref{eq:symans1}. Their exact forms are irrelevant, hence omitted for ease of exposition. Now we can see that the answers from any $R=K_c(N+\ell-1)+N(X-1)+2B+1$ servers, whose indices are denoted as $s_1,s_2,\cdots,s_R$, are coded according to the following $R\times (R-2B)$ generator matrix of an MDS$(R-2B,R)$ code.
\begin{align}
    \begin{bmatrix}
\frac{1}{f_{1,1}-\alpha_{s_1}}&\frac{1}{f_{1,2}-\alpha_{s_1}}&\cdots&\frac{1}{f_{\ell,K_c}-\alpha_{s_1}}&1&\alpha_{s_1}&\cdots&\alpha_{s_1}^{R-2B-L-1}\\
\frac{1}{f_{1,1}-\alpha_{s_2}}&\frac{1}{f_{1,2}-\alpha_{s_2}}&\cdots&\frac{1}{f_{\ell,K_c}-\alpha_{s_2}}&1&\alpha_{s_2}&\cdots&\alpha_{s_2}^{R-2B-L-1}\\
\vdots&\vdots&\vdots&\vdots&\vdots&\vdots&\vdots&\vdots\\
\frac{1}{f_{1,1}-\alpha_{s_R}}&\frac{1}{f_{1,2}-\alpha_{s_R}}&\cdots&\frac{1}{f_{\ell,K_c}-\alpha_{s_R}}&1&\alpha_{s_R}&\cdots&\alpha_{s_R}^{R-2B-L-1}\\
    \end{bmatrix}.
\end{align}
Thus the user (decoder) can correct up to $(R-(R-2B))/2=B$ errors in the answers. Upon error correction, the user is able to recover desired evaluations, which appear along the dimensions spanned by the Cauchy part. This completes the proof of recovery threshold $R=K_c(N+\ell-1)+N(X-1)+2B+1$.

{\it Remark 1: }Because of the  $X$-secure constraint, the systematic construction presented in Section \ref{sec:systematic} cannot be applied to $N$-CSA codes for XSBNCDBC.

{\it Remark 2: }GCSA codes for coded distributed batch matrix multiplication presented in Section \ref{sec:gcsa} can similarly be generalized to allow $X$-secure and $B$-Byzantine settings. Such a generalization is straightforward, thus  omitted here.

\bibliographystyle{IEEEtran}
\bibliography{Thesis}
 
\end{document}